# Galactic Chemical Evolution: Hydrogen through Zinc

F. X. Timmes[1,2,3], S. E. Woosley[2,3], and Thomas A. Weaver[3]

[1] Laboratory for Astrophysics and Space Research
Enrico Fermi Institute, University of Chicago
Chicago, IL   60637

[2] Board of Studies in Astronomy and Astrophysics
UCO/Lick Observatory, University of California at Santa Cruz
Santa Cruz, CA   95064

[3] General Studies Division, Lawrence Livermore National Laboratory
Livermore, CA   94550

email contacts: fxt@flame.uchicago.edu

# ABSTRACT

Using the output from a grid of 60 Type II supernova models (Woosley & Weaver 1994) of varying mass ($11 \lesssim M/M_\odot \lesssim 40$) and metallicity (0, $10^{-4}$, 0.01, 0.1, and 1 $Z_\odot$), the chemical evolution of 76 stable isotopes, from hydrogen to zinc, is calculated. The chemical evolution calculation employs a simple dynamical model for the Galaxy (infall with a 4 billion year $e$-folding time scale onto a exponential disk and $1/r^2$ bulge), and standard evolution parameters, such as a Salpeter initial mass function and a quadratic Schmidt star formation rate. The theoretical results are compared in detail with observed stellar abundances in stars with metallicities in the range $-3.0 \lesssim$ [Fe/H] $\lesssim 0.0$ dex. While our discussion focuses on the solar neighborhood where there are the most observations, the supernovae rates, an intrinsically Galactic quantity, are also discussed.

Sampled 4.6 billion years ago at a distance of 8.5 kpc, we find a composition at the solar circle that is in excellent agreement with the solar abundances from hydrogen to zinc. Type Ia supernovae provide about one-third of solar iron abundance in this distribution. Oxygen comes from the massive stars, but carbon and nitrogen come chiefly from $M \lesssim 8$ $M_\odot$. The light metal and iron group elements, with the exception of titanium, are in generally good agreement with the stellar abundance data. We also find an age-metallicity relation, a G-dwarf distribution, and present day supernova rates that are in satisfactory agreement with observations. The neutrino process provides a good explanation for the origin of $^{11}$B and $^{19}$F, and the increase in $^{7}$Li over its canonical homogeneous Big-Bang value. In order to explain the observed helium to metal enrichment ($\Delta Y/\Delta Z \simeq 4$), we find a favored cut off in the mass of supernovae that eject all material external to the iron core of about 30 $M_\odot$, but this limit may be increased by considering the effects of mass loss. The robustness of the results to variations in the iron yields of the Type II supernova models are examined.

Subject headings: Stars: Supernovae: General —- Stars: Abundances —- Galaxy: Abundances —- Galaxy: Evolution —- Solar System: General Evolution

1. INTRODUCTION

During the past 40 years the theory of stellar nucleosynthesis has developed from the first pioneering explorations of Burbidge *et al.* (1957) and Cameron (1957) into a fully developed discipline capable of confronting, in detail, the vast data base of stellar abundance measurements. The fruition of this theory, still in the making, will be the ability to predict reliable abundances for all observationally accessible isotopes and the variation of these abundances with location and time in our Galaxy. This goal has been difficult to achieve while all that existed were a few stellar models (e.g., 15 and 25 M$_\odot$ stars of solar metallicity, models for SN 1987A, etc.) characterized by variable physics and questionable accuracy.

Recent advances now allow the construction of at least a preliminary solution of the desired sort. Weaver & Woosley (1994; henceforth Paper I) have followed the presupernova evolution and Woosley & Weaver (1994b; henceforth Paper II), the explosive evolution of intermediate mass isotopes for 77 supernovae in the mass range 11 to 40 M$_\odot$ for a grid of metallicities; 0, $10^{-4}$, 0.01, 0.1, and 1 times solar. In a number of these models the explosion energy and parameterization of the piston were also varied. While the models may be uncertain on various grounds – the simplicity with which the explosion mechanism was simulated, the neglect of mass loss, the use of a certain convective model, and residual irresolution in key nuclear cross sections – they form a standard set upon which a benchmark for Galactic chemical evolution may be based.

Our Galactic evolution model and the assumed input are presented in §2. In addition to the yields of massive stars, it is necessary to have estimates for the nucleosynthesis from the homogeneous Big Bang, Type Ia supernovae, and in the winds and planetary nebulae of stars lighter than 8 M$_\odot$. Standard references are employed for each. Cosmic ray nucleosynthesis, responsible for a few light isotopes ($^6$Li, $^9$Be, $^{10}$B) is not included. Some uncertainty exists in how to treat the Type Ib supernovae. If only the hydrogen envelope is lost in such stars, then the synthesis of isotopes other than a few light ones – $^4$He, $^{14}$N, $^{17}$O, and $^{23}$Na – is unaffected. If a major part of the helium core is lost as well, the overall nucleosynthesis can be considerably altered (Woosley, Langer & Weaver 1993, 1994). Because the models for Type Ib supernovae are still uncertain, in the interest of simplicity we have not considered the effects of mass loss in our calculations.

In §3 our results are compared to the solar composition (Anders & Grevesse 1989), the age-metallicity relationship in the solar vicinity, the stellar abundance determinations of 27 elements (plus isotopic ratios when available), the solar circle G-Dwarf distribution, and the Galactic supernovae rates. This comparison, and the generally good agreement between observations and theory, are the chief result of the paper. The sensitivity of the calculated abundance histories to variations in the iron yields of Type II supernova models is investigated. We also explore the relative production of heavy elements and helium in our calculation and implications that this may have for black hole formation in massive stars (Maeder 1992; 1993; Brown & Bethe 1994).

Comprehensive reviews of chemical evolution in the solar vicinity or Galaxy by Truran & Cameron (1971), Tinsley (1980), Lambert (1989), Wheeler, Sneden & Truran (1989), Trimble (1991), Wilson & Matteucci (1992) and van den Bergh & Hesser (1993) provide sufficient background and introductory material.

## 2. METHOD OF CALCULATION

The models used here to represent the dynamical and chemical evolution of the Galaxy are simple and standard. Each radial zone in the exponential disk begins with zero gas and accretes primordial or near-primordial material over a 4 Gyr $e$-folding time scale. The isotopic evolution at each radial coordinate is calculated using "zone" models (as opposed to hydrodynamic models) of chemical evolution (e.g Talbot & Arnett 1971). Standard auxiliary quantities such as a Salpeter (1955) initial mass function and a Schmidt (1959; 1963) birth rate function were used. However, we discuss the models in detail since some aspects of the calculation are unique.

### 2.1 THE DYNAMICAL MODEL

The total surface mass density of a single zone at a radius $r$ and time $t$ is

$$\sigma_{\rm tot}(r,t) \; = \; \sigma_{\rm gas}(r,t) \; + \; \sigma_{\rm stars}(r,t) \qquad \frac{\rm M_\odot}{\rm pc^2} \; . \qquad (1)$$

Changes in the total surface mass density mass, due to the accretion of gas (stars are not accreted), are taken to be

$$\dot{\sigma}_{\rm tot}(r,t) \; = \; A(r) \; \exp\left(\frac{-t}{\tau_{\rm disk}}\right) \qquad \frac{\rm M_\odot}{\rm pc^2 \; Gyr} \; . \qquad (2)$$

The exponential infall term is motivated by the assumption that the disk component of the Galaxy grew to many times its initial value during the early phase of the evolution. Beyond the time scale for disk formation, $\tau_{\rm disk}$, infall rapidly decreases, and finally attains a small, but non-zero, value in the present epoch (Chiosi 1980). One dimensional hydrodynamical models of the Galaxy's evolution suggest that the energy deposited by Type II supernovae delay the appearance of the thin disk by about 4 Gyr (Larson, 1976; Burkert, Truran, & Hensler 1992). This value of $\tau_{\rm disk}$ was used in the results shown in §3. The form of equation (2) is in accord with the paradigm developed by Eggan, Lynden-Bell & Sandage (1962) that the Milky Way formed when a single large, rotating gas cloud collapsed. Somewhat more difficult to assign to equation (2) is the picture developed by Toomre (1977) that galaxies form by the merger of several large pieces, or the somewhat different scenario suggested by Searle & Zinn (1978) that many small, chemically unique fragments coalesced.

Integration of equation (2) over time, combined with the initial condition

$$\sigma_{\text{tot}}(r, t = 0) = \sigma_0 = \text{constant} , \tag{3}$$

gives

$$\sigma_{\text{tot}}(r, t) = \sigma_0 + A(r) \, \tau_{\text{disk}} \left[ 1 - \exp\left(\frac{-t}{\tau_{\text{disk}}}\right) \right] . \tag{4}$$

The results of §3, assumed that each radial zone is initially devoid of gas, that is $\sigma_0 = 0$. In the present epoch the total surface mass density at any radius is $\sigma_{\text{tot}}(r, t = t_{\text{now}})$. Imposing this boundary condition gives the function $A(r)$ as

$$A(r) = \left[ \sigma_{\text{tot}}(r, t_{\text{now}}) - \sigma_0 \right] \left[ \tau_{\text{disk}} \left[ 1 - \exp\left(\frac{-t_{\text{now}}}{\tau_{\text{disk}}}\right) \right] \right]^{-1} . \tag{5}$$

Notice that if the total surface mass density at time zero $\sigma_0$ is equal to its present day value $\sigma_{\text{tot}}(r, t_{\text{now}})$, then the accretion rate is zero and the time evolution reduces to the proper expressions for closed box (i.e. no infall) "zone" models of Galactic chemical evolution. The results of §3 assumed $t_{\text{now}} = 15$ Gyr ; other values for the age of the Galaxy do not significantly affect the main conclusions.

The present day mass distribution of the Galaxy is not known in detail, but the Ansatz of an exponential disk may be a suitable first approximation (Binney & Tremaine 1987)

$$\sigma_{\text{tot}}(r, t_{\text{now}}) = K_{\text{disk}} \, \exp\left(\frac{-r}{r_{\text{disk}}}\right) \qquad 2 \leq r \leq 15 \text{ kpc} , \tag{6}$$

where $K_{\text{disk}}$ and $r_{\text{disk}}$ are two scale parameters that remain to be determined. Near the Galactic Bulge an exponential surface density is clearly not valid, so one may adopt the inverse square law

$$\sigma_{\text{tot}}(r, t_{\text{now}}) = \frac{K_{\text{bulge}}}{(r + r_{\text{bulge}})^2} \qquad 0 \leq r \leq 2 \text{ kpc} , \tag{7}$$

where $K_{\text{bulge}}$ and $r_{\text{bulge}}$ are two additional scale parameters to be determined.

For the Galactic center the total surface mass density is assumed to be

$$\sigma_{\text{tot}}(r = 0, t_{\text{now}}) = \sigma_{\text{center}} = 10^4 \, \frac{M_\odot}{\text{pc}^2} . \tag{8}$$

Fortunately, the results of §3 are not particularly sensitive the choice of this parameter (e.g. Chiosi 1980). For the solar neighborhood we used the value

$$\sigma_{\text{tot}}(r = r_\odot = 8.5 \text{ kpc} , t_{\text{now}}) = \sigma_\odot = 75 \, \frac{M_\odot}{\text{pc}^2} , \tag{9}$$

consistent with present estimates (Bahcall 1984; Kerr & Lyndell-Bell 1986; Gilmore, Wyse & Kuijken 1989; Kuijken & Gilmore 1989; Statler 1989; Gould 1990; Kuijken & Gilmore 1991). Demanding continuity and smoothness at the $r=2$ kpc juncture between the exponential disk and the inverse square bulge (equations (6) and (7)) yields a single transcendental equation for the scale parameter $r_{\text{bulge}}$:

$$\sigma_\odot \exp\left[\frac{2(r_\odot - 2)}{(2 + r_{\text{bulge}})}\right] = \sigma_{\text{center}} \frac{r_{\text{bulge}}^2}{(2 + r_{\text{bulge}})^2} \ . \tag{10}$$

Once this equation is solved, the three other scale parameters follow immediately as $K_{\text{bulge}} = \sigma_{\text{center}} r_{\text{bulge}}^2$, $r_{\text{disk}} = (2 + r_{\text{bulge}})/2$, and $K_{\text{disk}} = \sigma_\odot \exp(r_\odot/r_{\text{disk}})$.

Equations (1) - (10) provide simple analytical expressions for the total mass and gas accretion rate at any instant of time and for any radius. At the end of 15 Gyr the total baryonic mass of the Galaxy in this model is $1.7 \times 10^{11}$ $M_\odot$, and the accretion rate for the total Galaxy is 0.3 $M_\odot \text{yr}^{-1}$. Although direct evidence for the infall of near primordial material onto the Galactic disk is weak, this accretion rate is in reasonable agreement with the estimates provided by observations of the very high velocity H I clouds (Muller, Oort & Raimond 1963; Oort 1970; Bothun 1985; Tosi 1988; Mirabel 1989; Wakker & Schwarz 1991; Wakker 1991; Wakker & van Woerden 1991), or observations of tidal stripping from the Magellanic clouds (Jones, Klemola & Lin 1994ab).

## 2.2 THE ISOTOPIC EVOLUTION MODEL

First consider the case where only stellar birth, infall of primordial material, Type II nucleosynthesis, and intermediate-low mass star nucleosynthesis are included. Type Ia supernova will be added later. For "zone" models of chemical evolution, the rate of change of the surface mass density of isotope "i" in the gas is

$$\begin{aligned}
\frac{d\sigma_i}{dt} &= \text{death} - \text{birth} + \text{infall} + \text{decay} \\
&= \int_{0.08}^{40} B(t - \tau(m)) \ \Psi(m) \ X_i(t - \tau(m)) \ dm \\
&\quad - B(t) \frac{\sigma_i}{\sigma_{\text{gas}}} + \dot{\sigma}_{i,\text{gas}} + \frac{\sigma_i}{\tau_{1/2}} \qquad \frac{M_\odot}{\text{pc}^2 \ \text{Gyr}} \ .
\end{aligned} \tag{11}$$

The first term in equation (11) describes the enrichment of isotopes in the interstellar medium due to stellar processing, while the second term accounts for disappearance of isotopes from the gas due to star formation. The third term is the infall model discussed above, with the accreted gas assumed to have a homogeneous Big Bang composition. The fourth term tracks any possible radioactive decay into and out of an isotope. This term is important to for isotopes such as $^{40}$K, which has a half-life of $\tau_{1/2}$=1.2 Gyr.

The function $\tau(m)$ takes into account the fact that stars of different mass have different, nonzero main-sequence lifetimes. If one sets $\tau(m)=0$ (the instantaneous recycling approximation), then the system of integro-differential equations represented by equation (11) reduces to a simpler set of ordinary differential equations. Further assumptions about the behavior of the gas reduces the system to the classical, analytical models (Tinsley 1980). It is critical to take into account realistic main-sequence lifetimes if significant contributions of an isotope come from intermediate or low mass stars (e.g. carbon and nitrogen), or from Type Ia supernovae (e.g. $^{56}$Fe).

For the results shown in §3, we used the main-sequence lifetimes of the massive stars as given directly by the stellar evolution calculations (Paper I). For stars less massive than 11 $M_\odot$, we adopted the main-sequence lifetimes of Schaller *et al.* (1992). The solar metallicity main-sequence lifetimes from these two sources are shown in Figure 1. Together they form a smooth curve, and linear interpolation is used for masses not on the grid. Various analytical expressions and fits for the main-sequence lifetimes have been suggested in the literature (Talbot & Arnett 1971; Güsten& Mezger 1983; Greggio & Renzini 1983; Burkert & Hensler 1987; Bazan & Mathews 1990; Mathews, Bazan & Cowan 1992; Colin & Schramm 1993). There may also be time delays due cooling of the supernovae ejecta, molecular cloud fragmentation processes or the protostellar collapse. The effect of using these fits or including any extra time delays change some of the details of the calculations shown in §3, but they do not alter the principle conclusions.

Metallicity effects on the main-sequence lifetimes of the massive stars amounts to shifts of about $\simeq 5\%$ (Paper I). Schaller *et al.* (1992) determined the intermediate and low mass main-sequence lifetimes for stars of Z=$Z_\odot$ and Z=0.05 $Z_\odot$. Since metallicity effects on the massive stars lifetimes are small, and the metallicity grid so sparse for the lower mass stars, the values shown in Figure 1 were used for all stellar metallicities. The primary effect of using a metallicity dependent main-sequence lifetime is that intermediate and low mass stars of low metallicity contribute their nucleosynthetic products at an earlier time (Bazan & Mathews 1990). The effects of using a metallicity dependent main-sequence lifetime on the results presented in §3 are expected to be small.

Schmidt's star formation function (Schmidt 1959; Schmidt 1963)

$$B(t) \simeq \rho_{\text{gas}}^n \qquad (12)$$

employed the volume density of the gas, with a rough observational justification for an $n=2$ exponent. The formalism developed above uses the surface density of the gas. These two forms are equivalent when $n=1$ for any gas scale height, and they are identical for any value of $n$ provided the scale height is constant with galactocentric radius (Lacey & Fall 1985). Some chemical evolution studies have suggested that a linear star formation rate gives a slightly better fit to the solar neighborhood than a quadratic star formation rate (Matteucci & Francois 1989; 1992). The advantage of using the surface density is that many of the observed quantities are best expressed as a surface density as it does not require computation of the gas scale height (Talbot & Arnett 1975; Bahcall 1984; Lacey & Fall 1985). In the results to be presented in §3 a quadratic Schmidt function with a dependence on the total mass was adopted

$$B(t) = \nu\ \sigma_{\rm tot}(t)\ \left(\frac{\sigma_{\rm gas}(t)}{\sigma_{\rm tot}(t)}\right)^2 \qquad \frac{\rm M_\odot}{\rm pc^2\ Gyr}\ , \qquad (13)$$

where $\nu$ is the efficiency factor, and is one of three major free parameters (the other two are the exponent of the initial mass function and the amplitude of the Type Ia supernovae contributions, both of which are discussed below). In the results shown in §3, the value $\nu$=2.8 gave the best fit to the observed element evolutions and the Anders & Grevesse (1989) solar composition. This value of the parameter $\nu$ is consistent with the 3.0% estimates of the efficiency of star formation within a giant molecular cloud. Fortunately, the main results of §3 are relatively very robust with respect to variations about the nominal value of $\nu$. A few of the evolutions shown in §3 display the effects of changing the star formation efficiency factor. If however, the star formation is either extremely supressed ($\nu \simeq 0.1$) or enhanced ($\nu \simeq 10.0$) then agreement with the majority of the observations requires either changing the slope of the initial mass function or abandoning the quadratic dependence of the Schmidt function.

The functional form of the star formation rate in equation (13) is supported by measurements of H$\alpha$, H I, and CO emission from spiral galaxies which generally find good correlation between the estimated star formation rate, the estimated total baryonic mass of the galaxy, and the mean surface mass density (Dopita 1985; Kennicut 1989). These two forms (equations 12 and 13) of the Schmidt star formation rate have been used and investigated in many "zone" and hydrodynamical models of chemical evolution (Talbot & Arnett 1975; Chiosi 1980; Matteucci & Greggio 1986; Matteucci & Francois 1989; Ferrini et al. 1992; Burkert, Truran & Hensler 1992; Mihos, Richstone & Bothun 1992; Mihos & Hernquist 1994).

The initial mass function $\Psi(m)$ in equation (11) was normalized between 0.08 M$_\odot$ and 40 M$_\odot$. Initial mass functions from Miller & Scalo (1979), with coefficients from Shapiro & Teukolsky (1983), and Scalo (1986) were examined, but the results of §3 employed a Salpeter (1955) power law. The assumed constant exponent of initial mass function represents the second of three major free parameters in the calculation (the first free parameter is the efficiency of star formation which was discussed above, and the third is the amplitude of the Type Ia supernovae contributions which is discussed below). Best fits to the observed element evolutions and the Anders & Grevesse (1989) solar composition was obtained with the an exponent of -1.31, remarkably close to the Salpeter value of -1.35 (see §2.5 and Table 1). The fits to the observations (especially the iron-peak elements) are fairly robust with respct to changes in the initial mass function exponent. A few of the evolutions shown in §3 display the effects of changing the exponent of the initial mass function. However, values larger that -2.0 would require relinquishing the quadratic dependence of the Schmidt function, or changes in the amplitude of the Type Ia supernova contribution.

## 2.3 ADDITION OF TYPE Ia SUPERNOVAE

Following Matteucci & Greggio (1986) the integral in equation (11) is broken into 4 pieces. The rate of change of the mass of isotope "i" is now given by

$$\begin{aligned}
\frac{d\sigma_i}{dt} = & \int_{0.8}^{3} B(t-\tau(m)) \, \Psi(m) \, X_i(t-\tau(m)) \, dm \\
& + C \int_{3}^{16} \int_{\mu_m}^{0.5} B(t-\tau(m)) \, \Psi(m) \, f(\mu) \, X_i(t-\tau(m)) \, d\mu \, dm \\
& + (1-C) \int_{3}^{16} B(t-\tau(m)) \, \Psi(m) \, X_i(t-\tau(m)) \, dm \\
& + \int_{16}^{40} B(t-\tau(m)) \, \Psi(m) \, X_i(t-\tau(m)) \, dm \\
& - B(t) \frac{\sigma_i}{\sigma_{\text{gas}}} + \dot{\sigma}_{i,\text{gas}} + \frac{\sigma_i}{\tau_{1/2}} \qquad \frac{M_\odot}{\text{pc}^2 \, \text{Gyr}} \, .
\end{aligned} \qquad (14)$$

The second term on the right-hand side is new and represents those binary systems that undergo a Type Ia event, while the third term represents stars that are single or are members of a binary system that do not produce a Type Ia supernova. The free parameter $C$ adjusts the amplitude of the Type Ia contributions, with $C = 0$ reducing equation (14) to equation (11). In the results presented in §3, the parameter value $C$=0.007 was used. The methodology used in obtaining this particular value and the sensitivity of the results to this choice are discussed in §3.1.

The upper limit of the integrations in the second and third terms was based on the assumption that the maximum mass of the primary that produces a carbon-oxygen white dwarf is 8 $M_\odot$. This implies a maximum binary mass of 16 $M_\odot$. In order to insure that the accreting white dwarf eventually reaches the Chandrasekhar limit, the minimum mass (and lower integration limit) that could give a Type Ia event was taken to be 3 $M_\odot$.

With binary systems invoked for Type Ia supernovae progenitors, a minimum of one additional integral must be performed, in this case, over a binary distribution function $f(\mu)$. More complicated scenarios also integrate over a orbital separation distribution, including any gravitational radiation losses that accelerate the final merger. For simplicity, these factors are ignored. The results of §3 employed the binary distribution function of Greggio & Renzini (1983)

$$f(\mu) = 24\mu^2 \, , \qquad (15)$$

where $\mu$ is being the ratio of the secondary to the total binary mass, and $f(\mu)$ is normalized over the interval [0,1/2].

Two derived quantities (not input parameters) are the Type II supernovae rate by number

$$R_{II}(t) \;=\; \int_{16}^{40} B(t-\tau(m)) \; \Psi(m) \; \frac{dm}{m} \;+\; (1-C) \int_{11}^{16} B(t-\tau(m)) \; \Psi(m) \; \frac{dm}{m} \;, \quad (16)$$

and the Type I supernova rate by number

$$R_I \;=\; C \int_{3}^{16} \int_{\mu m}^{0.5} B(t-\tau(m)) \; \Psi(m) \; f(\mu) \; d\mu \; \frac{dm}{m} \;. \quad (17)$$

These equations show that the supernova rates are simply a birth rate evaluated at an earlier time (the $t-\tau(m)$ terms), convolved with an initial mass function. Integration of these equations over all radial zones gives the Galactic supernovae rates.

## 2.4 THE STELLAR YIELDS

Arguably the most important input into a Galactic chemical evolution calculation is the stellar yields. Three distinct components need to be specified; one for the massive stars ($11 \leq M_\odot \leq 40$) which become Type II supernova; one for intermediate to low mass stars ($M_\odot < 11$) which become planetary nebula; and one for the remnants of the intermediate to low mass stars that become Type Ia supernovae. These components are embodied in equations (11) and (14) as the $X_i$ terms; the mass fraction of a star of mass $m$ ejected in the form of isotope "i". In general, the mass fractions $X_i$ depend on the mass and initial metallicity of the star.

The nucleosynthetic yields of massive stars were taken from Papers I and II. An iterative loop operated between those calculations and the present study. The initial composition employed in Papers I and II was a homogeneous Big Bang composition: by mass 76% H, 0.0091% $^2$H, 0.0034% $^3$He, 24% $^4$He, and $8 \times 10^{-8}$ % $^7$Li (Walker *et al.* 1991). A second composition with a solar metallicity, Anders & Grevesse 1989 composition was also used. These were used to generate two sets of supernovae models. The chemical evolution calculation then interpolated between the two sets of yields until a total metallicity (which was dominated by oxygen) of $Z_{gas} = 10^{-4}$ $Z_\odot$ was reached. The abundances at this point were used as the initial compositions in another series of massive star and supernovae models. The yields from this series was incorporated into the chemical evolution calculations, and the process repeated at metallicities of $Z_{gas} = 0.01$ $Z_\odot$ and $Z_{gas} = 0.1$ $Z_\odot$. Such a procedure is more consistent than using stellar models whose initial abundances are simply scaled to the solar abundance set. However, the results of §3 are fairly robust with respect to deviations from this self-consistency. The abundances of carbon, nitrogen, oxygen and iron abundances are the most important, but the principle results are relatively insensitive to the precise values employed. Significant fluctutations in the other element abundances have no discernable effect of the final nucleosythetic yields. A total of 60 Type II supernova models were used to cover the mass – metallicity plane.

For stars heavier than 25 $M_\odot$ Paper II showed that the mass of heavy elements ejected was quite sensitive to "fall back", and thus to the uncertain explosion energy. We adopted as our standard set the models ending with the letter "A" in Paper II for stars in the mass range 11 – 25 $M_\odot$. At 30, 35, and 40 $M_\odot$, models "B" were used instead. This was felt to represent a reasonable compromise between models that had higher energy (the "B" models) than the canonical $\simeq 10^{51}$ erg usually adopted for such explosions (though not unrealistically so), and other lower energy models (the "A" models) in which all of the heavy elements fell back into a black hole. Although it would be reasonable in a future study to examine the effect of these unused 30, 35, and 40 $M_\odot$ massive star models, the first order effects of using these models are easy to see. As the explosion energy is lowered, less metals are ejected. This is especially true in the very metal-poor massive star models where the progenitors have a more compact structure than their solar metallicity counterparts (Paper I). In the extreme case of a very small explosion energy in the zero-metal stars, no iron is ejected (Paper II). Thus, if the smaller explosion energy models were employed in the chemical evolution calculations then the total metallicity in the interstellar medium would increase more slowly, and the total number of compact remnants in the Galaxy would increase. However, where the mass cut is placed is more critical than explosion energy. The farther out the mass cut is placed, the less mass there is (which is rich in iron-peak nuclei) to eject. In fact, the results of §3 suggest that the mass cut used in the present calculations may have been placed too far into the iron core (see also §4). The sensitivities of the models to the progenitor structure, the explosion energy and the mass cut are discussed in detail in Paper II.

Few other groups have systematically followed the main-sequence massive star and Type II supernova nucleosynthetic yields as a function of mass and metallicity. Maeder (1992; 1993) studied a large number of main-sequence massive stars (with and without mass loss), but the evolution was only followed up to the end of core carbon burning with a very small nuclear reaction network. Nomoto & Hashimoto (1988) and Hashimoto *et al.* (1993) examined the evolution of helium cores from helium burning to the onset of the iron core collapse. Thielemann, Nomoto & Hashimoto (1994) have recently extended these calculations by including a parameterized explosion. The only element present in the initial composition, besides helium, was nitrogen, where it was assumed that all the original CNO elements were converted into $^{14}N$ during core hydrogen burning (which was not followed). A minor difficulty in using these yields in a chemical evolution model is the uncertainty in mapping a main-sequence mass to a helium core mass. Hydrogen core and shell burning phases in stars of different mass can lead to significant variations in the final mass of the helium core (Bressan, Berelli & Chiosi 1981; Maeder 1992) and the results of a chemical evolution model are sensitive to variations in the adopted main-sequence to helium core mapping. More importantly, there is no indication of how the yields in these helium core studies change for massive stars of different initial metallicities.

The yields from intermediate to low mass stars and Type Ia supernova are taken from standard references. Renzini & Voli (1981) derived CNO yields for $1 \, M_\odot \leq M \leq 8 \, M_\odot$ stars. They took into account the convective dredge-up mechanism of Iben & Truran (1978), nuclear processing in the deepest layers of the convective envelope, and post main-sequence mass loss. Their $Z=0.004$ and $Z=Z_\odot$ case B tables were adopted, and linearly interpolated between these two metallicities. At smaller metallicities the $Z=0.004$ tables were used. The popular W7 model of Nomoto, Thielemann & Yokoi (1984) was adopted as representative of Type Ia supernovae, but the detailed nucleosynthesis was taken from Thielemann, Nomoto & Yokoi (1986).

## 2.5 SOME IMPLEMENTATION DETAILS

Each isotope in every zone is represented by an equation (14). Thus, evolution of 76 isotopes in a dozen radial zones requires the solution of about 1000 coupled, non-linear integro-differential equations. The computation is kept efficient, without compromising accuracy, by using Gaussian quadrature summation for the integrals (Press *et al.* 1992), and a Cash-Karp stepper (Press & Teukolsky 1992) for the explicit time integration. In addition, a general property of integro-differential equations is the display of memory; manifested by the $t - \tau(m)$ terms in equation (11) and (14). The solution today depends upon what the solution was in the past. This requires that the solution be stored as it is generated and interpolated as needed.

One convenience of our chemical evolution model is the adoption of a "solar masses in and solar masses out" formalism in lieu of a "production matrix" formalism (Talbot & Arnett 1973ab; Matteucci & Francois 1989; Ferrini *et al.* 1992) for describing the nucleosynthesis, the latter of which can become quite unwieldy as the number of isotopes evolved increases. Linear interpolation of the stellar yields in the mass - metallicity plane returns values for the yields that are not explicitly on the grid. Initial testing the chemical evolution code reproduced the results of several previous studies (Talbot & Arnett 1971, 1973ab, 1975; Chiosi & Matteucci 1982; Matteucci & Greggio 1986). Table 1 shows a summary of the parameters used in generating the results of §3, with the free parameter values being chosen on the basis of achieving the best fit to the solar composition and the elemental abundance histories. Table 2 compares some key present day solar vicinity quantities with the observational estimates. Overall, the agreement is satisfactory.

Figure 2 shows several dynamical quantities of the solar neighborhood that are independent of the composition of the gas or the nucleosynthetic yields. The total surface mass density, accretion rate, surface mass density of the gas, surface mass density of stars, and the stellar birthrate values indicated by the left y-axis. The total metallicity of the gas $Z_{\rm gas}$ is also shown, with values give by the right y-axis. The detailed breakdown of $Z_{\rm gas}$ into its 76 components will constitute the remainder of this paper.

## 3. RESULTS

The atmospheres of dwarf stars are assumed to represent a perfect record of the chemical composition of the interstellar medium at the time the stars formed. Although this critical assumption may be violated, for example through convective mixing of core and atmospheric material, dwarf stars will be systematically preferred to giant stars in the ensuing discussion and figures. In addition, the local thermodynamic equilibrium (henceforth LTE) hypothesis is presumably more valid for dwarfs than giants because of the larger density of the dwarf star atmospheres. However, in some cases there may still be good reasons to use field giants, if it can be shown that the *ab initio* abundance of the element in question is reasonably immune from contamination (Lambert 1989; Wheeler *et al.* 1989).

Cluster giants are sampled from the tip of the red giant branch downward (due to flux limitations) which emphasizes the most evolved giants. On the other hand, magnitude limited field giant surveys emphasize giants that are less evolved (due to the field luminosity function and flux limitations). Thus, it is safer to recover the *ab initio* abundances from field giants than from cluster giants (Kraft 1994). Only field giants, when giants are used at all, are considered in this paper.

The literature of stellar abundance determinations in relatively nearby stars is very large. In order to keep the discussion at a manageable length and the figures uncluttered, only 3 or at most 4 of the "best" observations are shown and discussed. "Best" is a subjective term and the definition varies for each element, but in general, studies were preferred that spanned a large metallicity range, included important physical effects (e.g. hyperfine structure or non-LTE effects), or were historically important in generating spirited debate.

The notation used to describe the abundance of "X" relative to the solar value is

$$[{\rm X}] \;=\; \log \frac{\rm X}{\rm X_\odot} \;=\; \log {\rm X} - \log {\rm X}_\odot \quad {\rm dex} \;. \qquad (18)$$

In a few cases it is desirable to compare the abundance of "X" relative to hydrogen by number,

$$N({\rm X}) \;=\; \log \left(\frac{\rm X}{\rm H}\right) + 12.0 \quad {\rm dex} \;. \qquad (19)$$

The observational data was not renormalized to a single set of solar abundances since all the stellar abundance determinations would have to be redone, especially if the original investigation employed a differential analysis. Fortunately, the vast majority of the stellar spectroscopic studies choosen in this paper are consistent with the Anders & Grevesse (1988) abundance set, perhaps with a slightly modified iron content (Biémont *et al.* 1991; Holweger *et al.* 1991; Hannaford *et al.* 1991). Nevertheless, differences in the solar abundance sets remains a potential source of error when comparing data from different sources or when comparing the theory to the observations.

### 3.1 COMPOSITION AT SOLAR BIRTH

Figures 3 - 5 show the stable isotopes from hydrogen to zinc present in the interstellar medium at a time when (4.6 Gyr ago) and place where (8.5 Kpc Galactocentric radius) the Sun was born. The calculated mass fractions are compared to the Anders & Grevesse (1989) solar values. In Figures 3 and 4 only contributions from massive stars, intermediate, and low mass stars are included. Figure 4 removes some of underproduced isotopes in Figure 3 and expands the y-axis scale. The effect of adding Type Ia supernovae is shown in Figure 5.

In each figure the x-axis is the atomic mass number. The y-axis is the logarithmic ratio of the model abundance to the Anders & Grevesse mass fraction. The most abundant isotope of a given element is marked by an asterisk and isotopes of the same element are connected by solid lines. If the calculation were perfect, all of the isotopes would have a constant ordinate (equal to unity in this case) and the solar composition would have been replicated. In each figure, the perfect case is denoted by a dashed horizontal line.

Part of the spread in Figures 3 - 5 is a consequence of uncertain physics in the models – the nuclear reaction rates, treatment of convection, and the parameterizations of the star formation rate and the explosion mechanism to name a few. Part of the spread also reflects uncertainty in the measured abundances themselves. While isotopic ratios are known precisely (except in a few cases, such as neon, argon, and potassium), elemental abundances are less certain. A formal analysis of the total uncertainty would be a daunting task, and we therefore adopt the convention that isotopes falling within a factor of two of their solar value are a "success". In each figure, the horizontal dotted lines denote this factor.

The solar abundances are most sensitive to the nucleosynthesis that occurred in stars born with a metallicity between $0.2~Z_\odot \leq Z_{\rm gas} \leq 0.4~Z_\odot$. This is consistent with the broad peak in the G-dwarf distribution in the solar vicinity (see §3.9.2). The contribution of Type Ia supernova, whose canonical nucleosynthesis is taken to be the carbon deflagration model of Thielemann *et al.* (1986), is determined by varying the free parameter $C$ in equation (14) until the solar $^{56}$Fe abundance is achieved. This procedure also gives results consistent with the observed age-metallicity relationship (§3.2), and Galactic supernova rates (§3.10). The value thus derived is $C=0.007$, which is consistent with the value found by Matteucci & Greggio (1986). The principle results are fairly insensitive to the precise value of $C$, for example, the isotope $^{56}$Fe goes above the success band in Figure 5 when $C=0.03$.

As the figures show, isotopes below calcium show less scatter than those isotopes above. This is because isotopes below calcium are chiefly produced by hydrostatic burning processes before the explosion, while heavier isotopes are sensitive to the uncertain modeling of the explosion mechanism (Weaver & Woosley 1993; Timmes, Woosley & Weaver 1993; Woosley, Timmes & Weaver 1993). Up to calcium, it generally makes little difference whether the presupernova or the exploded yields are used.

It is gratifying that the hydrogen production factor and the $^{16}$O production factor are both very nearly equal to one. The model makes the correct solar metallicity. The production of the rare and fragile isotopes, $^7$Li, $^{11}$B and $^{19}$F, by the neutrino process can account for a significant fraction, if not all, of their solar system abundance (certainly the majority of $^{11}$B). The carbon isotopes and $^{14}$N are underproduced in the massive star models by about a factor of 4. This is consistent with the bulk of these isotopes being produced during CNO processing and dredged-up material from helium shell flashes in intermediate and low mass stars (Iben & Truran 1978; Renzini & Voli 1981; Iben & Renzini 1983; Wheeler *et al.* 1989), which are included in the calculation.

Almost all of the isotopes below calcium, and nearly all of the most abundant isotopes of an element above calcium show good agreement with solar values. Oxygen is not overproduced at the expense of the any of the other elements, in contrast to some arguments found in the literature (Twarog & Wheeler 1982,1987; Wheeler *et al.* 1989). There is probably no systematic pattern to the scatter. Figures 4 and 5 indicate that Type II supernova produce 2/3's of the solar system iron abundance, with Type Ia supernova contributing the rest (see §3.2 for a more thorough discussion of this point). Inclusion of nucleosynthesis from Type Ia supernovae improves the fit of $^{50-53}$Cr, $^{55}$Mn, and $^{54-58}$Fe, while scarcely affecting the solar abundances of the other isotopes.

The "failures" are almost as interesting as the successes. Starting at low atomic mass, one first encounters deficiencies of $^6$Li, $^9$Be, and $^{10}$B. While there is some small production of these species by the neutrino process, the net action of massive stellar evolution is to decrease their abundance over what the star began with (Paper II). The amounts ejected are insufficient by a factor of a hundred in explaining their solar abundances. It is believed that the primary source of these isotopes is the spallation of CNO nuclei in the interstellar medium by cosmic rays (Reeves, Fowler & Hoyle 1970; Walker, Mathews & Viola 1985; Reeves *et al.* 1990; Reeves 1994). Some recent models of inhomogeneous Big Bang nucleosynthesis also appear capable of producing interesting amounts of some of these isotopes (Fuller, Mathews & Alcock 1988; Malaney 1993; Thomas *et al.* 1994; Jedamzik, Fuller

& Mathews 1994; Jedamzik *et al* 1994). Since nonstandard primordial nucleosynthesis or cosmic ray spallation scenarios are not included in Figure 3, it is not surprising that the chemical evolution model "fails" for the $^6$Li, $^9$Be, and $^{10}$B isotopes.

The production of $^{15}$N here is due exclusively to the neutrino process. The fact that it is within a factor of three of its solar value suggests that the relevant neutrino cross sections and spectra should be examined more closely. But for now, it seems that $^{15}$N must come from somewhere else. Hot hydrogen burning (Lazareff *et al.* 1979; Woosley 1986) has been suggested, with environments typified by classical novae systems.

The slightly large production of $^{40}$K may reflect its uncertain solar abundance.

The neutron rich isotopes $^{48}$Ca, $^{50}$Ti, and $^{54}$Cr are all deficient. They can be produced in neutron-rich nuclear statistical equilibrium (Hartmann, Woosley, & El Eid 1985). However, the neutron rich material in Type II supernovae is at too high an entropy and too small a mass is ejected to be the production site for these isotopes (Woosley *et al.* 1994). They are probably produced in deflagration models for Type Ia supernovae, especially those characterized by slow initial flame speeds and high ignition densities (Woosley & Weaver 1986; Timmes & Woosley 1992; Woosley & Weaver 1993; Livne & Arnett 1993; Khokhlov 1993; Arnett & Livne 1994ab). They are copiously produced in Chandrasekhar mass white dwarf systems that almost experienced accretion induced collapse, but managed to successfully explode from a very high central density ($\rho_c \gtrsim 4 \times 10^9$ g cm$^{-3}$; Woosley & Eastman 1994).

The isotopes $^{44}$Ca, $^{47}$Ti and $^{51}$V have historically presented problems (Woosley 1986), with a low $^{44}$Ca reflecting an inadequate production of $^{44}$Ti during an $\alpha$-rich freeze out. All three of these isotopes may have significant contributions from helium detonation in sub-Chandrasekhar mass models for Type Ia supernovae (Woosley, Taam & Weaver 1986; Livne & Glasner 1991; Shigeyama *et al.* 1992; Woosley & Weaver 1994a; Nomoto *et al.* 1994). The high production of $^{62}$Ni and $^{68}$Zn from the exploded massive star models may be a consequence of the fact that the nuclear reaction network employed in the stellar evolution calculations was terminated at germanium, or perhaps they reflect a symptom of how the explosion was parameterized.

In terms of absolute solar abundances, the stable isotopes from hydrogen to zinc range over some 10 orders of magnitude. That the detailed nucleosynthesis used in the chemical evolution calculations compresses the ratios to the Anders & Grevesse (1989) abundances down to around a factor of two is very encouraging. However, the Sun samples only one point in time and space in our Galaxy's history. There is potentially much more information in the time history of each element in the solar neighborhood, where there exists a wealth of stellar abundances determined from spectroscopic studies of nearby stars. This is the topic to which we now turn.

## 3.2 THE AGE-METALLICITY RELATION

Stellar abundance determinations are traditionally discussed in terms of an elemental abundance relative to iron [X/Fe] as a function of iron to hydrogen [Fe/H] ratio, primarily because [Fe/H] is a relatively easy quantity to measure in stars. The [Fe/H] ratio represents a chronometer in that the accumulation of iron in the interstellar medium increases monotonically with time (Wheeler *et al.* 1989). Calibration of the [Fe/H] ratio as a function of time forms the basis of the age-metallicity relationship. It is critical for Galactic chemical evolution models to reproduce this relationship as it links the variety of physical time scales involved (e.g. infall rate, star formation rate, and stellar lifetimes).

Type II and Type Ia supernovae provide two sources of iron production, each operating on distinctly different time scales, and each injecting very different amounts of iron. Figure 6 shows the $^{56}$Fe yields from the standard set of explosion models described in §2.4. The points labeled with the symbol "z" represent zero initial metallicity stars, "u" for $1.0^{-4}$ $Z_\odot$, "t" for $1.0^{-2}$ $Z_\odot$, "p" for 0.1 $Z_\odot$, and "s" for 1.0 $Z_\odot$. These iron yields are not monotonic with respect to either mass or metallicity. Variations are caused by the uncertainty in modeling the explosion mechanism, the amount of iron which may fall back onto the compact remnant, and the sensitivity of the pre-supernova models to the interaction of the various convective zones during the later stages of nuclear burning (Weaver & Woosley 1993; Paper I).

A metallicity independent W7 model (Nomoto *et al.* 1984; Thielemann *et al.* 1986) implies that 0.63 $M_\odot$ of $^{56}$Fe is ejected by all Type Ia supernova. This estimate is uncertain from both theoretical and observational standpoints. Carbon deflagration models for Type Ia events have different amounts of iron synthesized depending on the assumptions made about the speed and acceleration of the nuclear flame front (Nomoto *et al.* 1984; Woosley, Axelrod & Weaver 1984; Thielemann *et al.* 1986; Khokhlov 1991ab; Arnett & Livne 1994ab; Woosley & Weaver 1994a). Observationally the expectation of a constant $^{56}$Fe yield has been challenged by measurements that show a range of $^{56}$Fe ejected by different Type Ia events (Ruiz-Lapuente & Filippenko 1993a, Ruiz-Lapuente *et al.* 1993b).

A potential concern is that the W7 model omits any sensitivity to the initial composition. When the very low metallicity massive stars are becoming Type II supernovae, W7 already has its full complement of $^{22}$Ne and other heavy element seeds that influence its yields. The W7 model of Thielemann *et al.* (1986) consists $^{12}$C and $^{16}$O in equal parts and a neon mass fraction (the initial white dwarf "metallicity") of X($^{22}$Ne)=0.025. The initial $^{22}$Ne abundance (and thus the initial metallicity) has a relatively mild effect on the neutron excess in those regions where the iron group isotopes are produced (Bravo *et al.* 1992), electron capture providing most of the neutrons for these species. However, it does affect the composition of iron and nickel somewhat (higher metallicity leads to increased $^{54}$Fe and $^{58}$Ni production) and could affect the composition of lighter elements even more.

Fortunately, the contribution of W7 to elements such as sodium, aluminum and secondary isotopes early in the chemical evolution model were found to be small, so the assumption of metallicity independent W7 model may be safe. We note in passing that the metallicity would have a greater effect on the nucleosynthesis, especially of iron group species, in sub-Chandrasekhar mass models for Type Ia supernovae (e.g, Woosley & Weaver 1994a).

Figure 7 shows the solar neighborhood age-metallicity relationship. Calculations with and without the iron contributions from Type Ia supernovae are shown, along with the observational data of Twarog (1980), Carlberg *et al.* (1985), Meusinger, Reimann & Stecklum (1991), and Edvardsson *et al.* (1993). The curves do not display the raggedness of the iron yields shown in Figure 6 because repeated integration (in time) over an initial mass function erases irregularities in the yields. Figure 7 indicates that the inclusion of Type Ia events is important for reproducing the observed iron evolution, and suggests that about 2/3's of the solar iron abundance comes from Type II events and about 1/3 comes from Type Ia supernovae (see also Figures 4 and 5). Variations in these two fractions depend primarily on the iron yields of Type II supernovae, but the main-sequence lifetimes are also important. Fractions of 50% from Type Ia supernova and 50% from Type II supernova are well within the uncertainties associated with the iron yields from each type of supernovae, and is preferable given the fits to all the observed element evolutions (see §4 for a summary). It may be that 2/3 of the solar iron abundance comes from Type Ia events and 1/3 from Type II events without doing grave injustice to the stellar physics. Since there are discrepancies between between the various observations shown in Figure 7 (three of which used the same data set), the salient features of the observational analyses are examined.

The relationship between the color excesses and age of a star has been known for quite some time (Sandage 1962), but Twarog (1980) was the first to show an age-metallicity relationship for local disk stars. Twarog's data is based upon $uvby\beta$ photometry of two large surveys (1007 and 2742 stars respectively) that was used in constructing a color-magnitude diagram. Subsequent use of the Yale isochrones determined the age-metallicity relationship. The raw data displays considerable scatter, so averages were taken in age bins. Within the error bars of Twarog's averaged data, Figure 7 suggests [Fe/H] = -1.0 dex near the beginning of the disk's formation about 15 Gyr ago, and rises relatively smoothly to the solar value. In the 4.6 Gyr since solar birth, [Fe/H] has climbed by another 0.1 dex.

The Twarog data was reexamined by Carlberg *et al.* (1985). They only used the smaller survey, and then deleted about 50 of the low [Fe/H] stars. The VandenBerg (1983, 1985) isochrones and a different photometric calibration were employed, the combination tending to give larger [Fe/H] values to the cooler, generally older stars. The result of these selection effects and procedures led Carlberg *et al.* to suggest that [Fe/H] declines from its present value of 0.1 dex to only about -0.2 dex some 15 Gyr ago. This large [Fe/H] ratio at an early epoch is in substantial disagreement with Twarog, and sparked a number of investigations and lively debate.

Meusinger, Reimann & Stecklum (1991) also reanalyzed the Twarog data, but employed both samples of stars. They used the same VandenBerg (1985) isochrones as Carlberg *et al.*, a newer calibration for the metallicity index, and improved stellar atmosphere calculations. The shape of the age-metallicity relation that they found is similar to the original Twarog result.

Photometric surveys such as the Twarog data have the advantage of providing a large number of stars to reduce the statistical uncertainty. However, they have the disadvantage not possessing any kinematical information. Thus, one typically assumes that all stars with [Fe/H] > -1.0 dex are disk stars. Meusinger, Reimann & Stecklum (1991) used kinematical data that was by then available for the larger of Twarog's photometric surveys. They derived an age-velocity dispersion relation, which found that the velocity dispersion of disk stars rises continuously with age from about 15 km s$^{-1}$ for the youngest stars to about 40 km s$^{-1}$ for the oldest stars. The velocity ellipsoid axis ratios were found to be $\sigma_U : \sigma_V : \sigma_W = 1 : 0.74 : 0.78$. These results are consistent with previous kinematical studies, and suggested that the associated age-metallicity relationship they derived was self-consistent and reliable.

The observations of Edvardsson *et al.* (1993) are the only data points shown in Figure 7 that are not another reanalysis of the Twarog survey. From accurate kinematic data they selected 189 nearby F and G disk dwarfs from the northern and southern hemispheres, and derived abundances for O, Na, Mg, Al, Si, Ca, Ti, Fe, Ni, Y, Zr, Ba and Nd. Both the spectra and photometry were analyzed using LTE model atmospheres that incorporated blanketing from millions of spectral lines, enabling a high degree of internal consistency. Thus, they were able then to determine a direct, spectroscopically calibrated, age-metallicity relationship. Their results are probably the most accurate of the data sets shown in Figure 7. Using binned ages and distances, they find that the overall trend is in reasonable agreement the photometric survey of Twarog.

Two related and important points were made by Edvardsson *et al.* (1993). Based on their extensive survey, they claim that the sun is a quite typical [Fe/H] = 0 dex star for its age and galactic orbit. This adds justification to the comparison of the calculated abundances in the solar zone with the Anders & Grevesse (1989) solar abundances given in §3.1. They also claim that the scatter present in their unbinned [Fe/H] data is primarily real and not observational. While a real [Fe/H] scatter had been shown for young and old open clusters (Nissen 1988; Boesgaard 1989; Friel & Janes 1991), it had not been clearly demonstrated before for field stars. Almost all of the suggested mechanisms in the literature for achieving a real scatter in [Fe/H], or any abundance ratio, rely on relaxing the assumption that the nucleosynthetic products of stars are mixed instantaneous throughout the zone at a given galactocentric radius. Models which invoke instantaneous mixing, like ours, can only explain the average trends. Only a few models that relax this assumption in various ways have been attempted (Malinie, Hartmann & Mathews 1991; Malinie et al. 1993).

Finally, many workers have suggested that oxygen may be a better choice than iron for a Galactic chronometer (Wheeler et al. 1989). The main reason for this is that oxygen is synthesized from a single, well-defined source: hydrostatic helium burning in presupernova stars. However, oxygen abundances in stars are more difficult to measure, and the amount of oxygen synthesized is affected by uncertainties in the $C^{12}(\alpha,\gamma)O^{16}$ nuclear reaction rate. Nonetheless, oxygen may well be a superior chronometer.

### 3.3 $\nu$-PROCESS CONTRIBUTIONS

Paper II found that the most important regions for $\nu$-process nucleosynthesis are the helium, carbon, and neon shells where significant amounts of primary $^7$Li, $^{11}$B, and $^{19}$F respectively are synthesized (see also Woosley et al. 1990). These yields given in Paper II take into account any re-arrangement of nuclei due to passage of the shock wave through the star.

The choice of the $\mu$ and $\tau$ neutrino temperature strongly influences the results. Some have argued that the effective $\mu$ and $\tau$ neutrino temperature should be substantially less than 8 MeV employed in Paper II, perhaps more like 6 Mev (Janka & Hillebrandt 1989; Janka 1991). The problem is that the actual temperature distribution is not a blackbody distribution at any temperature. An additional difficulty is that few neutrino transport calculations have been carried to sufficiently late time (at least 3 seconds) or have sufficient neutrino energy resolution to see the hardening of the spectrum that occurs as the proto-neutron star cools (Janka 1991; Wilson & Mayle 1993). It is gratifying that the carbon-to-boron and fluorine-to-neon ratios are reasonably correct in Figure 5. These might be construed as arguing for a $\mu$ and $\tau$ neutrino temperature near 8 Mev.

#### 3.3.1 $^7$Li

The evolution of $^7$Li as a function of [Fe/H] is shown in Figure 8. The calculated $^7$Li abundance is shown as the solid line, and the dashed lines show factors of two variation in the $\nu$-process yields. Also shown are abundance determinations of $^7$Li in disk and halo dwarfs that have $T_{\text{eff}} > 5500$ K, and the severely depleted solar photospheric value (Anders & Grevesse 1989).

Spite & Spite (1982) examined the lithium abundances in 21 field dwarfs with high resolution ($\Delta\lambda \simeq 0.26$ Å), low noise (S/N$\simeq 150$) photodiode spectra of the lithium doublet at 6707 Å. The survey stars spanned the metallicity range -2.4 $\leq$ [Fe/H] $\leq$ -0.3 dex.

Effective temperatures, surface gravities and metal abundances from the literature were used, and a microturbulence velocity of 1 km s$^{-1}$ was adopted for all the survey stars. An LTE differential analysis determined the final abundances. Their results suggested that the halo dwarfs (i.e [Fe/H] $\lesssim$ -1.0 dex) had a constant lithium abundance of N(Li) = 2.1 dex, while the disk dwarfs spanned a large range, with some dwarfs having lithium abundances about twice as large as the halo dwarfs.

Boesgaard & Tripico (1986) studied the surface lithium abundances in 75 F field dwarfs with high resolution ($\Delta\lambda \simeq 0.11$ Å), very low noise (S/N$\simeq$ 500) photodiode spectra of the Li I line at 6707 Å. The stars in the survey covered the metallicity range -0.8 $\leq$ [Fe/H] $\leq$ 0.3 dex. Effective temperatures and surface gravities were found from Strömgren photometry and R-I color indices. A microturbulence velocity of 1.8 km s$^{-1}$ was used for all the survey stars. An LTE differential analysis, with particular care given to blending from nearby Fe I features, determined the final abundances. They found that about 1/3 of the survey stars had lithium abundances around N(Li) = 3.0 dex, and that these stars tended to be the youngest and hottest stars in the sample. Depletion factors of 3-10 were found for about 17% of the stars, and about 50% of the dwarfs had depletion factors between 10-150.

Rebolo, Molaro & Beckman (1988) determined the lithium abundances of 37 field dwarfs that covered the metallicity range -2.3 $\leq$ [Fe/H] $\leq$ -0.6 dex. The analysis was based on high resolution (R$\simeq$20,0000), moderate noise (S/N$\simeq$100) digital spectra of the 6707 Å Li I feature. Effective temperatures were established from calibrated V-K, R-I, and $b$-$y$ color indices. Surface gravities were generally found from calibrated $b$-$y$ and $c_1$ relations, and all the survey stars were assigned a microturbulence of 1.5 km s$^{-1}$. An LTE synthetic spectrum, with care given to nearby Fe I blending features, determined the final abundances. The results suggested a mean lithium abundance of N(Li) = 2.2 $\pm$ 0.15 dex for dwarfs with [Fe/H] $\leq$ -1.4 dex. For the less metal deficient dwarfs they reported a wide scatter in the lithium abundances.

Thorburn (1994) examined the atmospheric lithium content in 90 stars, which included both dwarfs and subgiants, using high resolution (R$\simeq$28,000), moderate noise (S/N$\simeq$100) digital spectra of the Li I 6707 Å doublet. The sample stars spanned the metallicity range -3.2 $\leq$ [Fe/H] $\leq$ -1.7 dex. The results of this large survey suggested a mean halo lithium abundance of N(Li) = 2.12 $\pm$ 0.03 dex, where the uncertainty represents the error in the mean. The scatter about the mean trend is 0.12 dex, suggesting that the scatter in the lithium abundances is real. Lithium abundances were calculated using a modern set of stellar atmosphere models, which systematically give abundances that are larger by 0.2 dex than those computed with older atmospheric models. In one of rare cases where the original analysis has been rectified, the Thorburn (1994) observations shown in Figure 8 have been "corrected" by -0.2 dex.

It is common practice, following Spite & Spite (1982), to assume that the lithium abundance of very metal-poor stars represents the primordial value. An alternative suggestion is that the primordial value is actually much higher, $N(^7Li) \gtrsim 3$, and all the values measured today are the results of depletion (Applegate, Hogan & Scherrer 1987, 1988; Alcock, Fuller & Mathews 1987; Malaney & Fowler 1988; Fuller, Mathews & Alcock 1988; Mathews, Alcock & Fuller 1990; Brown 1992). How the age, metallicity and mixing effects of stellar evolution conspire to yield a constant value for halo stars is not well understood in this alternative viewpoint.

Beyond the Spite plateau, the upper envelope of the observations smoothly increases to a maximum value that is an order of magnitude greater than those found in the most metal-poor dwarfs. Thus, if one assumes that the Spite plateau is the primordial $^7Li$ abundance, then one important consequence is that an important site (or sites) of lithium production has existed.

Below the upper envelope are dwarfs that span the entire range of measurable lithium abundances because it is very easy to destroy lithium in a star. As a dwarf star evolves the atmospheric lithium content is subject to diffusive, convective, and meridional circulation transport mechanisms that move the $^7Li$ into regions where the temperature is in excess of a few million degrees (Michaud, 1986, Vauclair 1988; Michaud & Charbonneau 1991), where it is quickly burned away by (p,$\gamma$) and (p,$\alpha$) reactions.

Agreement of the calculations with the Spite plateau is due strictly to the infall of primordial material with the homogeneous Big Bang composition of Walker *et al.* (1991). At metallicities larger that [Fe/H] $\simeq$ -1.0 dex, injections of freshly synthesized $^7Li$ into the interstellar medium by the $\nu$-process causes the lithium abundance to rise above the Spite plateau. The exploded massive star models are producing lithium prior to [Fe/H] $\simeq$ -1.0 dex, but the contributions are small compared to the infall values. Figure 8 shows that these $^7Li$ contributions can account for a significant fraction of the shape and amplitude of the upper envelope, and that Type II supernova contribute about 1/2 of the solar $^7Li$ abundance (also see Figure 4).

### 3.3.2 $^{11}B$

The evolution of elemental boron (whose dominant isotope is $^{11}B$) as a function of [Fe/H] is shown in Figure 9. The calculated boron abundance is shown as the solid line, while the dashed lines depict factors of two variation in the $\nu$-process yields. Boron is very difficult to measure in stars because it is a trace element, and all of its usable transitions are in the ultraviolet. As a result there have been very few boron abundance determinations in stars. The solar boron abundance (Anders & Grevesse 1989) is also shown in the figure.

Boesgaard & Heacox (1978) determined the atmospheric boron abundances in 16 B and A field stars from high resolution ($\Delta\lambda \simeq 0.05$ Å) phototube spectra of the singly-ionized 1362 Å feature. The effective temperatures and surface gravities of the survey stars were taken from the literature. Unfortunately, the iron abundances for the survey stars were not reported. In plotting their observations in Figure 9, the compilation of Cayrel de Strobel et al. (1992) was used to determine some of the [Fe/H] values. Preference was generally given to the most recent entry in the compilation. Boesgaard & Heacox (1978) generally found a solar abundance of boron in the survey stars.

Duncan, Lambert & Lemke (1992) examined the surface boron abundances in the halo dwarfs HD 19445, HD 140283, and HD 201891 using high resolution (R$\simeq$25,000), moderate-noise (S/N$\simeq$70) digital spectra of the ultraviolet B I resonance line near 2497 Å. The effective temperature, surface gravity, microturbulence, and metallicity of the three survey stars were taken from the literature. An LTE absolute analysis was used with experimentally measured B I oscillator strengths, and line blanketing from millions of iron-peak spectral lines were included in the analysis. Their results strongly suggest a linear relationship between boron abundances and [Fe/H] in the halo. The linearity implies that boron must be synthesized along with other metals (Walker et al. 1993; Fields, Schramm & Truran 1993).

Lemke, Lambert & Edvardsson (1993) studied the atmospheric boron abundances in the F dwarfs Procyon, $\theta$ Ursae Majoris, and $\iota$ Pegasi using high resolution (R$\simeq$80,000), low-noise (S/N$\simeq$100) digital spectra of the ultraviolet B I resonance line near 2497 Å. The abundance analysis was very nearly the same as Duncan, Lambert & Lemke (1992). They concluded that the boron abundance, N(B) $\simeq$ 2.3, of $\theta$ UMa and $\iota$ Peg were similar to the Boesgaard & Heacox (1978) abundances. Procyon, which has a nearly solar [Fe/H], was found to be depleted in boron by a factor of 3.

However, Kiselman (1994) and Edvardsson et al. (1994) have recently reinvestigated the formation of the neutral boron resonance lines taking into account various departures from LTE (i.e deviations from detailed balance). They find that the blue and ultraviolet radiation field in solar-type dwarfs has a hotter radiation temperature than the local electron temperature, which drives the line source function above the Planckian value. This makes the absorption lines weaker than in LTE, hence increasing the derived boron abundance. The effect is insignificant for the Sun, but increases for more metal-poor stars. They suggest that the boron abundances of the 3 metal-poor dwarfs in Figure 9, along with Procyon, should have their boron abundances increased by roughly 0.3 dex. These non-LTE correction factors have been applied to the data sets shown in Figure 9.

Most of the attention for producing $^{10}$B and $^{11}$B has focused on the spallation events, even though the cross sections can be measured and calculated with much greater precision than the very uncertain flux history. Closed or leaky box cosmic ray models have had reasonable success at producing the isotope $^{10}$B, but these models generally predict a solar ratio of $^{11}$B/$^{10}$B = 2.5 (Reeves, Fowler & Hoyle 1970; Meneguzzi, Audouze, & Reeves 1971; Meneguzzi & Reeves 1975; Austin 1981; Read & Viola 1984; Walker, Mathews & Viola 1985; Abia & Canal; 1988; Walker *et al.* 1993; Reeves 1994). To obtain the correct solar ratio of $^{11}B/^{10}B = 4.6$ an unobserved, low energy component of the cosmic ray spectrum has been commonly postulated. The cross sections of $^{10}B$ and $^{11}B$ in this low energy regime differ enough to lead to a larger isotopic ratio. A difficulty with this *ad hoc* spallogenic origin of $^{11}B$ is that the resulting lithium production considerably exceeds the observed values in halo dwarfs (Prantzos, Cassé & Vangioni-Flam 1993). The difficulties in producing the observed $^{11}B/^{10}B$ ratio have often prompted suggestions that there may be an alternative nucleosynthetic origin site for $^{11}B$.

The fit to the observations shown in Figure 9, which span over 2 orders of magnitude, and include the solar abundance, is encouraging. When comparing the $^{11}B$ calculations to the observed elemental boron abundances, one is assuming that $^{11}B$ dominates over $^{10}B$ throughout the Galactic history. The doublet at 2090 Å has not been extensively observed and may be useful for studying the isotopic ratios of boron (Johansson *et al.* 1993). Using some of our preliminary $^{11}B$ yields, Olive *et al.* (1993) and Fields (1994) found similar results. In addition, they found that by including a standard cosmic ray model, without the unobserved low-energy component to the cosmic ray spectrum, that the solar $^{11}B/\ ^{10}B$ ratio was reproduced.

### 3.3.3 $^{19}F$

The evolution of fluorine to oxygen ratio [F/O] as a function of [Fe/H] is shown in Figure 10. The calculated abundance history is shown as the solid line, while the dashed lines depict factors of two variation in the $\nu$-process yields. Oxygen is used as the basis of comparison instead of iron because fluorine and oxygen are neighbors in the periodic table, and they are expected to be synthesized in close proximity to each other. Fluorine is very difficult to measure in stars primarily because it is the least abundant of all the stable $12 \leq A \leq 38$ nuclides. The solar abundance of $^{19}F$ is based on analysis of the hydrogen fluoride molecule in sunspots (Anders & Grevesse 1989).

Jorissen, Smith & Lambert (1992) examined the fluorine abundances for several classes of red giant stars from high resolution (R$\simeq$100,000) infrared spectroscopy of the rotation-vibration bands of hydrogen fluoride. Their observations constitute the only information about fluorine abundances outside the solar system. The survey stars included four K field giants that appear to be chemically normal, a few barium stars (which are binary systems), and an assortment of extreme carbon stars (MS, S, SC, N and J spectral types). The stellar atmosphere model parameters were taken from the literature, and synthetic spectra determined the final fluorine abundances.

The significant deviation from a solar [F/O] ratio at low values of [Fe/H] in Figure 10 is a direct reflection of the enhanced oxygen abundance at these metallicities (see §3.4.1)

Thus, our model makes the prediction that metal-poor dwarfs will have a subsolar [F/O] ratio. At larger metallicities the model fits the chemically normal K giant observations. The fit to red giants stars that are chemically peculiar is not good, nor should the model be expected to fit these spectral classes as they are dominated by the effects of low mass stellar evolution rather than chemical evolution. However, the fluorine abundances observed in the chemically peculiar red giants means that Galactic fluorine receives a significant contribution from asymptotic giant branch stars.

The subsolar fluorine to oxygen ratio ([F/O]=-0.5 dex) content of the most metal deficient ([Fe/H]=-0.6 dex) K giant ($\alpha$ Boötes, Arcturus), in contrast to the claim made by Jorissen et al. (1992), strengthens the claim that Type II supernovae are responsible for a significant fraction of Galactic fluorine production. The [F/O] ratio is subsolar at this metallicity since [O/Fe] is enhanced (see §3.4.1), not because of supressed fluorine production. It may be that 1/2 of the solar fluorine abundance comes from Type II events and 1/2 from asymptotic giant branch stars.

### 3.3.4 SUMMARY OF $\nu$-PROCESS CONTRIBUTIONS

The sum of Figures 4, 8, 9 and 10 suggests that the $\nu$-process provides an an attractive site for the nucleosynthetic origin of the rare and fragile $^7$Li, $^{11}$B, and $^{19}$F isotopes. The major evolutionary features and a significant fraction of their solar abundances are accounted for. It is also gratifying that the agreement is achieved for $\mu$ and $\tau$ neutrino temperatures in the range 6 to 8 MeV, which just the range suggested by SN 1987A and preferred by those who model core collapse of massive stars. The $\nu$-process, cosmic ray spallation, and homogeneous Big Bang nucleosynthesis are complimentary, and together they yield a comprehensive set of prescriptions for the evolution of the light (A $\leq$ 11) elements.

## 3.4 THE CNO NUCLEI

The most abundant elements in the Galaxy after hydrogen and helium are the CNO triad. They are important in stellar interiors as opacity sources, and in energy production through the CNO cycle. Thus, observations of the CNO abundances in main-sequence stars are valuable for recording what types of stars have been responsible for CNO nucleosynthesis during different Galactic evolutionary phases.

### 3.4.1 OXYGEN

Evolution of the oxygen to iron ratio [O/Fe] as a function of [Fe/H] is shown in Figure 11. The solid line shows the calculation, the dashed lines indicate factors of two variation in the iron yields from massive stars and the dotted line shows the results when Type Ia supernovae are excluded. The observations shown in the figure employ the [O I] doublet. The O I triplet at 7700Å is a very high excitation feature (9.15 eV) and quite sensitive to the temperature structure of the adopted model atmosphere, especially in stars for which the Boltzmann factor $kT \simeq 0.4$ ev. Thus, oxygen abundances derived from the triplet lines are generally unsafe, and most investigators prefer to base results on the forbidden lines.

The convective envelope of low mass stars near the base of the red giant branch were predicted not to reach down deep enough into material where significant ON processed material could be dredged-up to the surface (Iben 1964, 1967; Iben & Renzini 1983). Thus, surface oxygen abundances should not be affected in G and K field giants, an effect confirmed by Leep & Wallerstein (1981) and Lambert & Ries (1981). Field giants then, and not just dwarf stars, may be valuable tools for analyzing *ab initio* oxygen abundances.

Gratton (1985) used photographic spectra to determine the Fe and Ti abundances in 23 field giants, and were combined with a weighted mean of 6300 Å [O I] equivalent widths found in the literature to derive the oxygen abundances. The survey stars spanned the metallicity range of $-0.25 \leq$ [Fe/H] $\leq 0.16$ dex. Effective temperatures were derived from an LTE model atmosphere analysis of the excitation temperatures of Fe I, and surface gravities were found from the ionization equilibrium of titanium. The microturbulence parameters were extracted from the requirement that Fe I abundances be independent of equivalent width, and a differential analysis was adopted. The resulting oxygen abundances compared well with those from whom the equivalent widths were taken, but tended to be systematically lower. This was attributed to the unusual choice of using Ti for the surface gravities, which give a smaller surface gravity scale than those based on iron.

Gratton & Ortolani (1986) examined oxygen abundances in 18 southern sky stars using high resolution (R$\simeq$100,000), low noise (S/N $\simeq$ 150) photodiode spectra of the forbidden [O I] feature at 6300 Å. The dwarfs and giants in the survey spanned a metallicity range of $-2.35 \leq$ [Fe/H] $\leq 0.37$ dex. The effective temperatures were derived from calibrated V-R, R-I and V-K color indices. Surface gravities were provided by the ionization equilibrium of iron, and the microturbulences were found by avoiding any trend of the derived abundances with the equivalent widths. A differential analysis was used, and the oscillator strengths were derived from an inverse analysis of the solar spectrum. They found that [O/Fe] $\simeq 0.5$ dex in metal-poor stars with a smooth decline down to a metallicity of [Fe/H] $\simeq -0.8$ dex. At larger metallicities the oxygen abundances rapidly declined to solar values, but with a fair amount of scatter.

Barbuy & Erdelyi-Mendes (1989) used high resolution (R$\simeq$=60,000), low noise (S/N $\simeq$ 200) photodiode and digital spectra of the forbidden [O I] line at 6300 Å to determine the oxygen abundances in 24 southern sky dwarfs and field giants. Based on an analysis of the space velocities and eccentricities, the survey stars included both both halo and thick disk populations. Effective temperatures were found from calibrated (i.e corrected for interstellar extinction) B-V, *b-y*, B2-V1, R-I, V-R and V-K color indices. The surface gravity was found fitting the Sc II line at 6300.6 Å with hyperfine structure effects taken into account. This unusual technique was motivated by suggestions that [O I] and Sc II lines are formed in similar regions of the atmosphere. They essentially imposed the condition that [Sc/Fe] = 0.0 dex and then derived [O/Fe]. A differential analysis was adopted with the oscillator strengths derived from an inverse solar analysis. Their analysis suggested that oxygen is overabundant relative to iron at the factor of three level for [Fe/H] $\leq$ -0.8 dex. For -0.8 $\leq$ [Fe/H] $\leq$ -0.5 dex there was a spread in the [O/Fe] ratio, which the authors suggested as further evidence of a thick disk stellar population.

Peterson, Kurucz & Carney (1990) determined the atmospheric oxygen abundances for the extremely metal-poor giants HD 184711 and HD 122563, with metallicities of [Fe/H] = -2.64 and -2.93 dex respectively. Their analysis was based on high resolution (R=15000), low noise (S/N $\simeq$ 100) digital spectra of the [O I] line at 6300 Å. Effective temperatures were primarily determined from a calibrated V-K color index, and verified by examining the J-K color index, along with iron abundances determined through the weak Fe I lines. Surface gravities were found from the condition that the same iron abundance result from Fe I and Fe II transitions. Microturbulent velocities were extracted from the condition that abundances deduced from both weak and strong Ca I and Fe I lines exhibit no dependence on the equivalent width. An absolute analysis was adopted, with the majority of the *gf*-values coming from terrestrial furnace experiments. They found that HD 184711 possessed [O/Fe] = 0.80 dex, and HD 122563 had [O/Fe] = 0.18 dex.

Oxygen is the dominant element ejected by Type II supernovae. Figure 11 shows the [O/Fe] ratio has its largest value at extremely low metallicities, and the slowly decreases due to mass and metallicity variations, an effect anticipated by Tinsley (1980). Intermediate mass stars begin to dominate contributions to the interstellar medium later on, but because they produce very little oxygen and no iron the [O/Fe] ratio is not affected. Type Ia supernovae begin to inject large amounts of iron and negligible amounts of oxygen into the interstellar medium beyond about 1 Gyr, which causes the distinct downturn of [O/Fe] to the solar value. This is shown by comparing the solid line in the figure (which includes both Type II and Type Ia supernovae) with the dotted line (which excludes Type Ia supernovae). Determining observationally the metallicity at which Type Ia supernovae begin to effect the [O/Fe] ratio is very difficult, as the uncertainties in even the best of the observations is much larger than the theoretical effect (Wallerstein 1994). The figure suggests that the best fit to the [O/Fe] observations may be a systematic reduction of the Type II iron yields by a factor of two. The reduced iron yields are consistent with observations of several Type II supernovae including SN 1987A (see references and discussion in Eastman *et al.* 1994), and within the uncertainty in modeling the explosion. A factor of two increase in the iron yield is excluded.

Figure 12 shows the evolution of the less abundant oxygen isotopes as a function of the total metallicity [Z]. The isotope $^{17}$O comes from hydrogen burning by the CNO tricycle (e.g Rolfs & Rodney 1988) and is produced quite well in massive stars. The isotope $^{18}$O comes from $\alpha$ captures on $^{14}$N during helium burning. Since these oxygen isotopes come from different processes, the yields vary for stars of different masses and metallicities (Paper I). Interestingly, these two oxygen isotopes show nearly identical behavior over the metallicity range shown due to their direct dependence on the initial CNO abundance. The roughly linear [O18/O16] was also found in the study of Woosley & Weaver (1982a). However, Type Ib supernovae which have lost only their hydrogen envelope may affect the $^{17}$O abundance (Woosley et al. 1993, 1994). In addition, $^{17}$O is also produced by all intermediate–low mass stars after the first dredge-up and may survive further processing (Iben & Renzini 1983, Lattanzio 1994; Boothroyd, Sackmann & Wasserburg 1994; García-Berro & Iben 1994). Oxygen isotopic abundances have been measured through molecular CO emission in the interstellar medium (Wannier 1980; Penzias 1981) and in several types of carbon stars (Harris, Lambert & Smith 1985, 1987; Harris et al. 1987). The isotopic ratios found in the interstellar medium tend to be about 40% smaller than the solar ratio, while those observed in the asymptotic giant branch stars tend to be much larger than expected from canonical intemediate-low mass stellar evolution theories (although see Boothroyd et al. 1994).

### 3.4.2 CARBON

Evolution of the carbon to iron ratio [C/Fe] as a function of [Fe/H] is shown in Figure 13. The solid line shows the calculation,the dashed lines indicate factors of two variation in the iron yields from massive stars and the dotted line shows the results when Type Ia supernovae are excluded. The [C/Fe] ratio in halo and disk dwarfs has been observed to be roughly constant for a long time (Sneden 1974; Bell & Branch 1976; Clegg 1977; Peterson 1978; Peterson & Sneden 1978; Clegg, Lambert & Tomkin 1981; Tomkin & Lambert 1984). However, the number of stars and the metallicity range spanned in some of these early surveys are modest. Field giants are, unfortunately, not reliable indicators of the *ab initio* carbon abundance because of effects from the first dredge-up. The three surveys of carbon in nearby dwarf stars shown in Figure 13 employ a relatively large number of unevolved stars and span a large range of metallicity.

Laird (1985) determined carbon abundances in 116 dwarfs using intermediate resolution ($\Delta\lambda = 1$ Å) image tube spectra of 4290-4328 Å band features at molecular CH. The survey stars spanned the metallicity range -2.5 $\leq$ [Fe/H] $\leq$ 0.5 dex, with the majority of the stars having [Fe/H] > -2.0 dex. Effective temperatures were found from calibrated R-I, $b$-$y$ and V-K color indices. Surface gravities were derived from the spectra and Strömgren photometry, supplemented with gravities based on parallax data and estimated masses. A differential analysis was adopted, and equivalent widths of the Fe I lines were used to determine the iron abundances. Since no individual CH lines could be detected in the spectra, LTE synthetic spectra determined the final carbon abundances. Laird found [C/Fe] = -0.22 dex, with an intrinsic scatter of 0.10 dex, over the entire metallicity range. An analysis of the [C/Fe] ratio as a function of the effective temperature indicated a systematic offset, so a correction factor of 0.20 dex was applied to all the [C/Fe] ratios.

Tomkin, Sneden & Lambert (1986) examined of 32 halo dwarfs that spanned the metallicity range -2.5 ≤ [Fe/H] ≤ -0.7 dex. They used high resolution ($\Delta\lambda$ = 0.25 Å), low noise (S/N > 100) photodiode spectra centered on the 4330 Å G-band feature of the CH molecule. Effective temperatures were the mean values derived from calibrated V-K, $b$-$y$, and R-I color indices. The surface gravities were found from a log $g$ - $T_{eff}$ relationship calibrated for metal-poor main-sequence stars. A constant 1.0 km s$^{-1}$ microturbulence was used for all stars, equivalent widths of Fe I determined the iron abundances, and a differential analysis was adopted. LTE spectrum synthesis of the CH molecule was used to derive the final carbon abundances. Their analysis found the carbon to iron ratio was [C/Fe] = -0.2 ± 0.15 dex for [Fe/H] > -1.8 dex. At lower metallicities, they reported a trend of increasing [C/Fe] – ending with [C/Fe] = 0.2 dex at [Fe/H]=-2.5 dex. In addition, they noted that carbon abundances deduced from C I lines (which are only measurable for [Fe/H] > -1.0 dex) appeared to be systematically larger by 0.3 dex than abundances determined from molecular CH features.

Carbon *et al.* (1987) derived carbon abundances for 83 dwarfs from low resolution ($\Delta\lambda$ = 8 Å), moderate noise (S/N ≃ 50) image tube spectra centered on the 4300 Å feature of the CH molecule. The survey stars spanned the metallicity range -2.5 ≤ [Fe/H] ≤ -0.6 dex, with most of the stars having [Fe/H] ≥ -1.5 dex. Effective temperatures were found from calibrated B-V and H$\delta$ indices, and surface gravities were determined through isochrone fits in the log $g$ - $T_{eff}$ plane. A roughly constant 1.25 km s$^{-1}$ microturbulence parameter was adopted, and a differential analysis was used. LTE spectrum synthesis of the CH molecular features determined the carbon abundances. They found a solar [C/Fe] ratio over most of the metallicity range surveyed, with a star-to-star scatter of 0.18 dex. They also reported an upturn in the [C/Fe] values at the very lowest metallicities. However, they showed that the upturn of carbon is quite sensitive to the choice of assumed oxygen abundance, due to formation of molecular CO among dwarfs with $T_{\text{eff}}$ ≤ 5500 K. They showed that the upturn of carbon is significantly reduced when [O/Fe] ratios smaller than 0.6 dex are adopted.

Wheeler *et al.* (1989) reanalyzed the three surveys, and attempted to bring them onto a common effective temperature and abundance scale. To first order, the overall trend toward larger [C/Fe] ratios in the most metal poor-stars does not disappear. However, they caution that oxygen abundances for most of the stars in the three surveys are missing, so that possibly important effects from molecular CO are uncertain. Additional evidence for the slight upturn of [C/Fe] may come from extragalactic H II regions. With the nebulae, one has to contend with several stages of ionization in the ultraviolet, with some of the stages only being determined from International Ultraviolet Explorer or *Hubble Space Telescope* observations (Edmunds 1989).

A solar and relatively flat [C/Fe] ratio is quite interesting because carbon and iron are synthesized by very different processes at different stages in the Galaxy's evolution. Carbon is produced by the triple-alpha process during hydrostatic helium burning in massive, intermediate and low mass stars. Iron on the other hand, is synthesized in explosive burning conditions by Type II and Type Ia supernova (see §3.2). The relative contributions and evolutionary time scales for each of these origin sites of carbon and/or iron are quite different.

Figure 13 shows that massive star synthesis of primary carbon and iron are sufficient to explain the metal-poor halo dwarf observations. The small undulations are due to mass and metallicity variations. The uncertainties associated with these variations, combined with the smallness of the suggested observational effect, do not allow definitive statements concerning enhancement of the [C/Fe] ratio at the lowest metallicities. Two competing sources come into play at [Fe/H] $\simeq$ -0.8 dex. Intermediate and low mass stars begin depositing large amounts of carbon but no iron, while Type Ia supernovae start injecting significant amounts of iron but no carbon. The interplay between these two sources, such that [C/Fe] $\simeq$ 0 dex, places tight constraints on chemical evolution models (e.g lifetimes and amplitude of Type Ia contributions). Figure 13 shows that the calculation has intermediate mass stars raising the [C/Fe] ratio slightly before Type Ia events depress it, but the amplitude of these excursions are well within the intrinsic scatter of the observations. This point is illustrated by comparing the solid line in the figure (which includes both Type II and Type Ia supernovae) with the dotted line (which excludes Type Ia supernovae).

Both the $^{12}$C and $^{13}$C isotopes are underproduced in the solar metallicity massive star models by about a factor of 3, so that explanation of the solar system abundance is consistent with their proposed origin in dredged-up material from helium shell flashes, and incomplete carbon burning in intermediate-mass stars on the giant branches (Iben & Truran 1978; Renzini & Voli 1981). Additionally, if low mass stars are the principal source of $^{13}$C, then more of it should be found towards the center of the Galaxy where more stellar processing has occurred. The observed $^{13}$C/$^{12}$C ratio shows this radial behavior (Wilson & Matteucci 1992; Wilson & Rood 1994).

### 3.4.3 NITROGEN

Evolution of the nitrogen to iron ratio [N/Fe] as a function of [Fe/H] is shown in Figure 14. The solid line shows the calculation, and the dashed line shows a calculation with a larger amount of convective overshoot (Weaver & Woosley; unpublished). Field giants are, unfortunately, not reliable indicators of nitrogen history because of effects from the first dredge-up. The two surveys of nitrogen in nearby dwarf stars shown in the figure are based on low resolution spectra, but have the advantage of possessing a large number of stars, and spanning a large range of metallicity. However, the apparent agreement between the surveys belie the great difficulty in determining [N/Fe] ratios in dwarfs.

Laird (1985) determined nitrogen abundances in 116 dwarfs using intermediate resolution ($\Delta\lambda = 1$ Å) image tube spectra on the 3360 Å band heads of molecular NH. The survey stars covered the metallicity range $-2.5 \leq$ [Fe/H] $\leq 0.5$ dex, with most of the stars having [Fe/H] $> -2.0$ dex. Some details about the type of analysis used and how the stellar atmosphere parameters were established were discussed in §3.4.2. Laird found [N/Fe] = $-0.67 \pm 0.2$ dex over the entire metallicity range. Some of the survey stars had been studied previously, and indicated that [N/Fe] $\simeq = 0$ dex for these stars (Sneden 1974; Clegg 1977; Clegg, Lambert & Tomkin 1981; Tomkin & Lambert 1984). While searching for some systematic offset, Laird applied a zero-point correction factor of 0.65 dex to all the [N/Fe] ratios.

Carbon *et al.* (1987) derived nitrogen abundances for 83 dwarfs from low resolution ($\Delta\lambda \simeq 8$ Å), moderate noise (S/N $\simeq 50$) image tube spectra of the NH molecular band head at 3360 Å. The survey stars covered the metallicity range $-2.5 \leq$ [Fe/H] $\leq -0.6$ dex, with the majority of the stars having [Fe/H] $< -1.5$ dex. Some details about the type of analysis used and how the stellar atmosphere parameters were established were discussed in §3.4.2. Their careful analysis found an average [N/Fe] ratio of -0.18 dex, with a small trend of decreasing [N/Fe] at lower metallicities. However, they showed that systematic changes in [N/Fe] with the adopted effective temperature of the most metal-poor stars caused the trend to disappear.

A small number of halo dwarfs in the Laird (1985) and Carbon *et al.* (1987) surveys have large nitrogen overabundances. Spite & Spite (1986, 1987) suggested that the high [N/Fe] ratios in 4 of these stars cannot be due to internal CN cycle processing because it would disrupt the observed lithium abundances. Various internal mixing mechanisms apparently cannot induce significant nitrogen enhancements without turning the stars into a blue stragglers (Da Costa & Demarque 1982), although they only considered more metal-rich stars. Some halo binaries might have the appropriate masses and orbital separations to undergo the evolution required to produce metal-poor, but nitrogen rich stars.

If nitrogen production depends on the initial CNO abundance, then the rough trend [N/Fe] $\simeq$ [Fe/H] would be expected. If, however, nitrogen production is chiefly primary, then the [N/Fe] ratio would be nearly independent of the [Fe/H] ratio. The Laird (1985) and Carbon *et al.* (1987) studies strongly suggest that [N/Fe] is constant over the entire metallicity range, although some uncertainty exists in the precise value of the constant. The main conclusion, stated in both these surveys, is that nitrogen has a strong primary component. In addition, an independent source of evidence for a partial source of primary nitrogen comes from the radial gradients of CNO abundances in extragalactic H II regions (Edmunds & Pagel 1978; Alloin *et al.* 1979; Pagel & Edmunds 1981; Dufour, Schiffer & Shields 1984; Edmunds 1989).

The solid line in Figure 14 shows that no primary nitrogen is produced in the standard massive star models. However, if the numerical parameters governing convective overshoot are artificially enlarged, then primary nitrogen is produced in all the low metallicity massive stars heavier than 30 $M_\odot$ (Weaver & Woosley, unpublished). The synthesis of primary nitrogen in these low metallicity massive stars occurs as the convective helium burning shell penetrates into the hydrogen shell with violent, almost explosive, consequences. No primary production is found, using the same enhanced convective prescription, in massive stars of initially solar metallicity. The dashed line shows the results of using the enhanced convection prescription. It must be stressed that the exact amount of primary nitrogen is not determined very well computationally as it is sensitive to the zoning employed and the treatment of convection. However, the amount of convective overshoot necessary to produce primary nitrogen is whithin the uncertainties of convection theory. Turning the situation around, such sensitivity may indicate that the observed [N/Fe] ratios are an excellent discriminant of various treatments of convection.

Nitrogen is underproduced in the solar metallicity massive star models by about a factor of 3, which is consistent with the bulk of its production attributable to intermediate and low mass stars (Iben & Truran 1978; Renzini & Voli 1981). The abundance of the neutron rich isotope of nitrogen, $^{15}$N, is enhanced by the neutrino process in massive stars (see §3.1). The fact that it is within a factor of three of its solar value suggests that the relevant neutrino cross sections and temperature spectrum should be examined more closely. But for now, it seems that the dominant production site $^{15}$N must come from somewhere else (also see Matteucci & D'Antona 1991). Hot hydrogen burning (Lazareff *et al.* 1979; Woosley 1986) has been suggested, with environments typified by classical novae systems, although how much $^{15}$N is made in novae or intermediate-low mass stars in unknown.

### 3.5 THE LIGHT METALS

Sodium, magnesium and aluminum are adjacent to each other in the periodic table, and have large production factors in the ashes of hydrostatic carbon and neon burning.

The $\alpha$-chain isotope $^{24}$Mg has two stable neutron rich sisters, $^{25}$Mg and $^{26}$Mg, while sodium and aluminum have only one stable isotope, $^{23}$Na and $^{27}$Al, since even-A odd-Z isotopes are generally unstable (Fermi 1950, Evans 1955). Perhaps even more importantly, all these isotopes can be observed in the spectra of G and K dwarfs – allowing a direct comparison to the theoretical calculations.

### 3.5.1 MAGNESIUM

Evolution of the magnesium to iron ratio [Mg/Fe] as a function of [Fe/H] is shown in Figure 15. The solid line shows the calculation, the dashed lines shows variations of two in the iron yields from massive stars and the dotted lines show variations in the exponent of the initial mass function. Magnesium has been studied intensively due to its low ionization potential and a number of significant transitions in the optical region of the spectrum.

Gratton & Sneden (1988) derived magnesium abundances for 13 extremely metal-poor field giants from digital spectra of the sodium 5172, 5183, 5528, and 5711 Å Mg I lines. Resolution of the spectra was R $\simeq$ 20,000, with signal-to-noise ratios usually in excess of 150. The survey stars ranged in metallicity from [Fe/H] = -3.0 to -1.3 dex. Effective temperatures of the stellar atmospheres were taken to be the mean value of calibrated V-R and V-K color indices. Corrections for interstellar reddening were derived from the strength of the sodium D lines. Surface gravities for field giants fraught with difficulties, and great care was given in deriving a mean value from ionization equilibrium of iron, titanium, chromium, and the wings of strong iron lines. Microturbulent velocities were found from the requirement that there be no trend of the iron abundances derived from Fe I lines with equivalent width. An absolute analysis was adopted, with experimental measurements or theoretical calculations providing the required oscillator strengths. The results of the analysis suggested a mean magnesium to iron ratio of [Mg/Fe] = 0.27 dex, with a star-to-star scatter of 0.14 dex, with a slight trend of increasing [Mg/Fe] with decreasing [Fe/H].

Magain (1989) determined the magnesium to iron ratio in 20 extremely metal-poor halo dwarfs from digital spectra of the 8806 Å Mg I line. The survey stars spanned the range -1.4 $\leq$ [Fe/H] $\leq$ -2.9 dex. Effective temperatures were found from calibrated V-K and $b$-$y$ color indices, with no reddening correction because the dwarfs are nearby. Surface gravities were determined by forcing the Fe I and Fe II lines to give the same iron abundance. Microturbulent velocities were found from the requirement that Fe I lines with equivalent widths between 10 mÅ and 70 mÅ yield the same iron abundance. An absolute analysis was used, with most of the oscillator strengths derived from laboratory furnace experiments. The results of the survey showed that [Mg/Fe] = 0.48 $\pm$ 0.09 dex, with the low scatter determined entirely by the observational uncertainties.

Edvardsson et al. (1993) used high resolution (R≃60,000), low noise (S/N≃200) photodiode and digital spectra of the 8712 Å and 8717 Å Mg I lines to determine magnesium abundances in 189 F and G stars in the solar vicinity. The northern and southern hemisphere dwarfs in this extensive survey were shown to have disk-like kinematics, and covered the metallicity range -1.0 ≤ [Fe/H] ≤ 0.2 dex. Effective temperatures of the stellar atmospheres were computed from calibrated $b$-$y$ and $\beta$ color, with $\beta$ having the advantage of being unaffected by interstellar reddening but not being as sensitive to the effective temperature as the $b$-$y$ color index. Surface gravities were estimated from the Balmer discontinuity, calibrated with zero age main-sequence relations for stars of solar metallicity. The microturbulence parameters were calculated from an empirically derived relationship between the microturbulence, surface gravity and effective temperature. Blanketing effects from millions of spectral lines were taken into account in the LTE model atmospheres. A differential analysis relative to the sun was used, with oscillator strengths determined from an inverse analysis of clean solar lines. Their results suggest that the magnesium to iron ratio [Mg/Fe] is approximately 0.5 dex at [Fe/H] = -1.0 dex, and then decreases steadily to near solar values at about [Fe/H] = 0.0 dex. At larger metallicities, the [Mg/Fe] ratio seems to turn back up to supersolar values. The authors suggest that part, if not most, of the star-to-star scatter is real and intrinsic. In addition, they note that a few of the disk dwarfs have anomalously large [Mg/Fe] ratios (and large sodium and aluminum abundances).

Massive star contributions to the magnesium abundance account completely for the halo dwarf observations (Figure 15). The undulations and gradual decline in the halo is due to small metallicity and mass effects. Intermediate or low mass star produce no magnesium or iron in the model, so the magnesium to iron ratio does not change when they dominate contributions to the interstellar medium. Later in the evolution the injection of iron from Type Ia supernovae becomes large enough to drive the [Mg/Fe] ratio down to the solar value. The best fit to the [Mg/Fe] observations may be a systematic reduction of the iron yields from massive stars by a factor of two (see also §3.4.1). Even with a reduced iron yield the model tends to underestimate the [Mg/Fe] ratio in the disk, which may be indicative of a small magnesium contribution originating from another source.

The dotted line which lies below the standard calculation (solid line; initial mass function exponent of -1.31) for most of the [Fe/H] evolution corresponds to an initial mass function exponent of -1.61. This evolution supresses the most massive stars, which decreases the average magnesium yield (and slightly enhances the iron yields). For the opposite reasons the dotted line (initial mass funtion exponent of -1.01) lies above the standard calculation. Figure 15 shows the general property of the $\alpha$-chain elements that the sensitivity of the results to variations in the exponent of the initial mass function is much less than the uncertainties associated with modeling the explosion mechanism (i.e placement of the mass cut and the explosion energy).

### 3.5.2 $^{25}$Mg and $^{26}$Mg

Evolution of the magnesium isotopes as a function of [Fe/H] is shown in Figure 16. The solid line represents the [$^{25}$Mg/$^{24}$Mg] ratio, and the dashed line is for the [$^{26}$Mg/$^{24}$Mg] ratio. Only the observed [$^{25}$Mg/$^{24}$Mg] ratios are plotted since splitting the differences between the observed $^{25}$Mg and $^{26}$Mg abundances is difficult.

Tomkin & Lambert (1980) estimated the magnesium isotopic abundances in five G and K dwarfs ($\mu$ Cas, $\epsilon$ Eri, 61 Cas A and B, and Gmb 1830) using the absorption lines of the A $^2\Pi$ - X $^2\Sigma^+$ system of the magnesium hydride molecule ($^{24}$MgH, $^{25}$MgH, and $^{26}$MgH) near 5130 Å. Resolution of the photodiode spectra was 0.08 Å, with a signal-to-noise ratio of about 100. Effective temperatures were derived from calibrated Johnson colors, and a constant surface gravity of 4.5 was assumed. Solar abundances with a microturbulence of 1 km s$^{-1}$ was adopted for $\epsilon$ Eri and 61 Cyg A and B, and a scaled 1/10 solar composition with zero microturbulence was used for $\mu$ Cas and Gmb 1830. Synthetic spectra determined the final abundances. Their analysis suggested that the disk dwarfs $\mu$ Cas, $\epsilon$ Eri, 61 Cas A and B, which all have [Fe/H] > -1.0 dex, were not significantly different from the solar ratio of $^{24}$Mg : $^{25}$Mg : $^{26}$Mg = 76.07 : 10.0 : 11.46 (Anders & Grevesse 1989). Gmb 1830, which was the most metal-poor star ([Fe/H] = -1.37 dex), displayed a significantly deficient ratio of $^{24}$Mg : $^{25}$Mg : $^{26}$Mg = 94 : 3 : 3.

Barbuy (1985, 1987) used high resolution (R=100,000), low noise ($\Delta\lambda$=0.05 Å) photodiode spectra of the A $^2\Pi$ - X $^2\Sigma^+$ magnesium hydride absorption line system near 5130 Å to estimate the magnesium isotope ratios in 16 southern hemisphere dwarfs which spanned the metallicity range -1.8 $\leq$ [Fe/H] $\leq$ 0.4 dex. The survey also included about an equal number of giant stars, but only the dwarf stars are shown in Figure 16 since slow neutron additions (e.g $^{22}$Ne($\alpha$,n) $^{25}$Mg) may distort the magnesium isotopic ratios. Effective temperatures were found from a weighted mean of calibrated R-I, V-R, V-K, B2-V1, B-V, and *b-y* color indices. Surface gravities were determined from color-color diagrams, and a constant 1.0 km s$^{-1}$ microturbulent parameter was used for all the stars. It was assumed that the compositions obeyed the scaling relation [Mg/H] = [Fe/H], and synthetic spectra determined the final isotopic ratios. The analysis suggested the striking result that the magnesium isotopic ratios in dwarfs were essentially solar across the entire metallicity range.

McWilliam & Lambert's (1988) examined the 5130 Å MgH band features in the atmospheres of 4 disk dwarfs with high resolution (R$\simeq$100,000), low noise (S/N$\simeq$ 80) photodiode spectra. The survey also included a few disk giants, a weak G-band giant and a barium rich star, but only the dwarf stars are shown in Figure 16. Model atmosphere parameters were taken from the literature, and synthetic spectra determined the magnesium isotopic ratios. The analysis suggested that the isotopic ratios in dwarfs stars with metallicities between -1.0 ≤ [Fe/H] ≤ 0.4 dex appear to be constant and solar.

For [Fe/H] ≳ -1.0 dex, the calculations agree fairly well with the magnesium isotope ratios found in disk dwarfs (Figure 16). The observed ratios in halo dwarfs are less cohesive, with some studies suggesting solar ratios while others indicate factors of 3 suppression. At metallicities of [Fe/H] ≃ -1.0 dex, the model is about a factor 2 smaller than any of the observations would suggest, which may indicate some production from another source. It is predicted that intermediate mass stars could produce significant amounts of the magnesium isotopes through the s-process (Truran & Iben 1977; Iben & Renzini 1983), but precise yields have not been determined.

### 3.5.3 SODIUM

Evolution of the sodium to iron ratio [Na/Fe] as a function of [Fe/H] is shown in Figure 17. Again the solid line shows the calculation; dashed lines indicate variations of two in the iron yields from massive stars. Using magnesium instead of iron as the basis for comparison eliminates Type Ia supernova from consideration because nearly all the sodium and magnesium are produced in massive stars. Any odd-even effects will be enhanced by using magnesium as the basis of comparison. Evolution of the [Na/Mg] ratio as a function of [Fe/H] is shown in Figure 18.

Peterson (1981) examined the atmospheric sodium abundances in 7 metal-poor field dwarfs from high resolution ($\Delta\lambda$ = 0.25 Å) image tube spectroscopy of the 5682 Å and 5688 Å Na I lines. The stars in the survey (which included a a few subgiants) spanned the metallicity range -0.5 ≤ [Fe/H] ≤ -2.7 dex. Effective temperatures of the stellar atmospheres were found from calibrated R-I color indices. Surface gravities were assumed to be 4.5 for dwarfs with effective temperatures below 5500 K, and 4.0 for those dwarfs above this effective temperature. This assumption was supported in some cases by the ionization equilibrium of iron and titanium. Microturbulence parameters were found by requiring that Fe I line abundances be independent of variations in equivalent width. A differential analysis relative to the Sun was adopted, with the line transition probabilities chosen to match solar equivalent widths. The results of the analysis suggested that sodium tends to become increasingly deficient with respect to either iron or magnesium as the overall metallicity declines.

Gratton & Sneden (1988) derived sodium abundances for 13 extremely metal-poor field giants from moderate resolution (R≃ 20,000), low noise (S/N≃150) digital spectra of the sodium D (5890 Å & 5896 Å) lines. The survey stars ranged in metallicity from [Fe/H] = -3.0 to -1.3 dex. Some details about the type of analysis used and how the stellar atmosphere parameters were established were discussed in §3.5.1. They found that the [Na/Fe] ratio was essentially solar over the metallicity range covered in their survey, but that the [Na/Mg] ratios were about 0.3 dex larger, on average, than previous surveys of halo dwarfs. However, abundances based on the sodium D lines should be viewed with caution because of difficulties in continuum placement (the equivalent width is largely in the wings on the line), and the large uncertainties in the damping constants (Wheeler et al. 1989).

Later surveys by Sneden et al. (1991) and Kraft et al. (1992) had three metal-poor field giants in common with the Gratton & Sneden (1988) survey, namely HD 166161, HD 26297, and HD 204543. These more recent studies employed four subordinate Na lines, two near 5685 Å and two near 6160 Å, with higher resolution and lower noise spectra than the Gratton & Sneden spectra. The same analysis tools are employed in all the investigations, so there is a common analysis procedure (Kraft 1994). In addition, the subordinate lines near 5685 Å have been shown to be formed in LTE (Drake, Smith & Suntzeff 1992). The Gratton & Sneden (1988) analysis suggested [Na/Fe] of 0.11, 0.00, and -0.01 dex for these three stars, whereas the later analyses indicated that values of -0.05, -0.35 and -0.40 dex, respectively. Thus, the Gratton & Sneden sodium abundances appear to systematically too large by about 0.3 dex. These "corrected" results are used in Figures 17 and 18. As a result, these stars fall much more in line with the Peterson (1981) results.

Edvardsson et al. (1993) used high resolution (R$\simeq$60,000), low noise (S/N$\simeq$200) photodiode and digital spectra of the 6154 Å and 6160 Å Na I lines to determine sodium abundances in 189 F and G stars. The northern and southern hemisphere dwarfs in this extensive survey were shown to have disk-like kinematics, and covered the metallicity range -1.0 $\leq$ [Fe/H] $\leq$ 0.2 dex. Some details about the type of analysis used and how the stellar atmosphere parameters were established were discussed in §3.5.1. Their analysis suggested that the [Na/Fe] ratio is solar and varies very little with metallicity. Hence, when magnesium is used as the basis of comparison the [Na/Mg] ratio showed a distinct odd-even effect.

The calculated evolution of [Na/Fe] in Figure 17 and [Na/Mg] in Figure 18 appear slightly smaller at low metallicities than the observations. However, if the "corrected" Gratton & Sneden (1988) trend continues, then the [Na/Fe] ratios for the most metal-poor stars should be lowered, bringing the observations into better agreement with the theoretical models. The calculated [Na/Mg] evolution is in good agreement with the disk dwarf observations, where a crisp odd-even effect is present.

For [Fe/H] $\gtrsim$ -1.0 dex the figures suggest that there might be small additional sources of sodium. Standard carbon deflagration models of Type Ia supernovae are not promising candidates. Type Ib supernovae which have lost only their hydrogen envelope may affect the $^{23}$Na abundance (Woosley et al. 1993, 1994). Some synthesis of sodium in intermediate mass stars can occur in the hydrogen burning shell through the neon-sodium cycle (Denisenkov 1989, Denisenkov & Denisenkova 1990). In the parameterized study of hydrogen burning in intermediate mass stars of Langer, Hoffman & Sneden (1993), they found that sodium can be produced by proton captures on $^{22}$Ne. On longer time scales, reactions chains involving the more abundant $^{20}$Ne also produced sodium. If the convective mixing time scale is small enough, then this sodium rich (but oxygen depleted) material can brought to the surface. Mass loss of the stellar envelopes would enrich the interstellar medium. This process is supported by observations of a few intermediate mass field supergiants that appear to be sodium rich (Boyarchuk & Lyubimkov 1985). Should these reaction network studies be confirmed by stellar evolution calculations, then this would cause the [Na/Fe] ratio to increase starting at [Fe/H] $\simeq$ -1.0 dex.

### 3.5.4 ALUMINUM

Evolution of the aluminum to iron ratio [Al/Fe] as a function of [Fe/H] is shown in Figure 19. Again the solid line shows the calculation; dashed lines indicate variations of two in the iron yields from massive stars. Using magnesium instead of iron as the basis enhances any odd-even effects, and eliminates Type Ia supernova from being a complicating factor because nearly all the aluminum and magnesium are produced in massive stars. Hence, evolution of the [Al/Mg] ratio as a function of [Fe/H] is shown in Figure 20.

Peterson (1981) examined the aluminum abundances in 18, metal-poor field dwarfs from high resolution ($\Delta\lambda = 0.25$ Å) image tube spectroscopy of the 3944 Å and 3961 Å Al I lines. The stars in the survey (which included a a few subgiants) spanned the metallicity range -0.5 $\leq$ [Fe/H] $\leq$ -2.7 dex. Some details about the type of analysis used and how the stellar atmosphere parameters were established were discussed in §3.5.3. The results suggested that aluminum tends to become increasingly deficient with respect to either iron or magnesium as the overall metallicity declines, but with significant star-to-star variations.

Analysis of the aluminum resonance lines in the near-ultraviolet requires care, because they lie in the crowded vicinity of the Ca II H and K lines, making placement of the continuum difficult. Additionally, the 3944 Å Al I line can be heavily blended with a molecular CH feature. Since the strength of the CH feature is less sensitive to abundance variations than 3944 Å Al I line, contributions of CH to the blend become larger in metal-poor stars (Arpigny & Magain 1983). As a result, aluminum abundances determined with the 3944 Å line tend to be spuriously large. However, at least one extremely metal-poor star shows no molecular CH in its spectrum, yet displays the same problems with the 3944 Å line (Norris, Peterson & Beers 1993 ; Peterson 1994).

Gathering various data sets on aluminum abundances on metal-poor dwarfs from the literature, Magain (1987) subjected 15 halo dwarfs to a homogeneous reanalysis. The stars spanned the metallicity range -1.0 $\leq$ [Fe/H] $\leq$ -2.9 dex. Effective temperatures were determined from calibrated V-K and/or $b$-$y$ color indices, and surface gravities was found from a theoretically calibrated $b$-$y$ versus $c_1$ color-color diagram. Microturbulent velocities were arbitrarily set at 1.25 km s$^{-1}$ for all stars. An absolute analysis was used, with most of the oscillator strengths derived from laboratory values. All equivalent widths larger than 70 mÅ were rejected because of insensitivity to the derived abundances, and a high sensitivity to the assumed microturbulence parameter. The analysis suggested that aluminum abundances, relative to iron or magnesium, decrease as [Fe/H] becomes smaller.

Gratton & Sneden (1988) derived atmospheric aluminum abundances for 13 extremely metal-poor field giants from from moderate resolution (R$\simeq$ 20,000), low noise (S/N$\simeq$150) digital spectra of the 3961 Å Al I feature. The survey stars ranged in metallicity from -3.0 $\leq$ [Fe/H] $\leq$ -1.3 dex. Some details about the type of analysis used and how the stellar atmosphere parameters were established were discussed in §3.5.1. Based on the ultraviolet Al I resonance line they suggested that [Al/Fe] becomes gradually more deficient with smaller [Fe/H] values, and that the [Al/Mg] ratio is more subsolar than that due to enhancement of [Mg/Fe] at low metallicities.

Magain (1989) used high resolution (R$\simeq$15,000), low-noise (S/N $\geq$ 100) digital spectra of the 3961 Å Al I ultraviolet resonance line to determined the aluminum abundances in 20 extremely metal-poor halo dwarfs. The survey stars spanned the range -1.4 $\leq$ [Fe/H] $\leq$ -2.9 dex. Effective temperatures were found from calibrated V-K and $b$-$y$ color indices, with no correction for reddening because the dwarfs are nearby. Surface gravities were determined by forcing the Fe I and Fe II lines to give the same iron abundance. Microturbulent velocities were found from the requirement that Fe I lines with equivalent widths between 10 mÅ and 70 mÅ yield the same iron abundance. An absolute analysis was used, with most of the oscillator strengths derived from laboratory furnace experiments. The results of the analysis indicated that aluminum is deficient with respect to iron and magnesium with a distinct trend towards lower aluminum abundances with decreasing metallicity. It was also reported that the dwarfs which appear to be relatively aluminum rich are also nitrogen rich (see §3.4.3).

Edvardsson *et al.* (1993) used high resolution (R$\simeq$60,000), low noise (S/N$\simeq$200) photodiode and digital spectra of the 8773 Å Al I feature to determine aluminum abundances in 189 F and G stars. The northern and southern hemisphere dwarfs in this extensive survey were shown to have disk-like kinematics, and covered the metallicity range -1.0 $\leq$ [Fe/H] $\leq$ 0.2 dex. Some details about the type of analysis used and how the stellar atmosphere parameters were established were discussed in §3.5.1. Their analysis suggested that aluminum tends to mimic the rise in [Mg/Fe] with decreasing [Fe/H], but at a reduced level. As a result, the [Al/Mg] ratio shows barely any odd-even effect over the metallicity range of the survey.

The calculated evolution of [Al/Fe] in Figure 19 is slightly larger than the observations of the most metal-poor stars, does not rise as fast through the [Fe/H] $\simeq$ -1.0 dex halo-disk transition point, and is slightly lower than the disk star observations. Figure 20 shows that these problems remain when magnesium is used as the basis for comparison instead of iron. In the parameterized study of Langer et al. (1993; see §3.5.4) small enhancements of $^{27}$Al were also found. If their reaction network studies are confirmed by stellar evolution calculations, this would increase the computed [Al/Fe] ratio starting at [Fe/H] $\simeq$ -1.0 dex, and may be sufficient to explain the increase through the halo-disk transition.

### 3.5.5 $^{26}$Al

The radioactive decay of $^{26}$Al in the interstellar medium to $^{26}$Mg (half life of $1.1 \times 10^6$ years) gives rise to the 1809 keV and 1130 keV gamma - ray lines. Observations of the diffuse emission from these lines provide an estimate of about 2 M$_\odot$ of $^{26}$Al in the Galaxy (Mahoney et al. 1984; Share et al. 1985; Diehl et al. 1993). Production in the massive stars generally occurs at the base of the CNO hydrogen burning shell through the operation of the magnesium - aluminum cycle, and throughout the neon burning regions (Woosley & Weaver, 1980, Hoffman et al. 1994). Timmes et al. (1994a) find that 2.0 $\pm$ 0.5 M$_\odot$ of $^{26}$Al has been injected into the interstellar medium in the last million years when the massive star models of the present study are used. The quoted error estimate takes variations of the total mass of the Galaxy, the steepness of radial gradients, the slope of the initial mass function, and the star formation efficiency into account.

## 3.6 THE INTERMEDIATE METALS

In this group we count the isotopes that have 14 $\leq$ Z $\leq$ 20, that is, from silicon to calcium. This group is the last set of isotopes that produced chiefly by hydrostatic burning processes before the explosion, and thus are not as affected by uncertain modeling of the explosion mechanism.

### 3.6.1 SILICON

Evolution of the silicon to iron ratio [Si/Fe] as a function of [Fe/H] is shown in Figure 21. The solid line shows the calculation; the dashed lines indicate factors of two variation in the iron yields from massive stars and the dotted line shows the results when Type Ia supernovae are excluded. Silicon displays many of the trends typical of $\alpha$-chain nuclei. The abundance determinations shown in the figure employ both dwarfs and field giants, and are primarily determined from neutral line transitions.

Magain (1987) reanalyzed various literature data sets on silicon abundances determined from Si I transitions in metal-poor dwarfs. The stars spanned the metallicity range -1.0 $\leq$ [Fe/H] $\leq$ -2.9 dex, and some details on how the stellar atmosphere parameters were established were discussed in §3.5.4. The results of the analysis indicated a mean silicon to iron ratio [Si/Fe] = 0.39 ± 0.14 dex, with the star-to-star scatter entirely explainable by the observational uncertainties.

Gratton & Sneden (1988) derived silicon abundances for 13 metal-poor field giants from moderate resolution (R≃20,000), low noise (S/N ≥ 150) digital spectra of 7 Si I transitions. The survey stars ranged in metallicity from [Fe/H] = -1.3 dex to [Fe/H] = -3.0 dex. Some details about the type of analysis used and how the stellar atmosphere parameters were established were discussed in §3.5.1. They found a mean silicon to iron ratio of [Si/Fe] = 0.51 dex, with a star to star scatter, some of which may be real, of 0.24 dex.

Edvardsson *et al.* (1993) used high resolution (R≃60,000), low noise (S/N≃200) photodiode and digital spectra of several unblended Si I lines around 6145 Å, 7800 Å and 8735 Å to determine silicon abundances in 189 F and G stars. The northern and southern hemisphere dwarfs in this extensive survey were shown to have disk-like kinematics, and covered the metallicity range -1.0 ≤ [Fe/H] ≤ 0.2 dex. Some details about the type of analysis used and how the stellar atmosphere parameters were established was discussed in §3.5.1. Their results showed [Si/Fe] ≃ 0.25 dex at [Fe/H] = -1.0 dex, and the decreases to solar values at about [Fe/H] = -0.1 dex. At larger metallicities the [Si/Fe] ratio appears to increase to slightly supersolar values. They suggest that part, if not most, of the star-to-star scatter is real and intrinsic.

The calculated evolution of the [Si/Fe] ratio is in overall agreement with the observations. The departure from classical $\alpha$ element abundances at [Fe/H] ≃ -2.5 dex is primarily due to uncertainties in the low metallicity 30 $M_\odot$ and larger exploded massive star models. Intermediate and low mass stars contribute no silicon or iron in the calculations. The silicon produced by the Z ≥ 0.1 $Z_\odot$ massive star models is balanced by the iron produced from Type Ia supernovae, which keeps the [Si/Fe] ratio relatively flat in Population I stars. This effect is shown by comparing the solid line in the figure (which includes both Type II and Type Ia supernovae) with the dotted line (which excludes Type Ia supernovae). Production of all the stable silicon isotopes by the massive star models is sufficient to explain the solar abundances (see Figure 4). Evolution of the silicon isotopes in the interstellar medium and its relationship to meteoritic silicon carbide grains is discussed in Timmes *et al.* (1994b).

### 3.6.2 SULFUR

Evolution of the sulfur to iron ratio [S/Fe] as a function of [Fe/H] is shown in Figure 22. The solid line shows the calculation; the dashed lines indicate factors of two variation in the iron yields from massive stars. Sulfur has been seldom studied primarily because potentially useful lines in the near-infrared portion of the spectrum are weak and blended, but it appears to have many of the characteristics of an α-chain element.

Clegg, Lambert & Tomkin (1981) used high resolution ($\Delta\lambda$ =0.1 Å) photodiode spectra of the 8693 Å and 8694 Å Si I lines to determined the atmospheric sulfur abundances in 20 F and G main-sequence stars. Their survey stars covered the metallicity range -0.9 ≤ [Fe/H] ≤ 0.4 dex, and included three binaries (μ Cas, η Boo and 85 Peg) which are not shown in Figure 22. Effective temperatures was determined from a weighted mean of calibrated V-R and V-I color indices, Hβ and b-y photometry, and a calibrated Fe I excitation temperature scale. The survey stars were nearby so accurate luminosities could be determined, and their masses could be estimated from their positions in the Hertzsprung-Russell diagram. A standard theoretical relationship then allowed the surface gravity of the stars to be estimated. A microturbulence velocity of 1 km s$^{-1}$ was adopted for most of the survey stars, and an LTE differential analysis was used. The results suggested a mean sulfur to iron ratio of [S/Fe] = 0.10 dex, but with a large star-to-star scatter of ± 0.12 dex. A trend towards larger supersolar [S/Fe] ratios as [Fe/H] decreased was reported.

Atmospheric sulfur abundances in 25 dwarfs were studied by Francois (1987; 1988) using high resolution (R = 80,000), low noise (S/N ∼ 100) photodiode spectra of the 8693 and 8694 Å Si I lines. The survey stars spanned the metallicity range -1.6 ≤ [Fe/H] ≤ 0.2 dex. The model atmosphere parameters were adopted from the compilation of Cayrel de Strobel *et al.* (1985). A differential analysis was used with an inverse solar spectrum procedure determining the oscillator strengths. Normalization of each spectra was referenced to a hand-drawn smooth continuum. Since only weak Si I lines were used, the main source of systematic error was the adopted effective temperature, with changes of 100 K resulting in sulfur abundance changes of about 0.1 dex. The analysis indicated a mean sulfur to iron ratio of [S/Fe]=0.6 dex in the halo with a clear trend down to solar values. However, Lambert (1989) suggested that about 0.2 dex of this large halo value may have been due to the method used for determining the Si I oscillator strengths. In one of the few cases where violence has been done to the original data, the sulfur abundances of Francois (1987; 1988) shown in Figure 22 have been "corrected" by the suggested -0.2 dex.

The calculated evolution of the [S/Fe] ratio is generally in agreement with the observations, but is smaller at low values of [Fe/H] by about a factor of two (Figure 22). The depression to a solar value at [Fe/H] ≃ -2.5 dex is primarily due to uncertainties in the M ≥ 30 M$_\odot$ extremely metal-poor massive star models (see discussion in §2.4). Intermediate and low mass stars contribute no sulfur or iron in the calculations. The sulfur produced by the Z ≥ 0.1 Z$_\odot$ massive star models strikes a balance with the iron produced from Type Ia supernovae, which keeps the [S/Fe] ratio relatively constant in Population I stars. Production of all the stable sulfur isotopes by the massive star models is sufficient to explain the solar abundances (see Figure 4).

### 3.6.3 NEON, PHOSPHORUS, CHLORINE, & ARGON

Elemental neon and argon are noble gases that only have optical transitions from very high excitation levels, and apparently have no stellar abundance determinations in either dwarfs or field giants. Lines from these elements show up in the hot winds of planetary nebulae, classical novae and Wolf-Rayet stars (Peimbert & Torres-Peimbert 1983; Osterbrock 1989). Measurements of H II regions in the solar vicinity tend to give solar neon and argon abundances (Jenkins 1987; Meyer 1988). Molecular forms of chlorine and phosphorus show only very weak lines in synthetic stellar spectra models, and abundance determinations in either dwarfs or field giants could not be found in the literature. Measurements of HI regions in the solar vicinity tend to give slightly subsolar chlorine and phosphorus abundances (Meyer 1988).

Figure 23 shows the predicted [Ne/Fe], [P/Fe], [Cl/Fe], and [Ar/Fe] ratios in the interstellar medium as a function of the metallicity [Fe/H]. Both the odd-Z elements, phosphorus and chlorine, are subsolar at low metallicities and slowly climb to their solar values in the disk. Neon and argon, being $\alpha$-chain elements, are almost the complete opposite, showing supersolar values in the halo and then slowly decreasing to solar ratios in the disk.

### 3.6.4 POTASSIUM

Evolution of the potassium (dominated by $^{39}$K) to iron ratio [K/Fe] as a function of [Fe/H] is shown in Figure 24. The solid line shows the calculation; the dashed lines indicate factors of two variation in the iron yields from massive stars.

Gratton & Sneden (1987) made the (so far) only attempt to measure the atmospheric potassium abundances in stars. They surveyed 23 northern and southern hemisphere stars, which included both dwarfs and field giants, that spanned the metallicity range -2.3 $\leq$[Fe/H] $\leq$ 0.3 dex. The analysis was was based on high resolution ($\Delta\lambda = 0.2$ Å), low-noise (S/N $\simeq$ 100) photodiode spectra of the 7699 Å resonance lines of K I. Effective temperatures were found from $b$-$y$, R-I, V-R, and V-K color indices. Surface gravities were obtained from a combinations of distances, absolute visual magnitudes, bolometric corrections, stellar masses, and ionization equilibrium of iron. The microturbulence velocity was established by avoiding any trend of the derived abundances with equivalent width. A differential analysis, with oscillator strengths derived from the solar spectrum, was used. The results suggested that [K/Fe] $\geq$ 0.0 dex in metal-poor stars, but with a very large scatter.

The calculated [K/Fe] evolution is in agreement with the observations for [Fe/H] $\gtrsim$ -0.6 dex. At smaller metallicities the calculations suggest one might expect [K/Fe] $\lesssim$ 0.0 dex, which is in sharp contrast to the observations. Gratton & Sneden (1987) note that the K I resonance lines are often heavily blended with extremely strong lines of atmospheric molecular oxygen. The low excitation energies of the K I lines may be strongly susceptible to overionization or strong hyperfine structure effects. Production of all the stable potassium isotopes by the massive star models is sufficient to explain the solar abundances (see Figure 4).

### 3.6.5 CALCIUM

Evolution of the calcium to iron ratio [Ca/Fe] (essentially $^{40}$Ca) as a function of [Fe/H] is shown in Figure 25. The solid line shows the calculation, the dashed lines indicate factors of two variation in the iron yields from massive stars and the dotted lines show variations in the efficiency of star formation. It has been known for sometime that the [Ca/Fe] ratio is supersolar in mildly as well as extremely metal-poor halo dwarfs (Wallerstein 1962), but the precise evolutionary history of calcium was delayed until the advent of relatively low-noise of digital spectra and more reliable oscillator strengths for the relevant transitions.

Calcium abundances of 21 dwarfs that spanned the metallicity range $-2.5 \leq$ [Fe/H] $\leq -1.0$ dex were investigated by Hartmann & Gehren (1988). The abundance determinations were based high resolution ($\Delta\lambda = 0.2$ Å), modest noise (S/N$\simeq$40) photographic spectra of the Ca I multiplets. The Ca II H and K resonance lines were disregarded due to lack of reasonable solar data. Effective temperatures were found from the Balmer line profiles and a calibrated R-I color index, and surface gravities were determined from the ionization equilibrium of iron and titanium. The microturbulence velocity was chosen by minimizing the variation of single line iron abundances with equivalent width, and an LTE differential analysis was adopted. Their results suggested a mean halo calcium to iron ratio of [Ca/Fe] $= 0.28$, with a star-to-star scatter of 0.18 dex. No significant trend with metallicity was reported.

Zhao & Magain (1990) determined the calcium abundances in 20 southern hemisphere dwarfs with metallicities that ranged between $-1.4 \leq$ [Fe/H] $\leq -3.0$ dex. The abundance determinations were based on high resolution (R$\simeq$20,000), low noise (S/N $\sim$ 100), digital spectra of several Ca I lines in the wavelength range 5000 - 6000 Å. The effective temperatures were determined from calibrated $b$-$y$ and V-K color indices. Zero reddening was assumed. Surface gravities were extracted from the condition that the iron abundance as determined from the Fe II line equal that as determined from the Fe I line. The microturbulent velocity was forcing all the Fe I lines with equivalent widths between 10 mÅ and 70 mÅ to give the same abundance. An LTE absolute analysis was chosen, and experimental oscillator strengths was used. The analysis yielded a supersolar mean calcium to iron ratio [Ca/Fe] of 0.41 $\pm$ 0.08 dex, and fairly flat over the entire metallicity range of their study. The authors suggested that because of correlations between calcium abundances determined from blue and green portions of the spectra that part of the scatter in the calcium abundances was real, and not due to random errors in the measured equivalent widths.

Gratton & Sneden (1991) analyzed the atmospheric calcium abundances of 11 dwarfs and 8 field giants from the southern sky that uniformly spanned the metallicity range -2.7 ≤ [Fe/H] ≤ -0.2 dex. The analysis employed high resolution (R=50,000), low-noise (S/N ≃ 150) digital spectra of 7 Ca I lines. Effective temperatures were found from several colors indices gathered from the literature. No reddening was assumed for the dwarf stars, and reddening for the giant stars were estimated from near field $uvby\beta$ photometry of field stars. Surface gravities of the stellar atmospheres was determined from the ionization equilibrium of iron, and microturbulent velocities were obtained by avoiding abundance trends from individual Fe I lines with line strength. An absolute analyses was employed, with some oscillator strengths coming from terrestrial experiments and the rest coming from an inverse solar analysis. Their results showed a mean value of [Ca/Fe] = 0.29 ± 0.06 dex, with no distinct trend over the metallicity range studied.

Edvardsson *et al.* (1993) utilized high resolution, low noise digital spectra of the Ca I lines at 5867 Å and 6166 Å to determine the calcium abundances in 189 northern and southern F and G disk dwarfs in the solar vicinity. The stars in this extensive survey spanned the metallicity range -1.0 ≤ [Fe/H] ≤ 0.2 dex. Some details about the type of analysis used and how the stellar atmosphere parameters were established were discussed in §3.5.1. The results showed that [Ca/Fe] ≃ 0.25 dex at [Fe/H] = -1.0 dex, and then decreases to solar values at about [Fe/H] = -0.2 dex. At larger metallicities the [Ca/Fe] ratio appears to increase to slightly supersolar values. They suggest that part, if not most, of the star-to-star scatter is real and intrinsic.

The calculated evolution of the [Ca/Fe] ratio is in overall agreement with the observations of this $\alpha$ element (Figure 25). The departure from classical $\alpha$ element abundances at [Fe/H] ≃ -2.5 dex is primarily due to uncertainties in the extremely low metallicity M ≥ 30 $M_\odot$ massive star models. Intermediate and low mass stars contribute no calcium or iron in the calculations shown. The calcium produced by the Z ≥ 0.1 $Z_\odot$ massive star models is balanced by the iron produced from Type Ia supernovae, which keeps the [Ca/Fe] ratio relatively flat in Population I stars. The dotted lines show the evolutions for significant variations in the efficiency of star formation ($\nu$=0.8 and 5.8; see Table 1), and indicate the general property that $\alpha$-chain evolutions are very robust with respect to modifications in this free parameter.

## 3.7 THE IRON PEAK NUCLEI

In this group we count the isotopes that have 21 ≤ Z ≤ 28, that is, from scandium to nickel. Uncertainties in the parameters of the explosion and the amount of material that falls back onto the neutron star strongly affects the absolute mass of the elements that are ejected. However, factors of two variation in the explosion energy sometimes produce small variations in element ratios, which implies that the calculated ratio histories may be robust. In addition, two-dimensional simulations of the explosion mechanism suggest that the mixing of material by hydrodynamical instabilities is a vigorous process (Arnett, Fryxell & Müller 1989; Herant, Benz & Colgate 1992; Burrows & Fryxell 1992; Janka & Müller 1993; Colgate, Herant & Benz 1993; Shimizu, Burrows & Fryxell 1993; Yamada & Sato 1993; Janka & Müller 1994; Herant et al. 1994; Herant & Woosley 1994, 1995). This may represent further evidence for robustness of the calculated elemental ratios.

Hyperfine splitting is a source of potential serious systematic error in the observations of the odd-Z elements, where the imbalance of neutrons and protons gives rise to a large nuclear magnetic moment. Hyperfine structure effects effectively desaturate the curve of growth at intermediate line strengths. Accurate abundance histories of the odd-Z elements requires that this physical process be accounted for in the analysis (Lambert 1989).

The existence of an enhanced odd-even effect among the iron group elements in metal-poor stars has a rich background ( Burbidge et al. 1957; Cameron 1957; Helfer, Wallerstein & Greenstein 1959; Arnett 1971; Truran & Arnett 1971; Pardo, Couch & Arnett 1974; Woosley & Weaver 1980; Woosley & Weaver 1982a; Woosley 1986; Lambert 1989; Wheeler et al. (1989); Paper II). Discussion of the calculations and observations in the this section will generally, but with a few important exceptions, show the magnitude of the expected odd-even effects.

### 3.7.1 SCANDIUM

Evolution of the scandium to iron ratio [Sc/Fe] as a function of [Fe/H] is shown in Figure 26. The solid line shows the calculation; the dashed lines indicate factors of two variation in the iron yields from massive stars. Scandium has only one stable isotope ($^{45}$Sc), and it is important to include hyperfine structure effects in the observational analysis since it is an odd-Z nucleus.

Zhao & Magain (1990) examined the surface scandium abundances in 20 southern hemisphere dwarfs from high resolution, low noise (S/N $\sim$ 100), digital spectra of the weak 5526 Å and 5657 Å Sc II lines. The survey stars spanned the metallicity range -1.4 $\leq$ [Fe/H] $\leq$ -2.3 dex. Some details about the type of analysis used and how the stellar atmosphere parameters were established were discussed in §3.6.5. Hyperfine structure effects were taken into account in an approximate manner, with each line receiving an equivalent width dependent correction. Combining their data with two stronger Sc II lines observed in the blue region of the spectrum by Magain (1989), their analysis yielded a surprisingly large supersolar mean scandium to iron ratio [Sc/Fe] of 0.27 $\pm$ 0.10 dex. The authors attributed the discrepancy with previous scandium abundance determinations to the systematic errors in the differential analysis.

Gratton & Sneden (1991) studied the scandium abundances of 11 field dwarfs and 8 field giants from moderate resolution (R=50,000), low-noise (S/N $\geq$ 150) digital spectra of number of Sc I and Sc II lines. The southern hemisphere survey stars covered the metallicity range -0.2 $\leq$ [Fe/H] -2.7 dex. Some details about the type of analysis used and how the stellar atmosphere parameters were established were discussed in §3.6.5. Detailed hyperfine structure effects were taken into account in the synthetic spectrum computations, with the relative strengths being deduced from theoretical work as high quality experimental oscillator strengths were not available. The analysis suggested a mean value of [Sc/Fe] = -0.03 $\pm$ 0.09 dex was reported, with abundances deduced from the neutral and singly ionized lines agreeing to -0.04 dex. The results of this survey are probably the more accurate of surveys shown in Figure 26.

Gratton & Sneden (1991) suggested that the most probable reason for the large overabundance of scandium found by Zhao & Magain (1990) was due to how the transition probabilities and branching ratios for the Sc II lines were determined. They showed that using theoretical *gf* values on the Zhao & Magain data yielded abundances that are on average 0.23 dex smaller.

The calculated evolution is in good agreement with the observations of a flat [Sc/Fe] ratio (Figure 26), although perhaps systematically smaller than the observations by about a factor of 1.5 at metallicities characteristic of the halo population. However, the solar abundance of $^{45}$Sc is also too small by the about same factor (see Figures 4 and 5). Intermediate and low mass stars contribute no scandium or iron in the calculations shown. The scandium produced by the Z $\geq$ 0.1 Z$_\odot$ massive star models becomes nearly balanced with the iron produced from Type Ia supernovae, which keeps the [Si/Fe] ratio relatively flat in Population I stars.

### 3.7.2 TITANIUM

Evolution of the titanium (essentially $^{48}$Ti) to iron ratio [Ti/Fe] as a function of [Fe/H] is shown in Figure 27. The solid line shows the calculation; the dashed lines indicate factors of two variation in the iron yields from massive stars. It has been known awhile that the rise in the [Ti/Fe] ratio rises to supersolar in moderately metal-poor disk stars (Wallerstein 1962), but the precise evolutionary history of titanium became better known with the advent of relatively low-noise of digital spectra and more reliable oscillator strengths for the relevant transitions. The abundance determinations shown in the figure for titanium employ both dwarfs and giants, and are primarily determined from neutral line transitions.

Titanium abundances in 20 southern hemisphere halo dwarfs with metallicities that ranged between -1.4 $\leq$ [Fe/H] $\leq$ -2.3 dex were measured by Magain (1989). The abundance determinations were based on high resolution (R$\simeq$15,000), low-noise (S/N $\geq$ 100) digital spectra of several Ti I lines in the wavelength range 3700 - 4700 Å. Some details about the type of analysis used and how the stellar atmosphere parameters were established was discussed in §3.5.1. The results of the analysis gave a mean titanium to iron ratio [Ti/Fe] of 0.42 ± 0.09 dex, with a slight trend towards decreasing [Ti/Fe] as the metallicity increased.

Gratton & Sneden (1991) determined the surface titanium abundances of 11 field dwarfs and 8 field giants from moderate resolution (R=50,000), low-noise (S/N ≥ 150) digital spectra of large number of Ti I and a few Ti II lines. The southern hemisphere survey stars covered the metallicity range -2.7 ≤ [Fe/H] ≤ -0.2 dex. Some details about the type of analysis used and how the stellar atmosphere parameters were established were discussed in §3.6.5. Their results showed a distinct trend for titanium abundances provided by neutral lines to be systematically lower than those determined using singly ionized lines in metal-poor giants, which they attributed to significant departures (overionization) from LTE. Using only their titanium abundances derived from the Ti II lines, they derived a mean [Ti/Fe] ratio of 0.28 ± 0.10 dex. However, they showed that the titanium abundances are sensitive to the input atmospheric parameters, continuum placement and adopted solar model.

Edvardsson et al. (1993) used high resolution (R≃60,000), low noise (S/N≃200) photodiode and digital spectra of the 5087, 5113, 5866 and 6126 Å Ti I lines to determine titanium abundances in 189 F and G dwarfs. The northern and southern hemisphere dwarfs in this extensive survey were shown to have disk-like kinematics, and covered the metallicity range -1.0 ≤ [Fe/H] ≤ 0.2 dex. Some details about the type of analysis used and how the stellar atmosphere parameters were established were discussed in §3.5.1. Their analysis suggest that the titanium to iron ratio [Ti/Fe] is supersolar in the most metal-poor disk stars, with a distinct trend towards decreasing [Ti/Fe] as the metallicity [Fe/H] is increased. The star-to-star scatter about the mean is fairly large, and they suggested that part of, if not most of, the scatter is real and intrinsic.

The calculated abundance for elemental titanium in Figure 27 is not in good agreement with the observations, and the solar abundance of $^{48}$Ti, its most abundant isotope, is deficient by a factor of 2 (see Figure 4). Observationally titanium displays many of the $\alpha$ element hallmarks, while the calculations suggest it should scale with iron. However, both the $^{48}$Ti yield and the ratio [Ti/Fe] are sensitive to the parameters of the explosion and the amount of material that falls back onto the neutron star (Paper II). For example, the 30 $M_\odot$ (Z = $Z_\odot$) explosion with a kinetic energy at infinity of 1.13 × 10$^{51}$ erg had [$^{48}$Ti/$^{56}$Fe] = 0.0 dex, while when the same star exploded with a larger kinetic energy (2.01 × 10$^{51}$ erg), [$^{48}$Ti/$^{56}$Fe] = -0.35 dex. The larger energy was used in Figure 27. Similar ratios are obtained for other masses and initial metallicities.

Evolution of the titanium isotopic ratios is shown in Figure 28. The isotope $^{46}$Ti is produced in sufficient quantity by massive stars to explain the solar abundance, while $^{47}$Ti is underproduced by about a factor of 5. Historically, pinning down the origin of this isotope has proved elusive (Burbidge et al. 1957; Woosley 1986). The probable origin sites for the deficient isotopes $^{47}$Ti, $^{49}$Ti, and $^{50}$Ti are discussed in §3.1. Standard deflagration models of Type Ia supernova (Thielemann et al. 1986) contribute very little to the titanium isotopes (compare Figures 4 and 5).

### 3.7.3 VANADIUM

Evolution of the vanadium to iron ratio [V/Fe] (predominantly $^{51}$V) as a function of [Fe/H] is shown in Figure 29. The solid line shows the calculation; the dashed lines indicate factors of two variation in the iron yields from massive stars. Vanadium is the least abundant of the iron group elements.

Gratton & Sneden (1991) determined the atmospheric vanadium abundances of 11 field dwarfs and 8 field giants from moderate resolution (R=50,000), low-noise (S/N ≥ 150) digital spectra of large number of V I lines. The southern hemisphere survey stars covered the metallicity range -2.7 ≤ [Fe/H] ≤ -0.2 dex. Some details about the type of analysis used and how the stellar atmosphere parameters were established were discussed in §3.6.5. Full account of hyperfine structure effects on the neutral vanadium lines were taken into account during the spectrum synthesis calculations. The analysis suggested a mean vanadium to iron ratio of [V/Fe] = -0.03 ± 0.11 dex, but moderately sensitive to the adopted solar model and atmospheric parameters. Additional checks were made by using the two weak V II lines at 5303 Å and 4234 Å, which are not subject to important hyperfine structure effects. Although the analysis of each of these singly ionized lines of vanadium presented some difficulties, they determined that the mean difference between abundances determined with the V I feature versus the V II feature was small (-0.06 ± 0.06 dex).

The calculated evolution is in good agreement the observations of a flat [V/Fe] ratio (Figure 29), but are systematically smaller than the observations by about a factor of three. However, elemental vanadium is dominated by the isotope $^{51}$V, whose solar abundance is underproduced by slightly more than a factor of two (see Figures 4 and 5). The probable origin of the isotope $^{51}$V is discussed in §3.1. Intermediate and low mass stars produce no vanadium or iron in the calculation shown. The vanadium produced by the Z ≥ 0.1 Z$_\odot$ massive star models is balanced by the iron produced from Type Ia supernovae, which keeps the [V/Fe] ratio relatively flat in Population I stars.

### 3.7.4 CHROMIUM

Evolution of the chromium (dominated by $^{48}$Cr) to iron ratio [Cr/Fe] as a function of [Fe/H] is shown in Figure 30. The solid line shows the calculation, the dashed lines indicate factors of two variation in the iron yields from massive stars and the dotted lines show variations in the exponent of the initial mass function.

Chromium abundances in 20 southern hemisphere dwarfs were determined from high resolution, low noise digital spectra of several Cr I lines in the blue and green spectral region by Magain (1989) and Zhao & Magain (1990). The survey stars spanned the metallicity range -1.4 ≤ [Fe/H] ≤ -2.3 dex. Some details about the types of analysis used and how the stellar atmosphere parameters were established were discussed in §3.5.1 and §3.6.5. The analysis indicated a mean chromium to iron ratio of [Cr/Fe] = 0.01 ± 0.08 dex. Both studies tended to show a larger scatter as decreasing metallicities, especially in the blue region of the spectrum where a single neutral line was all that could be measured.

Gratton & Sneden (1991) determined the chromium abundances of 11 field dwarfs and 8 field giants from moderate resolution, low-noise digital spectra of several Cr I and 4 Cr II lines. The southern hemisphere survey stars covered the metallicity range -2.7 ≤ [Fe/H] ≤ -0.2 dex. Some details about the type of analysis used and how the stellar atmosphere parameters were established were discussed in §3.6.5. They reported that chromium abundances deduced from the neutral species are on average 0.02 dex smaller than those given by the Cr II lines, with a trend for the differences to get larger in the most metal-poor stars. They suggested that this may be due to some overionization (i.e non-LTE effects). Based on the Cr II line analysis, they found a mean chromium to iron ratio of [Cr/Fe] = -0.04 ± 0.05 dex, and good agreement was found between abundances deduced from dwarfs and field giants.

It is gratifying that the calculated evolution of this abundant iron group nucleide is in excellent agreement with observations (Figure 30). All the solar abundances of the chromium isotopes, except the rare $^{54}$Cr (which is discussed in §3.1), are well accounted for by the massive star models, although inclusion of the nucleosynthesis from Type Ia supernovae improves the fit to the solar abundances of the isotopes $^{50-53}$Cr (see Figures 4 and 5). Intermediate and low mass stars contribute no chromium or iron in the calculations shown. The chromium produced by the Z ≥ 0.1 Z$_\odot$ massive star models becomes nearly balanced with the iron produced from Type Ia supernovae, which keeps the [Cr/Fe] ratio relatively flat in Population I stars.

The dotted line which lies below the standard calculation (solid line; initial mass function exponent of -1.31) for most of the [Fe/H] evolution corresponds to an initial mass function exponent of -1.61, while the dotted line with an initial mass funtion exponent of -1.01 lies mostly above the standard calculation. Figure 30 shows the general property that evolution of the iron peak elements results are robust with respect to variations in the exponent of the initial mass function.

### 3.7.5 MANGANESE

Evolution of the manganese to iron ratio [Mn/Fe] as a function of [Fe/H] is shown in Figure 31. The solid line shows the calculation; the dashed lines indicate factors of two variation in the iron yields from massive stars and the dotted line shows the results when Type Ia supernovae are excluded. Manganese has only one stable isotope ($^{55}$Mn), and it is important to include hyperfine structure effects in the observational analysis since it is an odd-Z nucleus.

Beynon (1978ab) studied the atmospheric manganese abundances in 24 F and G dwarfs using low resolution photographic spectra of 7 Mn I transitions near 4400 Å and 4700 Å. The survey stars spanned the metallicity range -0.7 ≤ [Fe/H] 0.2 dex. The analysis pioneered the inclusion of hyperfine structure effects, albeit crudely, in the differential analysis of the curves of growth. The analysis suggested the trend [Mn/Fe] $\simeq$ 0.2 [Fe/H].

Gratton (1989) examined the surface magnesium abundances in 25 southern hemisphere stars, including both dwarfs and field giants, using high resolution (R$\simeq$50,000), low noise (S/N$\simeq$ 200) digital spectra of 10 weak and intermediate strength Mn I transitions between 4453 Å and 6021 Å. The survey stars uniformly covered the range -2.4 ≤ [Fe/H] ≤ -0.1 dex. Effective temperatures were found from a variety of calibrated visual and near-infrared color indices. No reddening was assumed for the dwarf stars, but small reddening corrections were necessary for the field giants. Surface gravities were determined by forcing the lines of Fe I and Fe II to give the same mean abundances. Microturbulence velocities were generally assumed to be 0.8 km s$^{-1}$ for dwarfs and 1.5 km s$^{-1}$ for giants. An LTE differential analysis that properly treated the hyperfine structure effects was used. The results showed that for [Fe/H] < -1.0 the mean magnesium to iron ratio was [Mn/Fe] = -0.34 ± 0.07 dex. At larger metallicities, the [Mn/Fe] ratio increased in a roughly linear manner to solar values.

The calculated [Mn/Fe] evolution is in good agreement with the shape of the observations, but is deficient at low metallicities by roughly a factor of two (Figure 31). Intermediate and low mass stars contribute no manganese or iron in the calculations shown. The solar metallicity exploded massive star models have about a factor of 5 larger manganese yield than the models with smaller initial metallicities (Paper I; Paper II). This overwhelms the iron contributions from Type Ia supernovae, and causes the increase in the calculated [Mn/Fe] ratio for [Fe/H] > -1.0 dex. This effect is shown by comparing the solid line in the figure (which includes both Type II and Type Ia supernovae) with the dotted line (which excludes Type Ia supernovae). The amplitude of the jump is decreased if more Type Ia supernovae are included. Inclusion of the nucleosynthesis from Type Ia supernovae improves the fit to the solar abundances of manganese (see Figures 4 and 5).

### 3.7.6 COBALT

Evolution of the cobalt to iron ratio [Co/Fe] as a function of [Fe/H] is shown in Figure 32. The solid line shows the calculation, the dashed lines indicate factors of two variation in the iron yields from massive stars and the dotted lines show variations in the efficiency of star formation. The only stable isotope of cobalt is $^{59}$Co, and it is an odd-Z element.

Gratton & Sneden (1991) determined the atmospheric cobalt abundances of 11 field dwarfs and 8 field giants from moderate resolution, low-noise digital spectra of 10 unblended Co I lines. The southern hemisphere survey stars covered the metallicity range -2.7 ≤ [Fe/H] ≤ -0.2 dex. Some details about the type of analysis used and how the stellar atmosphere parameters were established were discussed in §3.6.5. Full account of hyperfine structure effects on the neutral vanadium lines were taken into account during the spectrum synthesis calculations. Their results suggested a small deficiency of cobalt in metal-poor stars ([Co/Fe] = -0.12 ± 0.05 dex), with a slight trend for larger deficiencies in more metal-poor stars. They showed this result be robust with respect to the model atmosphere parameters or the solar model employed.

The calculated evolution is in reasonable agreement with the observations of a relatively flat [Co/Fe] ratio, although the computed history appears to be smaller than the observations by about a factor of two at particular low metallicities (Figure 32). The undulations of the curve shown in is due to mass and metallicity effects in the exploded massive star models. Intermediate and low mass stars contribute no cobalt or iron in the calculations shown. The cobalt produced by the Z ≥ 0.1 Z$_\odot$ massive star models is balanced by the iron produced from Type Ia supernovae, which keeps the [Co/Fe] ratio relatively flat in Population I stars. Production of cobalt by massive stars is sufficient to explain the solar abundance (see Figure 4). The dotted lines show the evolutions for significant variations in the efficiency of star formation ($\nu$=0.8 and 5.8; see Table 1), and indicate the general property that iron peak element evolutions are very robust with respect to modifications in this free parameter.

### 3.7.7 NICKEL

Evolution of the nickel to iron ratio, [Ni/Fe], as a function of [Fe/H] is shown in Figure 33. Again the solid line is the calculation and dashed lines indicate factors of two variation in the iron yields from massive stars. Nickel is the second most abundant element in the iron group but has fewer optical spectral lines. The field dwarf and giant star observations shown all use Ni I lines, and are expected to be rather robust since nickel and iron have similar first ionization potentials (7.87 eV and 7.64 eV respectively) and atomic structure. However, all the lines of Ni I longward of 3831 Å are very weak in most metal-poor stars, and this is an important issue for nickel abundance determinations in these stars.

Luck & Bond (1983; 1985) measured the nickel abundances of 36 metal-poor halo giants using high resolution ($\Delta\lambda \simeq 0.15$ Å), modest noise (S/N $\simeq$ 35) photographic spectra of Ni I transitions between 4700 Å and 5300 Å. The signal-noise-ratio set the limits for the equivalent width determinations. The survey stars spanned the metallicity range -2.6 ≤ [Fe/H] $\leq$ -0.6 dex, with the majority having [Fe/H] $\leq$ -1.0 dex. Effective temperatures were found from the condition that there be no dependence of iron abundances determined from Fe I lines upon excitation potential. Surface gravities were established by requiring Fe I and Fe II to yield the identical iron abundances, and the microturbulent velocities were determined by the demand that there be no dependence of Fe I abundance on equivalent width. An LTE differential solar analysis was used, with the oscillator strengths determined from an inverted solar analysis. Their analysis suggested that [Ni/Fe] = 0.15 $\pm$ 0.22 dex for [Fe/H] > -2.0 dex. At smaller metallicities a supersolar [Ni/Fe] ratios emerged, with a clear trend towards larger values as the metallicity declined. This result stimulated a number of investigations and lively debate on nickel abundances in halo stars.

Gratton & Sneden (1988) derived nickel abundances for 13 metal-poor field giants from moderate resolution, low noise digital spectra of various Ni I transitions. The survey stars ranged in metallicity from [Fe/H] = -3.0 to -1.3 dex. Some details about the type of analysis used and how the stellar atmosphere parameters were established were discussed in §3.5.1. They found that [Ni/Fe] = -0.06 $\pm$ 0.03 dex, which suggested that there was no nickel overabundance in very metal-poor Population II stars.

Gratton & Sneden (1991) determined the surface nickel abundances of 11 field dwarfs and 6 field giants from moderate resolution, low-noise digital spectra of 19 Ni I lines. The southern hemisphere survey stars covered the metallicity range -2.7 $\leq$ [Fe/H] $\leq$ -0.2 dex. Some details about the type of analysis used and how the stellar atmosphere parameters were established were discussed in §3.6.5. They found a mean nickel to iron ratio [Ni/Fe] of -0.04 $\pm$ 0.04 dex, which they claimed offered strong evidence for a solar [Ni/Fe] ratio for all stars.

Edvardsson *et al.* (1993) used high resolution, low noise photodiode and digital spectra of 20 Ni I transitions to determine magnesium abundances in 189 F and G stars in the solar vicinity. The northern and southern hemisphere dwarfs in this extensive survey were shown to have disk-like kinematics, and covered the metallicity range -1.0 $\leq$ [Fe/H] $\leq$ 0.2 dex. Some details about the type of analysis used and how the stellar atmosphere parameters were established were discussed in §3.5.1. Their results showed [Ni/Fe] = 0.0 $\pm$ 0.03 over the entire metallicity range of the survey.

The Peterson & Carney (1989) survey had two stars in common with the Luck & Bond (1983; 1985). None of the weak 22 Ni I lines with equivalent widths below 26 mÅ in the Luck & Bond survey were unambiguously detected by the later, better resolved, survey. Peterson & Carney suggested that the high nickel overabundances in the halo were probably caused by a systematic overestimate of line strengths near the detection limit. Gratton & Sneden (1991) showed that the 4786 Å and 5035 Å Ni I transitions have equivalent widths of 7 mÅ and 5 mÅ respectively, which is below the 20 mÅ detectability limit of the Luck & Bond survey. Using the 20 mÅ lower limit systematically decreases the [Ni/Fe] ratios in very metal-poor stars.

Elemental nickel is dominated by the two isotopes $^{58}$Ni and $^{60}$Ni, both produced in the deepest layers of massive stars (Paper II) and also abundantly in Type Ia supernovae of the deflagration variety. Although the deflagration model for Type Ia supernovae has a large $^{58}$Ni overproduction factor, most of the $^{58}$Ni in the chemical evolution model model comes from massive stars. It is gratifying that the calculated evolution of this abundant iron group nucleide is in good agreement with observations (Figure 33). In addition, the $^{58}$Ni and $^{60}$Ni solar abundances are well accounted for (see Figure 4). Uncertainties in the parameters of the explosion and the amount of material that falls back onto the neutron star clearly affects the absolute mass of nickel ejected. However, factors of two variation in the explosion energy of a solar metallicity 35 M$_\odot$ star produced variations of less than 10% in the [Ni/Fe] ratio (Paper II). This suggests that the calculated nickel to iron evolution is robust.

### 3.8 THE FIRST NUCLEI BEYOND THE IRON PEAK

In this group we examine copper and zinc, which are synthesized in significant amounts by the major nuclear burning stages in the massive star models (Paper I; Paper II). The arguments why the absolute mass ejected is strongly effected by the explosion, but the ratios to iron may be rather robust was discussed in §3.7 and applies here. Copper and zinc abundances in stars have not benefitted from a large number of spectroscopic investigations, simply because they do not have many suitable transitions in commonly observed wavelength regions. The neutral Zn I features possesses only two useful lines at 4722 Å and 4810 Å. There are a few more lines of Cu I, but only the 5105 Å and 5782 Å can be detected in the spectra of very metal-poor stars (Wheeler *et al.* 1989). No suitable ionized lines of either element exists in spectroscopically accessible wavelength regions.

#### 3.8.1 COPPER

Evolution of the copper to iron ratio [Cu/Fe] as a function of [Fe/H] is shown in Figure 34. The solid line shows the calculation; the dashed lines indicate factors of two variation in the iron yields from massive stars; and the dotted line shows the calculation when Type Ia supernovae are excluded. Copper has two stable isotopes, and because it is an odd-Z nucleus it is important to include hyperfine structure effects in the observational analysis.

Gratton & Sneden (1988) derived atmospheric copper abundances in 5 halo giants from moderate resolution, low noise digital spectra of the Cu I transitions at 5105 Å and 5782 Å lines. Some details about the type of analysis used and how the stellar atmosphere parameters were established were discussed in §3.5.1. Although the Cu I lines are strong enough in the sun that the hyperfine structure effects must be accounted for, in the LTE analysis the hyperfine structure effects were not included. However, they suggested that the errors induced by this omission, in an absolute abundance analysis, are small. To support this suggestion, they simulated the effect of including the hyperfine transitions by increasing the microturbulent velocity, which gave to copper abundances on average about 0.1 dex smaller. The results of their analysis suggested that [Cu/Fe] was deficient in these halo giants, and possibly becoming more deficient at lower metallicities.

Sneden & Crocker (1988) examined the Cu I transitions at 5105 Å and 5782 Å in 2 field dwarfs and 3 field giants with high resolution ($\Delta\lambda = \simeq 0.1$ Å), low noise (S/N $\simeq$ 350) digital spectra of 6 Cu I transitions between 4722 Å and 6615 Å. The survey stars covered the metallicity range -2.7 < [Fe/H] < -1.3 dex. Model atmosphere parameters from the literature was adopted, and an LTE absolute analysis with theoretical or laboratory oscillator strengths was used. Experimental hyperfine structure coefficients were employed during the reduction of the Cu I transitions. They found that that the copper to iron ratio declines smoothly below [Fe/H] $\simeq$ -1.3 dex, reaching [Cu/Fe] = -0.75 dex for the most metal deficient survey stars.

Sneden, Gratton & Crocker (1991) established the atmospheric copper abundances in 23 stars, including both field dwarfs and field giants, from high resolution (R$\simeq$50,000), low-noise (S/N$\simeq$ 150) digital spectra of the 5105 Å and 5782 Å Cu I lines. Much less emphasis was put on the 5782 Å line since it becomes unmeasurably weak in the most metal deficient stars, and has an uncertain and possibly substantial contribution from a blending Sc I feature. The survey stars spanned the metallicity range -2.9 $\leq$ [Fe/H] -0.4 dex. Model atmosphere parameters from the literature were adopted, and an absolute analysis with theoretical or laboratory oscillator strengths were used. Experimental hyperfine structure coefficients were employed during the reduction of the Cu I transitions. They found that copper is deficient in all metal-poor stars and appears to vary linearly with metallicity: [Cu/Fe] $\simeq$ 0.38 [Fe/H] + 0.15. The robustness of the results was demonstrated by showing that variations in the derived copper abundances with changes in the the stellar atmosphere parameters were small.

The calculated evolution is in reasonable agreement with the observations (especially when the Type II iron yields are reduced by a factor of two) of a deficient [Cu/Fe] ratio in metal-poor stars, and then increasing past the disk-halo transition point at [Fe/H] $\simeq$ -1.0 dex (Figure 34). The undulations of the curve is due to small mass and metallicity effects in the exploded massive star models. Intermediate and low mass stars contribute no copper or iron in the calculations shown, and standard carbon deflagration models for Type Ia supernovae produce little copper. The solar metallicity exploded massive star models have about a factor of 5 larger copper yield than the models with smaller initial metallicities (Paper I; Paper II). This overwhelms the iron contributions from Type Ia supernovae, and causes the increase in the computed [Cu/Fe] ratio for [Fe/H] > -1.0 dex. This is shown by comparing the solid line in the figure (which includes both Type II and Type Ia supernovae) with the dotted line (which excludes Type Ia supernovae). The amplitude of the jump is decreased if more Type Ia supernovae are included. The solar abundances of the two stable copper isotopes are well accounted for by the presupernova and exploded star models (see Figures 4 and 5).

### 3.8.2 ZINC

Evolution of the zinc to iron ratio [Zn/Fe] as a function of [Fe/H] is shown in Figure 35. The solid line shows the calculation, and the dashed lines indicate factors of two variation in the iron yields from massive stars.

Sneden & Crocker (1988) examined the atmospheric zinc abundances in 2 field dwarfs and 3 field giants with moderate resolution ($\Delta\lambda = 0.15$ Å), low noise (S/N $\sim$ 350) digital spectra of the Zn I lines at 4722 Å and 4810 Å. The 5 stars in this survey covered the metallicity range from -1.3 > [Fe/H] > -2.7 dex. Some details about the type of analysis used and how the stellar atmosphere parameters were established were discussed in§3.8.1. The zinc transitions are strong enough in the sun that hyperfine structure effects are important (primarily from $^{67}$Zn). Hyperfine effects were are not accounted for because the Zn I features are sufficiently weak that hyperfine splitting should not significantly change the equivalent widths. They found an average zinc to iron ratio of [Zn/Fe]=-0.03 $\pm$ 0.11 dex, but the small number statistics precluded any definite statements about the trends of [Zn/Fe] with metallicity. The authors caution that the relatively high 9.4 eV ionization potential of zinc means that a significant fraction of the zinc in the line forming layers is neutral, and the derived abundances might be susceptible to non-LTE effects.

Sneden, Gratton & Crocker (1991) studied the zinc abundances of 23 field dwarfs and 18 field giants using high resolution (R$\simeq$50,000), low noise digital spectra of the Zn I lines at 4722 Å and 4810 Å. The survey stars spanned the metallicity range -2.9 $\leq$ [Fe/H] $\leq$ -0.1 dex. Some details about the type of analysis used and how the stellar atmosphere parameters were established were discussed in §3.8.1. Hyperfine effects were are not accounted for because the Zn I features are sufficiently weak that hyperfine splitting should not significantly change the equivalent widths. They found that the zinc to iron ratio is solar with a very small star-to-star scatter over the entire metallicity range; [Zn/Fe] = 0.04 $\pm$ 0.01 dex. They showed this result is robust with respect to variations in the effective temperatures and surface gravities.

The calculated evolution is in reasonable agreement with the observations (especially when the Type II iron yields are reduced by a factor of two) of a solar [Zn/Fe] ratio at low metallicities, although the computed history is smaller than the observations by about a factor of two (Figure 35). However, elemental zinc is dominated by the isotope $^{64}$Zn which is unfortunately deficient in the exploded massive star models (see Figures 4 and 5). This may be a consequence of the fact that the nuclear reaction network employed in the stellar evolution calculations was terminated at germanium, or an artifact of how the explosion was parameterized. Intermediate and low mass stars contribute no zinc or iron in the calculations shown, and standard carbon deflagration models for Type Ia supernovae produce little zinc. The solar metallicity massive star models have about a factor of 5 larger zinc yield than the models with smaller initial metallicities (Paper I; Paper II). This overwhelms the iron contributions from Type Ia supernovae, and causes the increase in the computed [Zn/Fe] ratio for [Fe/H] > -1.0 dex. The amplitude of the jump is supressed if more Type Ia supernovae are included.

Other possible sources of zinc (and copper) production are a weak s-process component operating during core helium burning in massive or supermassive stars (Couch, Schmiederkamp & Arnett 1974; Lamb *et al.* 1977; Meynet & Arnould 1993; Baraffe, El Eid & Prantzos 1993), or a classical s-process operating in low mass asymptotic giant branch stars that are undergoing thermal pulsation (Iben 1982; Iben & Renzini 1983; Boothroyd & Sackmann 1988). Injection of zinc (and copper) into the interstellar from these massive stars would primarily affect the halo population, while these low mass stars would contribute chiefly to disk abundances.

### 3.9 OTHER SOLAR VICINITY QUANTITIES

#### 3.9.1 THE HELIUM TO METAL ENRICHMENT RATIO

Figure 36 shows some of the observed helium abundances versus the O/H ratios found in Galactic and extragalactic H II regions (Piembert 1986; Pagel, Terlevich & Melnick 1986; Pagel *et al.* 1992). Only the estimated error bars for the Piembert (1986) survey are shown; the other surveys have similiar error bars. Reduction of the spectrographic data is especially demanding in these observations. The principle sources of uncertainty are the removal of contamination from Wolf-Rayet stars, correction for the amount of neutral helium inside the H II regions, problems from telluric and Galactic absoption lines, disagreement on the theoretical recombination coefficients, corrections for the high population of the metastable He I 2 $^3$S state, transformation of the O/H ratios into total metallicity values, and the effects of oxygen depletion through grain formation (Piembert 1986; Osterbrock 1989; Pagel *et al.* 1992). For example, the figure shows Orion has O/H = $4.4 \times 10^{-4} \simeq$ Z/25, so that the metallicity of the gas is Z=0.011. However, when the estimates of the amount of oxygen locked up in silicate cores and polymer mantles is included ($\simeq 20 - 50\%$), the total metallicity can climb to solar (in agreement with stellar abundance determinations). Furthermore, the hypothesis of a linear relationship between the helium content and the metallicity content is often invoked

$$Y = Y_p + \left(\frac{\Delta Y}{\Delta Z}\right) Z \ . \tag{20}$$

Hence, the observed abundances critically influence estimates of the primordial helium $Y_P$ originating from Big Bang nucleosynthesis (the intercept), and the amount of nuclear processing done by stars (the slope). However, the assumption of linear relationship between the helium abundance and the metallicity is only valid at low metallicities. At metallicities larger than $\gtrsim 0.1$ $Z_\odot$, the contributions from Type 1a supernovae and low mass stars distort the linear relationship. Figure 36 shows the results of equation (20) for various slopes $\Delta Y/\Delta Z$ that that start from a primordial helium abundance of $Y_P$=0.23. Although the variability is large, values of $\Delta Y/\Delta Z = 4.0 \pm 1.0$ are taken to be representative.

The quantity $\Delta Y/\Delta Z$ is sensitive to the maximum mass of a star that will give its nucleosynthetic products back to the interstellar medium. Intermediate and low mass stars shed their mass, mostly hydrogen and helium, on the giant branches before becoming white dwarfs. Typically these stars have $\Delta Y/\Delta Z \simeq 4$ (Renzini & Voli 1981). Massive stars produce some helium, but mostly metals, of which oxygen comprises the vast bulk. The metal-poor exploded massive star models have $\Delta Y/\Delta Z \simeq 1.3$ (Paper II). It is clear from these values that if all the material in massive stars is returned to the interstellar medium, then it will be difficult to satisfy the inferred $\Delta Y/\Delta Z$ ratio. Hence, it has been suggested that the most massive stars become black holes and swallow a good fraction, if not all, of their metals (Chiosi & Matteucci 1982; Pagel 1986; Maeder 1992, 1993; Brown & Bethe 1994).

The best constraints on the black hole mass cutoff using the $\Delta Y/\Delta Z$ ratio come from low metallicity regions, in which the influence from stellar winds and the cumulative effects of chemical evolution are small. For a total metallicity $Z < 0.01$, we find that the approximation given by equation (20) is very reasonable. Integrating over a Salpeter type initial mass function, Maeder (1992, 1993) suggested that the observed $\Delta Y/\Delta Z$ is reproduced if black holes are formed above about $\simeq 20$ $M_\odot$, while Brown & Bethe (1994) found $25 \pm 5$ $M_\odot$. With a chemical evolution model, Chiosi & Matteucci (1982) determined that the observed ratio could be reproduced if the slope of the initial mass function was significantly steeper in the high mass region.

Figure 37 shows the calculated $\Delta Y/\Delta Z$ ratio as a function of the black hole mass limit. The observational limits inferred from figure 36 are shown as the grey band. One curve shown is for a time early in the evolution when the total metallicity is $Z=10^{-3}$ $Z_\odot$ the another curve shown is after 10.4 Gyr when $Z=Z_\odot$. The low metallicity curve suggests that the observed $\Delta Y/\Delta Z$ ratio is reproduced when stars above $\simeq 30$ $M_\odot$ become black holes. The solar metallicity curve indicates a much lower value of $\simeq 17$ $M_\odot$. However, computing a $\Delta Y/\Delta Z$ ratio from the solar helium abundance and solar metallicity is not meaningful since after 10 Gyr of evolution contributions from intermediate and low mass stars (which increase $\Delta Y/\Delta Z$) and Type Ia supernovae (which decrease $\Delta Y/\Delta Z$) are significant and difficult to disentangle. The tentative conclusion, based on the low metallicity curve, is that the Galaxy could contain a large number of stellar mass black holes (Maeder 1992, 1993; Brown & Bethe 1994).

However, single stars above 30 $M_\odot$, or at least above 40 $M_\odot$, may enter into a Wolf-Rayet stage possibly with a large mass loss. Also a non-trivial fraction of massive stars may be members of a binary system and go through a similar mass losing Wolf-Rayet stage. The actual mass loss rate for Wolf-Rayet stars metallicity dependent and uncertain, but it may be large (Schaller et al. 1992; Maeder 1992). In a recent investigation, Woosley et al. (1994) find $\Delta Y/\Delta Z \simeq 4$ for Wolf-Rayet stars. In addition, it may be myopic to only look at satisfying the $\Delta Y/\Delta Z$ ratio and ignore the effects on the evolutionary histories of all the elements. These two concerns, combined with the scatter present in the observations of Figure 36 and the extraordinarily difficult reduction of the spectroscopic data, suggest that strong statements regarding the black hole formation mass that are based on $\Delta Y/\Delta Z$ should be viewed with caution.

### 3.9.2 THE G DWARF DISTRIBUTION

An important historical constraint on models of Galactic chemical evolution in the solar vicinity is the distribution of disk G dwarf stars as a function of metallicity. Relevance of the G Dwarf distribution was established when simple, analytical models of chemical evolution failed to reproduce the observed distribution, predicting too many metal-poor stars (van den Bergh 1962; Schmidt 1963; Pagel & Patchett 1975; Audouze & Tinsley 1976; Tinsley 1980). Stars of the G dwarf spectral type are of interest because they have a main-sequence lifetime of about 15 Gyr, comparable to the estimated age of the Galaxy. Therefore, they represent a sample of stars that has almost never been depleted by stellar evolution, continually accumulating since the first epoch of star formation (although see Bazan & Mathews 1990). Hence, a complete sample of these stars in the solar neighborhood carries memory of the star formation history.

There are two main components to the solution of the classical G dwarf problem; infall of primordial or near primordial gas early in the history of the Galaxy, and dynamical heating of stars in the disk. Infall has been discussed thoroughly in the literature; suffice to say that infall of essentially unprocessed material deviates from the constant mass assumption of the simple model, and continually dilutes the metallicity of the gas that constitutes the disk (Larson 1974, 1976; Tinsley 1980; Clayton 1984; Pagel 1989; Ferrini et al. 1992). The metallicity distribution of 132 disk G dwarfs enclosed in a sphere of radius $\simeq 25$ pc centered on the Sun was examined by Pagel (1989). The older a disk stellar population is, and hence the more metal-poor the population generally is, the larger the scale height and velocity dispersion perpendicular to the Galactic plane. Hence, the oldest and most metal-poor disk G dwarfs have probably escaped out of the volume sampled by Pagel (1989). Using the observational data on the surface density in the Galactic plane and the velocity dispersion in the vertical direction, Rana (1991) and Sommer-Larsen (1991) estimated the extent to which every metallicity bin in the Pagel (1989) survey is affected and replenished it accordingly.

The number $N$ of G dwarfs with iron abundances between $f_1 \leq [\text{Fe/H}] \leq f_2$ is obtained as

$$N(f_1, f_2) = \int_{t(f_1)}^{t(f_2)} B(t) \, \Psi_G(m) \, dt \;, \tag{21}$$

where $\Psi_G$ is the fraction of G dwarfs in a single stellar generation and $t(f)$ is the time at which $[\text{Fe/H}]=f$ (Pardi & Ferrini 1994). The dashed line histogram of Figure 38 shows the cumulative G dwarf data reconstruction by Rana (1991), while the solid line histogram shows the calculation. Both histograms are normalized to the total number of G dwarf stars. Given the relative success of the proceeding sections in reproducing the abundance histories of all the stable elements from hydrogen to zinc, it is perhaps not surprising that the calculated G dwarf distribution is quite close to the observed distribution. The combination of infall and dynamical heating appears to make the paucity of metal-poor disk stars less critical than before.

## 3.10 EVOLUTION OF THE GALACTIC SUPERNOVA RATES

A direct measurement of the Galactic supernova rate is difficult owing to possible incompleteness in historical observations, and uncertainty as to the fraction of the Galactic disk and altitude that are sampled. Indirect inference from supernova rates in similar galaxies is adversely affected by the imprecise value of the Hubble constant, and the uncertainty in estimating the total blue luminosity and morphological classification of our Galaxy. Systematic searches for extragalactic supernova are also hampered by the need to know the distance, luminosity, and Hubble class of the host galaxy, as well as the dates and limiting magnitude of each observation. Such detailed information is only available for a few dozen supernova through the visual search of Evans, van den Bergh & McClure (1989), and the photographic search of Cappellaro et al. (1993ab). Both of these supernova searches examined only a subset of the spiral galaxies in the Shapley-Ames catalog.

Based on these surveys, estimates of the core collapse (Type II and Ib) and thermonuclear driven (Type Ia) supernova rates were derived and discussed by van den Bergh, McClure & Evans (1987), van den Bergh & Tammann (1991), and Cappellaro (1993ab). These estimates assumed that the peak luminosity of each supernova class was a standard candle, and a large correction for edge-on spirals ($\sin i$ effect). Using these extragalactic estimates with a total Galactic blue luminosity of 2.3 × $10^{10}$ $L_\odot$, a Hubble constant of 75 km s$^{-1}$ Mpc$^{-1}$, and a Sbc Galactic morphology, the Galactic supernova rates from van den Bergh & Tammann (1991) are 4.1 per century for core collapse events and 0.56 per century for thermonuclear events.

Recently, van den Bergh & Mcclure (1994) have reanalyzed the Evans *et al.* observations. Citing SN 1987A and SN 1991bg as good evidence, they relaxed the assumption that supernova display small intrinsic variations at maximum light by investigating the effects of different luminosity functions. They also dropped the $\sin i$ correction term because the supernova survey of Muller *et al.* (1992) found no evidence for a dependence on inclination angle of the host galaxy. For the same Galactic parameters used above, they estimate that the present Type II + Type Ib rate is 2.4 - 2.7 per century while the Type 1a rate is 0.3 - 0.6 per century. Thus, the total rate is about 3 per century, and the ratio of core collapse to thermonuclear events is about 6.

The evolution of the supernova rates in our model is shown in Figure 39. These rates are calculated by integrating in radius across the model Galaxy (see equations 16 and 17). As the evolution of the Galaxy begins, a large abundance of gas gets turned into a large number of stars. The death of these massive stars gives a correspondingly large Type II + Ib supernova rate. As the amount of gas decreases due to the formation of compact remnants and long-lived low mass stars the Type II +Ib supernova rate decreases with time. The onset of Type Ia supernovae is delayed due to longer lifetime of the intermediate mass star which produces a white dwarf.

After 15 billion years, the calculated Type II + Ib and Type Ia rates are in excellent agreement with the estimates of van den Bergh & Mcclure (1994). In addition, within a region 4 kpc from the Sun about 8% of the gas gets turned into massive stars. Hence, the Type II + Ib supernova rate within this region should be 2 per 1000 years. This is very close to the observed number that are known to have occurred within 4 kpc of the Sun (SN 185, SN 1054, SN 1181, SN 1670) during the last 2000 years (van den Bergh & Mcclure 1994). However, the statistical uncertainty in the frequency with which supernova of different types occur in galaxies of different Hubble class is large, and this agreement may be fortuitous.

## 4. CONCLUSIONS

The chemical evolution of 76 stable isotopes, from hydrogen to zinc, has been calculated using the nucleosynthetic yields from a grid of 60 Type II supernova models of varying mass (11 $\lesssim$ M/M$_\odot$ $\lesssim$ 40) and metallicity (0, $10^{-4}$, 0.01, 0.1, and 1 Z$_\odot$). The chemical evolution calculation employed a simple dynamical model for the Galaxy (infall with a 4 Gigayear e-folding time scale onto a exponential disk and 1/r$^2$ bulge), and standard chemical evolution parameters, such as a Salpeter initial mass function and a quadratic Schmidt star formation rate. Self-consistency was obtained by operating a feedback loop between the massive star models and the chemical evolution model. The element evolutions were shown to be robust with respect to variations in the efficiency of star formation, the exponent of the initial mass function, the amplitude of the Type Ia contributions and small deviations from self-consistency of the stellar evolution and chemical evolution calculations.

The theoretical results have been compared in detail with the observed stellar abundances of solar vicinity stars with varying metallicity in the range -3.0 $\lesssim$ [Fe/H] $\lesssim$ 0.0 dex. The robustness of the results to variations in the iron yields of the Type II supernova models was examined. The principal results of this paper are:

(1) Excellent agreement with the Anders & Grevesse (1989) solar abundances from hydrogen to zinc, when the calculation is sampled at a time 4.6 Gyr years ago at a distance of 8.5 kpc (Figures 3-5). The stable isotopes from hydrogen to zinc range over some 10 orders of magnitude. That the calculations agree with the Anders & Grevesse (1989) abundances to within a factor of 2 is very encouraging.

(2) Type Ia supernovae are important for reproducing the observed age-metallicity relationship, with Figure 7 indicating that about 1/3 of the solar iron abundance is due to the thermonuclear disruption of white dwarfs. Better agreement with most of the observed abundance evolutions occur if the Type II iron yields of Figure 6 are systematically reduced by a factor of two (the upper dashed curves in Figures 11 – 35). The reduced iron yields are consistent with observations of several Type II supernovae (including SN 1987A), and within the uncertainty in modeling the explosion. Factors of two increase in the iron yields are excluded. If the iron yields from Type II supernovae are reduced by a factor of two, then Type Ia supernovae contribute 1/2 of the solar iron abundance. It may be that 2/3 of the solar iron abundance comes from Type Ia events and 1/3 from Type II events could be accomplished without doing grave injustice to the stellar physics.

(3) Neutrino-process nucleosynthesis provides an adequate explanation for the origin $^{11}$B, $^{19}$F, and perhaps about half of the increase in $^{7}$Li over its canonical, homogeneous Big-Bang value (Figures 8, 9 and 10). The agreement is achieved for $\mu$ and $\tau$ neutrino temperatures in the range 6 to 8 MeV, which is just the range suggested by SN 1987A, and preferred by those who model core collapse of massive stars. The $\nu$-process, cosmic ray spallation, and homogeneous Big Bang nucleosynthesis are complimentary, and together they yield a comprehensive set of prescriptions for the evolution of the light (A $\leq$ 11) elements.

(4) Oxygen production in the massive star models is sufficient to completely explain the [O/Fe] observations across the entire metallicity range, although slightly smaller iron yields fit the observations a bit better (Figure 11). Carbon is adequately produced in very low metallicity Type II supernova to explain the relatively flat carbon-to-iron ratios observed in the halo (Figure 13). Primary nitrogen is not produced in the very low metallicity Type II supernova models, although the exact amount produced is not well determined numerically (Figure 14). Type II supernovae models with initial compositions $Z \gtrsim 0.1\ Z_\odot$ underproduce the solar carbon and nitrogen isotopes by about a factor of three, with intermediate and low mass stars providing the difference.

(5) Observations of sodium, magnesium, and aluminum abundances are reasonably accounted for by the calculations (Figures 15 – 20). If the corrections to stellar abundances determined through the sodium D lines in very metal-poor field giants continue, then the [Na/Fe] observations would come into much better agreement with the theoretical models. The calculated evolution of [Al/Fe] is slightly larger than the observations of the most metal-poor stars, does not rise as fast through the [Fe/H] $\simeq$ -1.0 dex halo-disk transition point, and is slightly lower than the disk star observations.

(6) Silicon, sulfur, potassium, and calcium production by Type II supernovae are sufficient to explain the observed trends, remarkably so for the $\alpha$-chain nuclei (Figures 21 – 25). Phosphorus and chlorine are predicted to have odd-Z element characteristics in dwarfs, while neon and argon behave as $\alpha$-chain elements.

(7) Although difficulties in modeling the explosion mechanism make the absolute yields of the iron group nuclei (and beyond) more uncertain, the calculated evolution of the scandium, vanadium, chromium, manganese, cobalt nickel, copper, and zinc to iron ratios reproduce the zeroth order observed trends and may be robust (Figures 26 – 35). Titanium is co-produced with iron, which is at variance with the observed characteristics of a $\alpha$-element behavior. There is a general suppression of the stable odd-Z nuclei relative to their stable even-Z neighbors.

(8) In order to explain the observed helium to metal enrichment ratio ($\Delta Y/\Delta Z \simeq 4$), there may be a cut off in the mass of supernovae that eject all material external to the iron core of about 30 $M_\odot$. However, this limit may be increased by considering the effects of mass loss, and its effect on the other elements is uncertain (Figures 36 and 37).

(9) The supernova rates in our Galaxy are uncertain and difficult to assess, but the canonical values of about 3 per century for Type II supernovae and about 0.5 per century for Type Ia supernovae in the present epoch are reproduced quite well by the calculations (Figure 39).

This work has been supported by the National Science Foundation under grant number AST91-15367; NASA under grant number NAGW 2525, the US Department of Energy by the Lawrence Livermore National Laboratory under contract number W-7405-ENG-48, the California Space Institute under grant number CS86-92, and an Enrico Fermi Postdoctoral Fellowship (FXT).

The authors especially wish to thank Rob Hoffman for keeping our nuclear reaction cross sections and nuclear reaction network up to date. Interesting conversations on the supernova mechanism and hydrodynamics were held with Hans Bethe, Gerry Brown, and Marc Herant. Very helpful discussions on stellar abundance determinations were held with Bob Kraft and Ruth Peterson. Finally, rewarding talks about Galactic chemical evolution were held with Jim Truran and Don Clayton.

## TABLE 1

### Chemical Evolution Model Parameters

| Value | Quantity | Fitting Parameter? | Units |
|---|---|---|---|
| 2.8 | Schmidt star formation efficiency factor | Yes | $Gyr^{-1}$ |
| 2.0 | Schmidt star formation exponent | No | |
| -1.31 | Salpeter exponent | Yes | |
| 0.08 | Lower limit of IMF | No | $M_\odot$ |
| 40.0 | Upper limit of IMF | No | $M_\odot$ |
| $8.5 \times 10^3$ | Distance to solar neighborhood | No | pc |
| 75.0 | Solar vicinity surface density | No | $M_\odot\ pc^{-2}$ |
| $1.0 \times 10^4$ | Galactic center surface density | No | $M_\odot\ pc^{-2}$ |
| $2.0 \times 10^3$ | Extent of inverse square profile | No | pc |
| 15.0 | Age of galaxy | No | Gyr |
| 4.0 | Time scale for disk formation | No | Gyr |
| 11.0 | Lower limit for Type II rates | No | $M_\odot$ |
| 40.0 | Upper limit for Type II rates | No | $M_\odot$ |
| 3.0 | Lower limit of binary systems | No | $M_\odot$ |
| 16.0 | Upper limit of binary systems | No | $M_\odot$ |
| 0.007 | Type Ia amplitude factor | Yes | |

## TABLE 2

### Solar Vicinity Quantities In Present Epoch

| Calculated Value | Quantity | Observational Limits |
|---|---|---|
| 8.41 | Gas surface density ($M_\odot\ pc^{-2}$) | $6.6 \pm 2.5^a$ |
| 9.83% | Gas fraction | $10.0 \pm 3.0^a$ |
| 0.32 | Accretion rate ($M_\odot\ pc^{-2}\ Gyr^{-1}$) | $0.2 - 1.0^b$ |
| 4.12 | Stellar birthrate ($M_\odot\ pc^{-2}\ Gyr^{-1}$) | $2.0 - 10.0^c$ |
| 0.0201 | Gas metallicity at stellar birth | $0.021 - 0.019^d$ |

[a] Rana & Basu (1992), [b] See §2.1 of text, [c] Güsten & Mezger (1983), [d] Anders & Grevesse (1989).

**Figures**

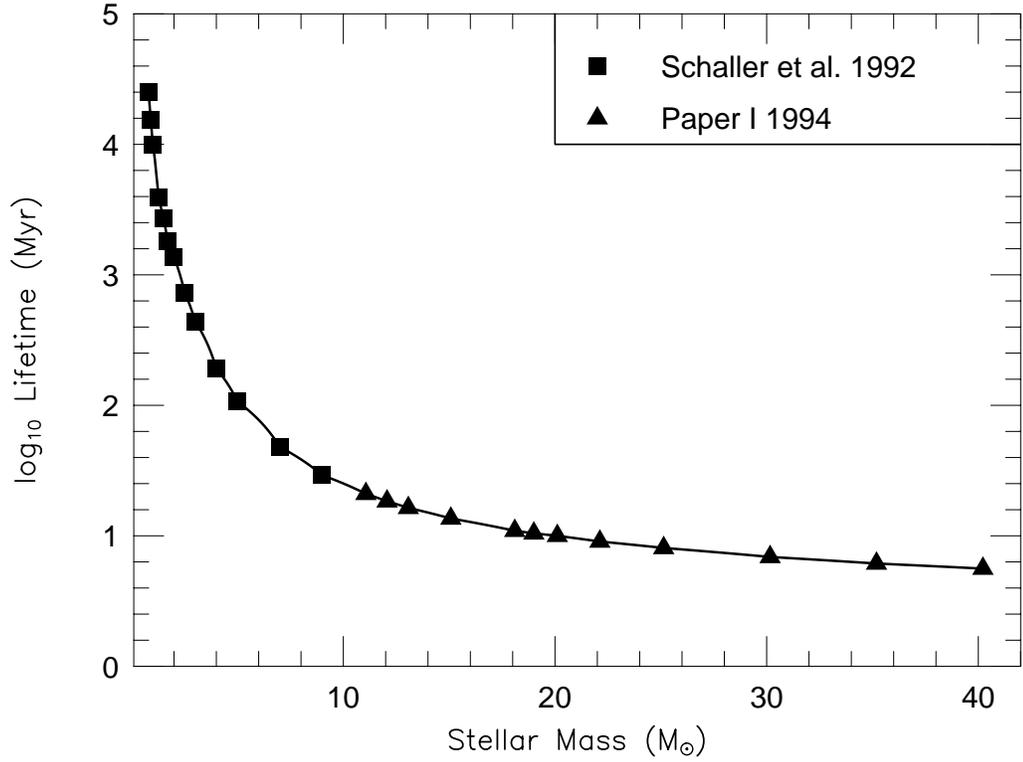

Fig. 1.— Main-sequence lifetimes as a function of mass for stars of solar metallicity. Lifetimes of the massive stars are given directly by the stellar evolution calculations of Paper I. For stars less massive than 11 $M_\odot$ the main-sequence lifetimes of Schaller *et al.* (1992) were used. Metallicity effects on the lifetimes of the massive stars amounts to shifts of about $\simeq 5\%$ (Paper I), and Schaller *et al.* (1992) determined the intermediate-low mass star main-sequence lifetimes for only two metallicities. Since the metallicity effects are small, and the metallicity grid sparse for the lower mass stars, the values shown in the figure were used for all stellar metallicities.

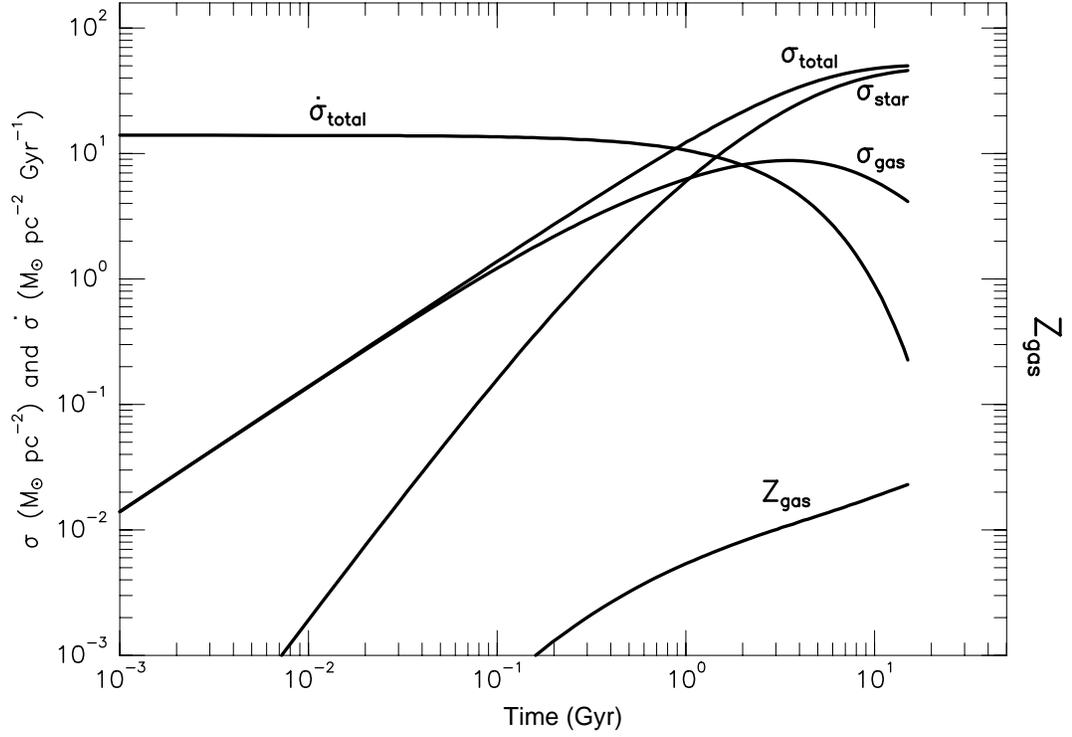

Fig. 2.— Dynamical quantities and total metallicity of the solar neighborhood as a function of time. The total surface mass density, accretion rate, surface mass density of the gas, surface mass density of stars, and the stellar birthrate are labeled and have values indicated by the left y-axis. These quantities are independent of the composition of the gas or the nucleosynthetic yields. The accretion rate is the decaying exponential given by equation (2). The total surface mass density is the sum of the gas and stellar mass densities (see equation (1)), whose value in the present epoch was chosen to be 75 $M_\odot$ pc$^{-2}$. The total metallicity of the gas $Z_{\rm gas}$ (which is dominated by oxygen) is also shown, with values give by the right y-axis.

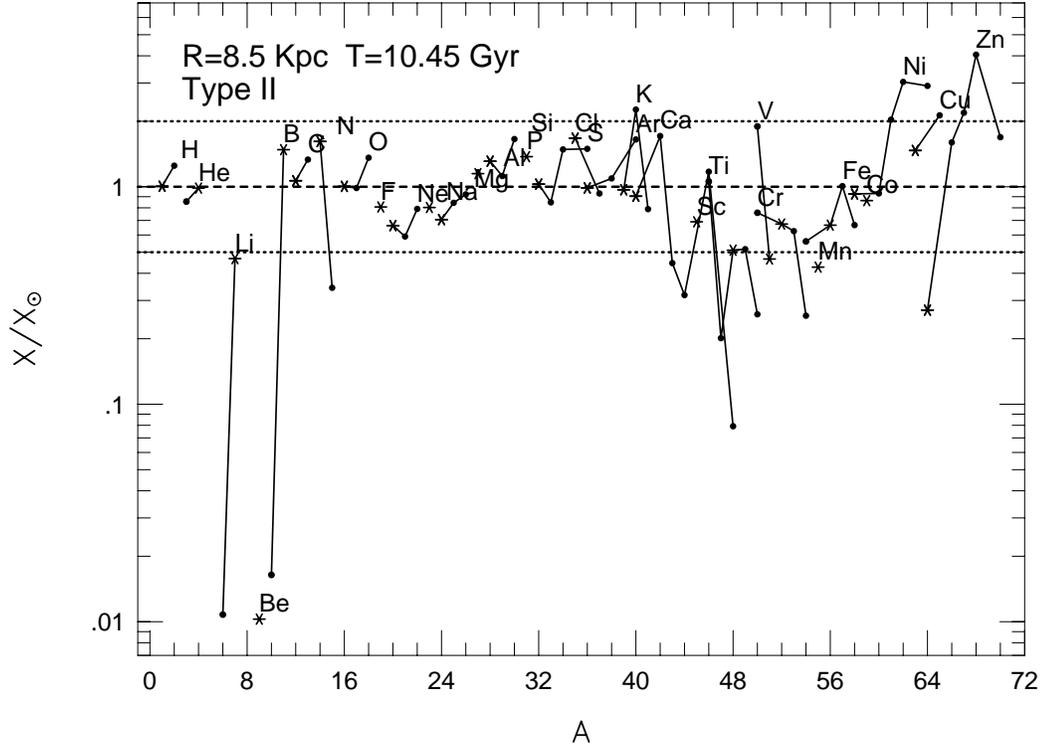

Fig. 3.— Stable isotopes from hydrogen to zinc present in the interstellar medium at a time when (4.6 Gyr ago) and place where (8.5 Kpc Galactocentric radius) the Sun was born. The x-axis is the atomic mass number. The y-axis is the logarithmic ratio of the model abundance to the Anders & Grevesse (1988) mass fraction. Only contributions from massive stars, intermediate, and low mass stars are included (see Figures 4 and 5). The most abundant isotope of a given element is marked by an asterisk and isotopes of the same element are connected by solid lines. If the calculation were perfect, all of the isotopes would have a constant ordinate (equal to unity in this case) and the solar composition would have been replicated. In each figure, the perfect case is denoted by a dashed horizontal line. A formal analysis of the total uncertainty would be a daunting task, and we therefore adopt the convention that isotopes falling within a factor of two of their solar value are a "success". In the figure, the horizontal dotted lines denote this factor.

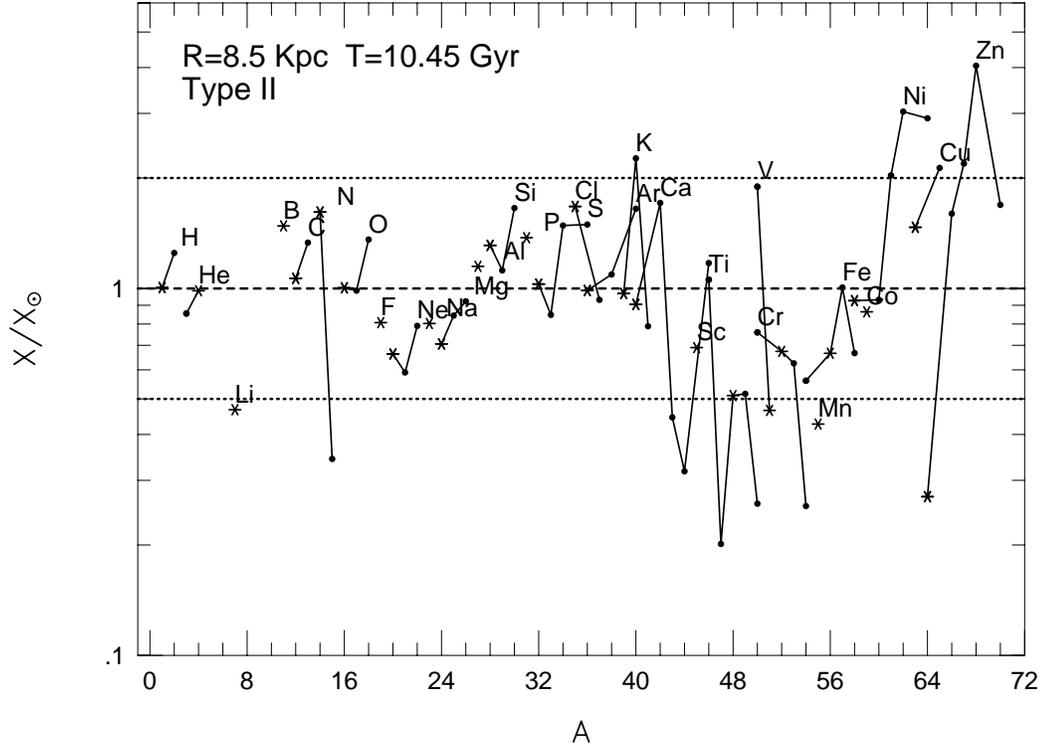

Fig. 4.— Same as Figure 3, except some of underproduced isotopes in Figure 3 are removed and the y-axis scale is expanded. The solar abundances are mostly determined by the nucleosynthesis that occurred in stars born with a metallicity between $0.2\ Z_\odot \leq Z_{gas} \leq 0.4\ Z_\odot$, which is consistent with the broad peak in the G-dwarf distribution in the solar vicinity (see §3.9.2). Isotopes below calcium show less scatter than those isotopes above because isotopes below calcium are chiefly produced by hydrostatic burning processes before the explosion, while heavier isotopes are sensitive to the uncertain modeling of the explosion mechanism. Note that the hydrogen, helium and $^{16}$O production factors are all very nearly equal to one.

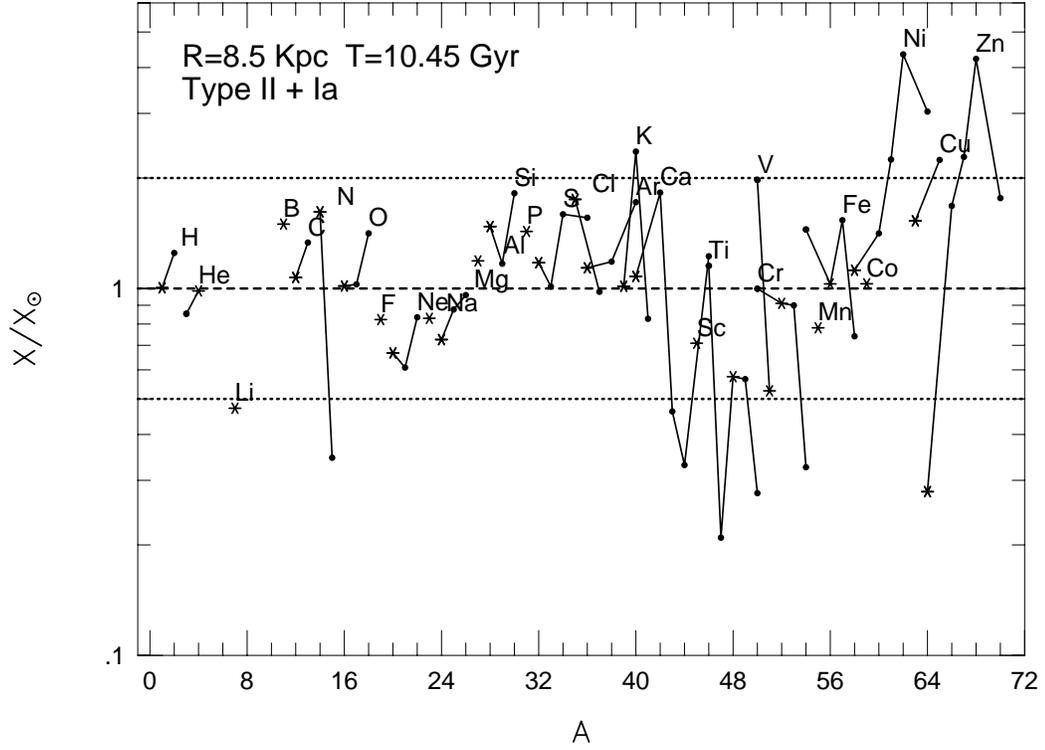

Fig. 5.— Same as Figure 4, but Type Ia supernovae are included. The contribution of Type Ia supernova, whose canonical nucleosynthesis is taken to be the carbon deflagration model of Thielemann *et al.* (1986), is determined by varying the parameter $C$ in equation (14) until the solar $^{56}$Fe abundance is achieved. This procedure gives results consistent with the observed age-metallicity relationship (§3.2), and Galactic supernova rates (§3.10). Type Ia supernova contribute 1/3 of the solar iron abundance, and improve the fit of the isotopes $^{50-53}$Cr, $^{55}$Mn, and $^{54-58}$Fe, while scarcely affecting the solar abundances of the other isotopes. Almost all of the isotopes below calcium, and nearly all of the most abundant isotopes of an element above calcium show good agreement with solar values.

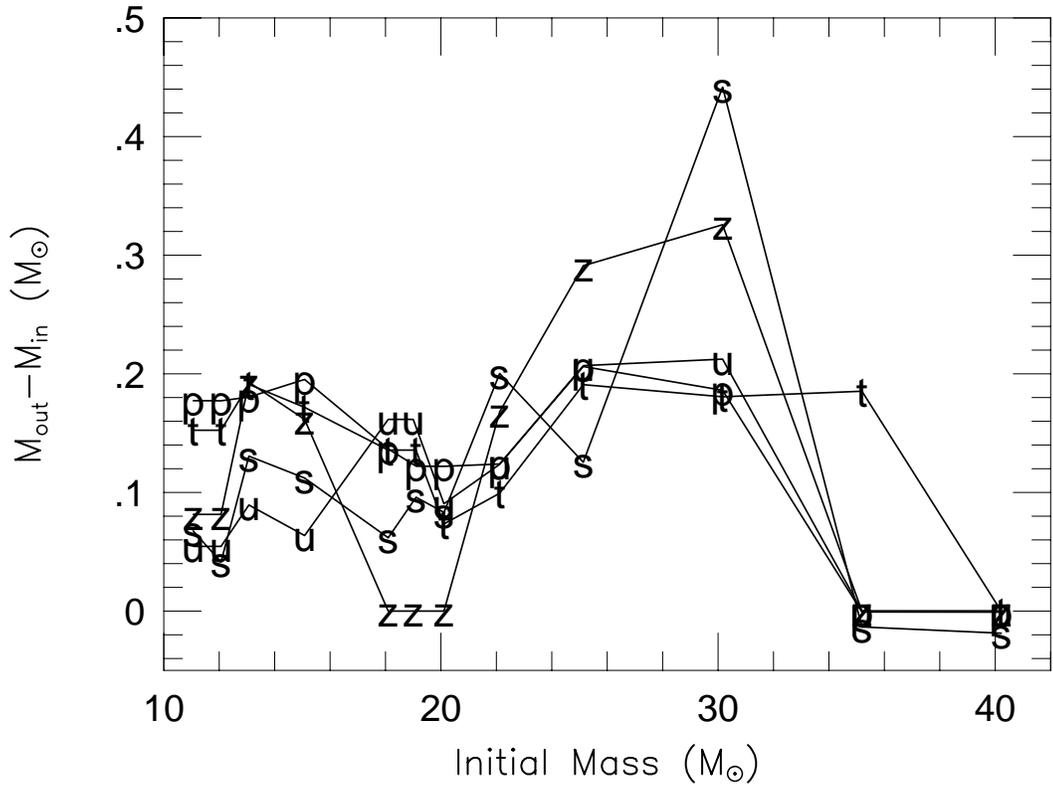

Fig. 6.— Amount of $^{56}$Fe produced from the standard set of exploded massive star models described in §2.4. The zero metallicity stellar models are labeled with the letter "z", "u" for $10^{-4}$ $Z_\odot$, "t" for $10^{-2}$ $Z_\odot$, "p" for 0.1 $Z_\odot$, and "s" for 1.0 $Z_\odot$. These iron yields are not monotonic with respect to either mass or metallicity. Variations are caused by the uncertainty in modeling the explosion mechanism, the amount of iron which may fall back onto the compact remnant, and the sensitivity of the progenitor models to the interaction of the various convective zones during the later stages of nuclear burning. The iron yields shown in the figure are consistent the Type II observations.

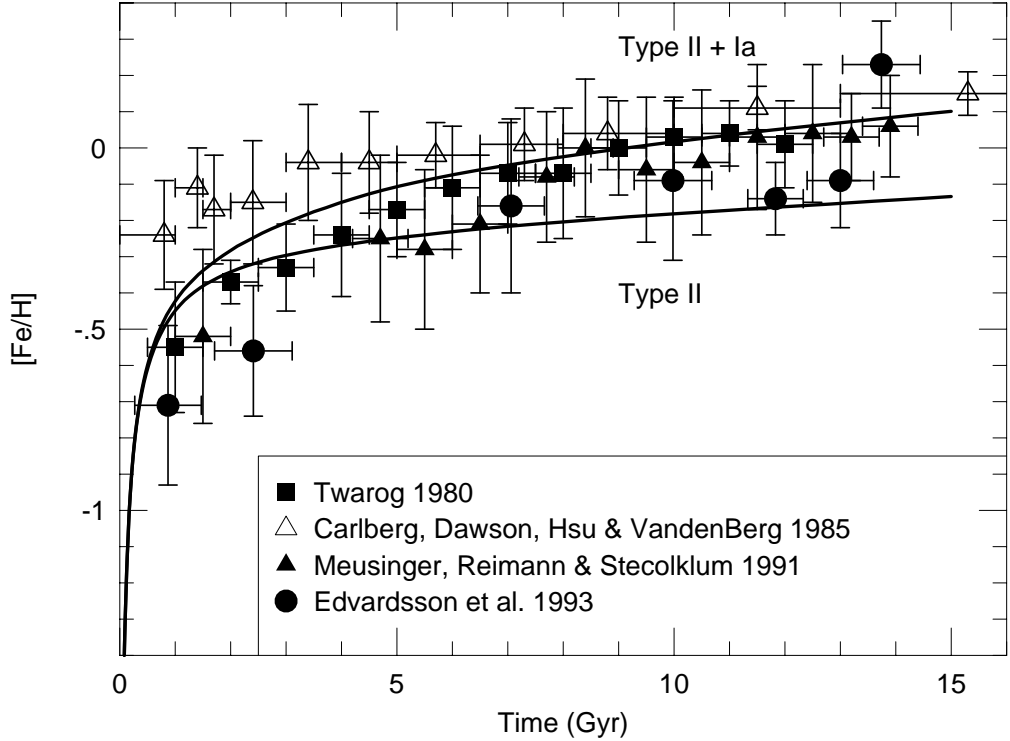

Fig. 7.— Solar neighborhood age-metallicity relationship. Calculations with and without the iron contributions from Type Ia supernovae are shown, along with the observational data. The error bars are not due to intrinsic uncertainties of a single observation, but represent the spread of many stars in the binned [Fe/H] data. The reanalysis of the Twarog (1980) data by Carlberg *et al.* (1985) gave larger [Fe/H] values at earlier times due to selection effects inposed on the data set. The data of Edvardsson *et al.* (1993) represent a spectroscopically calibrated, age-metallicity relationship, and their results are probably the most accurate of the data sets shown (see text). The calculated curves are not as ragged as the iron yields shown in Figure 6 because repeated integrations (in time) over metallicity and an initial mass function smooths out irregularities in the yields. The figure indicates that the inclusion of Type Ia supernovae is important for reproducing the observed iron evolution, and suggests that about 2/3's of the solar iron abundance comes from core collapse events and about 1/3 comes from Type Ia supernovae. If the Type II iron yields are reduced by a factor of two (which is well within the uncertainties of the explosion mechansim), these fractions become 50% from Type Ia supernova and 50% from Type II supernova. It may be that 2/3 of the solar iron abundance comes from Type Ia events and 1/3 from Type II events without doing grave injustice to the stellar physics.

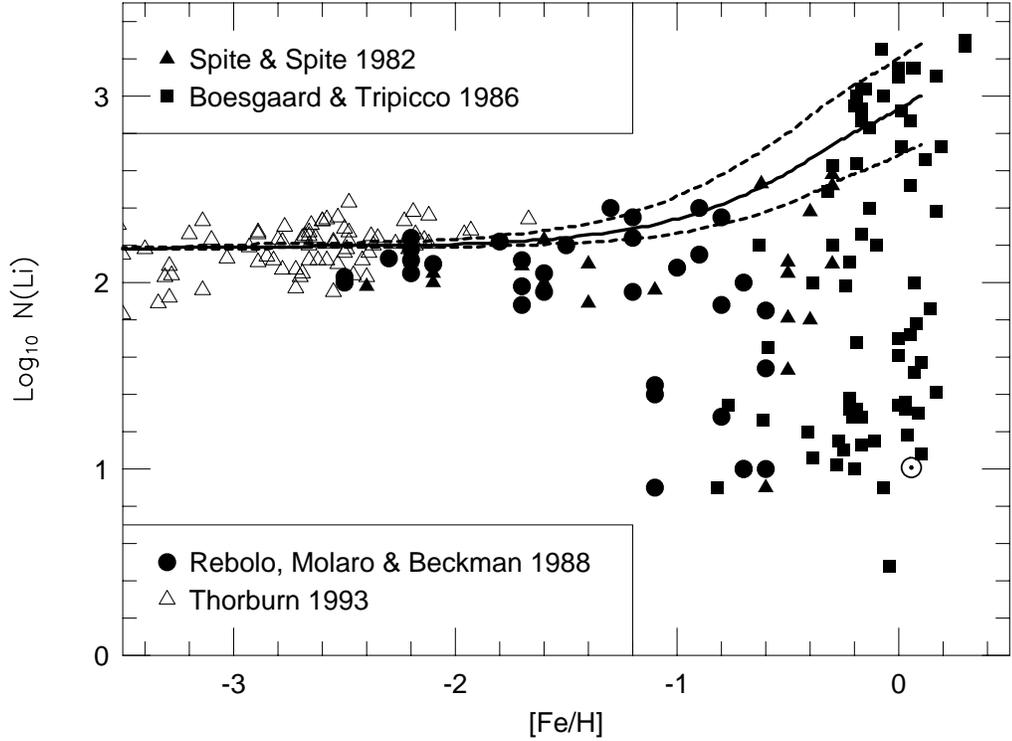

Fig. 8.— Evolution of $^7$Li relative to hydrogen as a function of the metallicity [Fe/H]. The calculated $^7$Li abundance is shown as the solid line, while variations of a factor of two in the $\nu$-process yields are shown as the dashed lines. Abundance determinations of $^7$Li in disk and halo dwarfs are shown, along with the severely depleted solar photospheric value. The Thorburn (1994) observations have been "corrected" by -0.2 dex due to differences in the stellar atmosphere models (see text). Agreement of the calculations along the Spite plateau (i.e. [Fe/H] < -1.4 dex) is due to the infall of primordial material with a homogeneous Big Bang composition. Beyond the Spite plateau, the upper envelope of the observations rises smoothly up to the maximum value found in Population I objects. The figure indicates that injections of freshly synthesized $^7$Li into the interstellar medium by the $\nu$-process can account for a a significant part of the shape and amplitude of the upper envelope, and that Type II supernova contribute about 1/2 of the solar $^7$Li abundance (also see Figure 4).

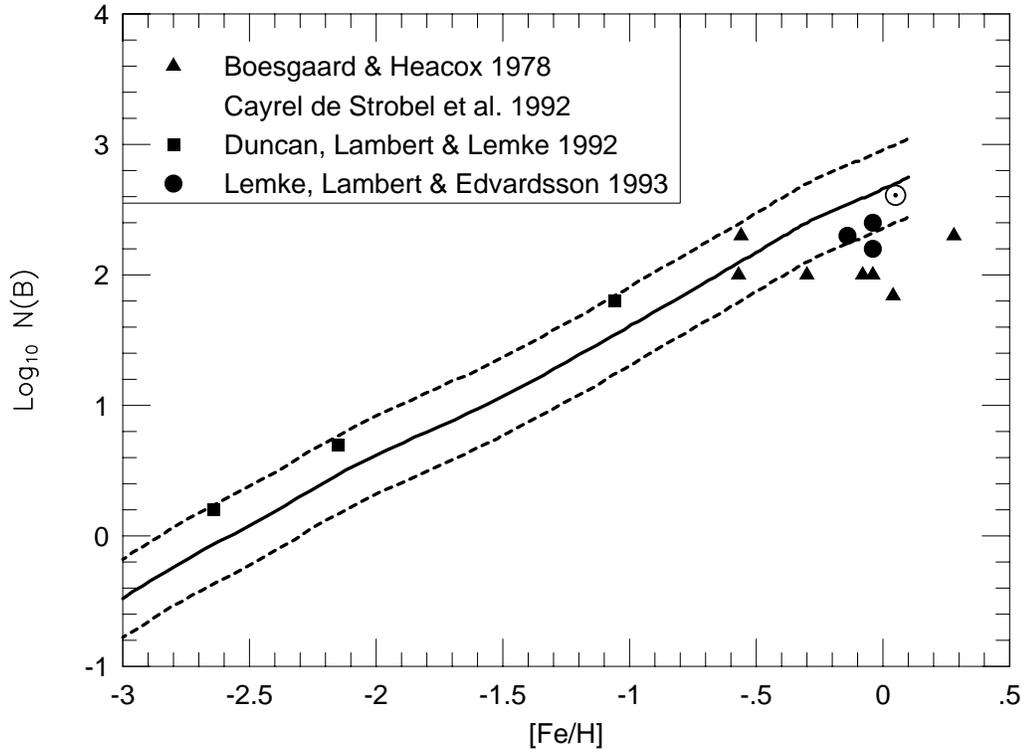

Fig. 9.— Evolution of boron relative to hydrogen as a function of the metallicity [Fe/H]. The calculated boron abundance is shown as the solid line, while the dashed lines depict factors of two variations in the boron yields. Boron abundances in stars is very difficult to measure because it is a trace element, and only detectable through its ultraviolet resonance lines. The few abundance determinations of boron in disk and halo dwarfs, and the solar value are shown in the figure (see text). Recent investigations have suggested non-LTE correction factors of about 0.3 dex for the 3 most metal-poor stars shown, along with Procyon. These corrections have been applied. Linearity of the boron abundances with [Fe/H] strongly suggests that boron is produced in lockstep along with the other metals. The fit to the observations span over 2 orders of magnitude, and includes the solar abundance (also see Figure 4).

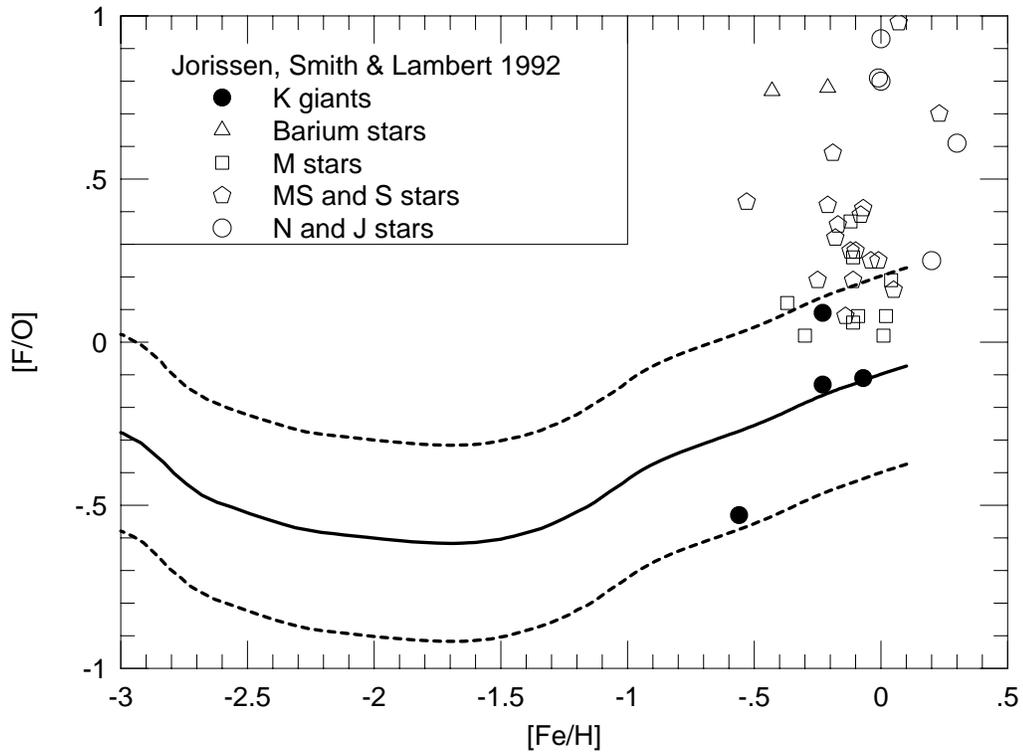

Fig. 10.— Evolution of the fluorine to oxygen [F/O] ratio as a function of the metallicity [Fe/H]. The solid line shows the calculated $^{19}$F abundance with the dashed lines indicating factors of two variation in the mass of $^{19}$F ejected by massive stars. Fluorine is compared to oxygen because they are neighbors in the periodic table, and are expected to be synthesized in close proximity to each other. Fluorine is the least abundant of all the stable $12 \leq A \leq 38$ nuclides, and very difficult to measure in stars. The stars of the Jorissen *et al.* (1992) survey included four K field giants that appear to be chemically normal, a few barium stars (which are binary systems), and an assortment of extreme carbon stars (MS, S, SC, N and J spectral types). The model fits the chemically normal K giant observations, and makes the prediction that metal-poor dwarfs will have a subsolar fluorine to oxygen ratio. The fit to the various types of extreme carbon stars is not good, nor should the model be expected to fit these spectral classes, which are dominated by the effects of stellar evolution rather than by chemical evolution.

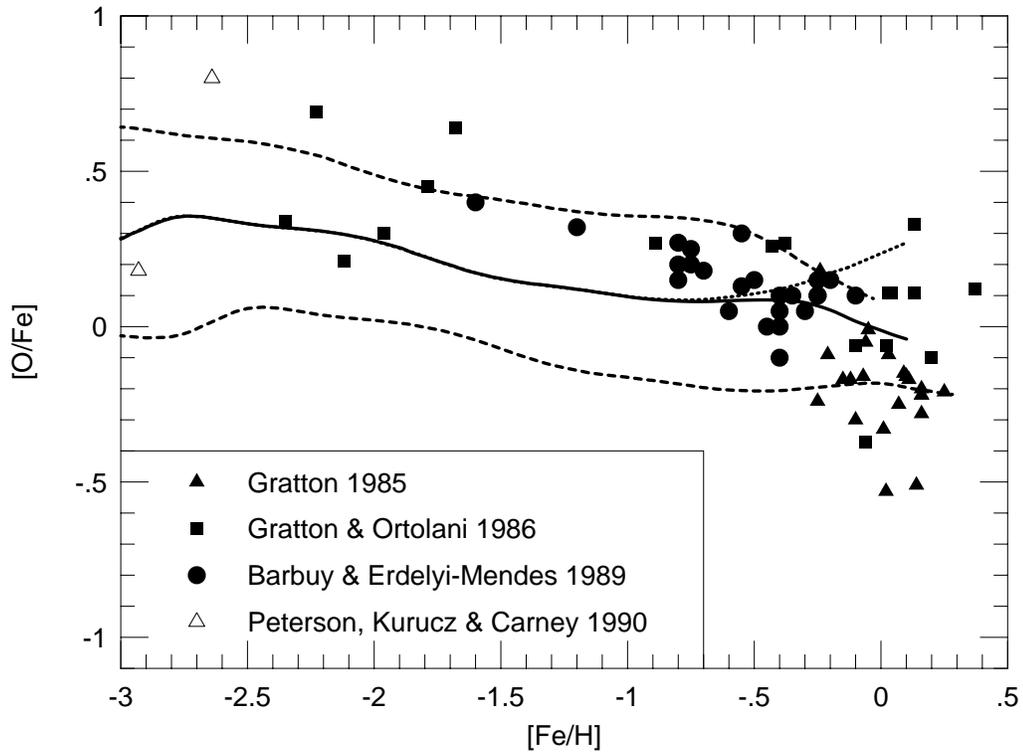

Fig. 11.— Evolution of the oxygen to iron ratio [O/Fe] as a function of [Fe/H]. The solid line shows the calculation, the dashed lines indicate factors of two variation in the iron yields from massive stars and the dotted line shows the results when Type Ia supernovae are excluded. Several sets of observations are shown in the figure and discussed in the text. The [O/Fe] is largest at very low metallicities, with the small dip at [Fe/H] $\simeq$ -3.0 dex is caused by the uncertainty in the iron yields of the M $\gtrsim$ 30 $M_\odot$ extremely metal-poor massive stars. The oxygen to iron ratio then slowly decreases due to small mass and metallicity effects. Intermediate mass stars begin to dominate contributions to the interstellar medium later on, but because they produce very little oxygen and no iron, the [O/Fe] ratio is not affected. Type Ia supernovae begin to inject large amounts of iron and negligible amounts of oxygen into the interstellar medium after $\simeq 10^9$ years, causing the downturn of the [O/Fe] ratio to its solar value. This effect is shown by comparing the solid line in the figure (which includes both Type II and Type Ia supernovae) with the dotted line (which excludes Type Ia supernovae). The best fit to the [O/Fe] observations may be a systematic reduction of the iron yields by a factor of two. The reduced iron yields are consistent with the SN 1987A observations, and the uncertainty in modeling the explosion. Factors of two increase in the iron yields are excluded.

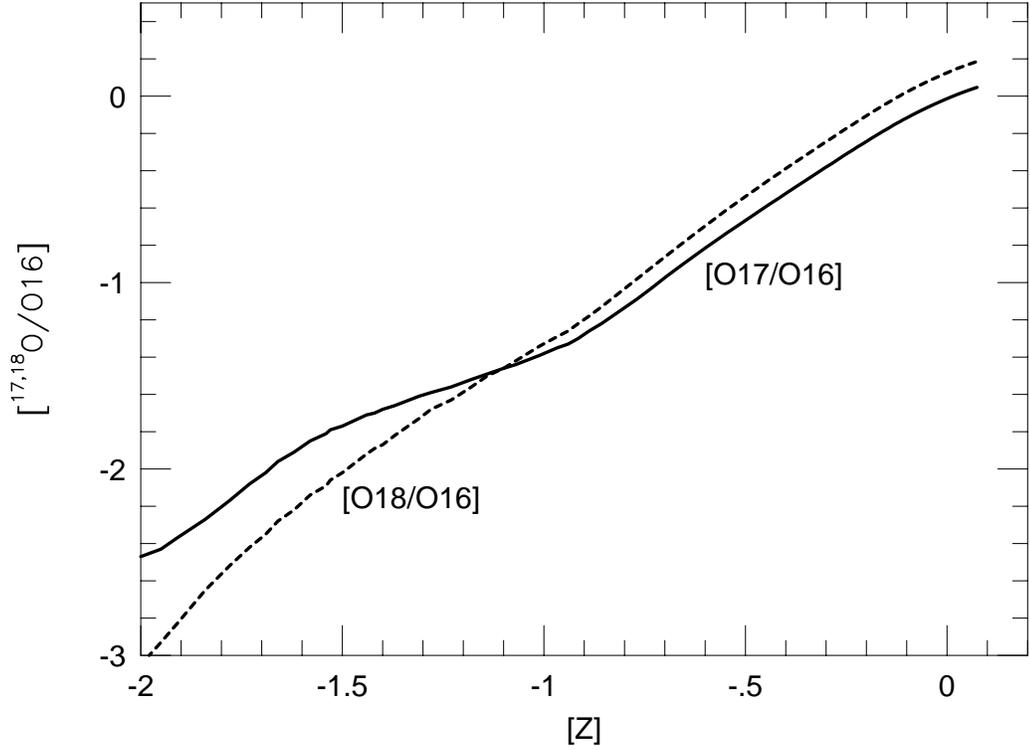

Fig. 12.— Evolution of [$^{17}$O / $^{16}$O] and [$^{18}$O / $^{16}$O] as a function of the total metallicity [Z]. Yields for $^{17}$O and $^{18}$O vary for stars of different masses and metallicities since $^{17}$O is produced primarily by CNO hydrogen burning, and $^{18}$O is generally synthesized from nitrogen reactions during helium burning. Nevertheless, these two oxygen isotopes show nearly identical linear variations over the metallicity range shown due to the direct dependence of both isotopes on the initial CNO abundances. Oxygen isotopic ratios inferred through molecular CO emission in the interstellar medium tend to be about 40% smaller than the solar ratio, while isotopic ratios observed in several types of carbon stars tend to be much larger than expected from low mass stellar evolution theories (see text).

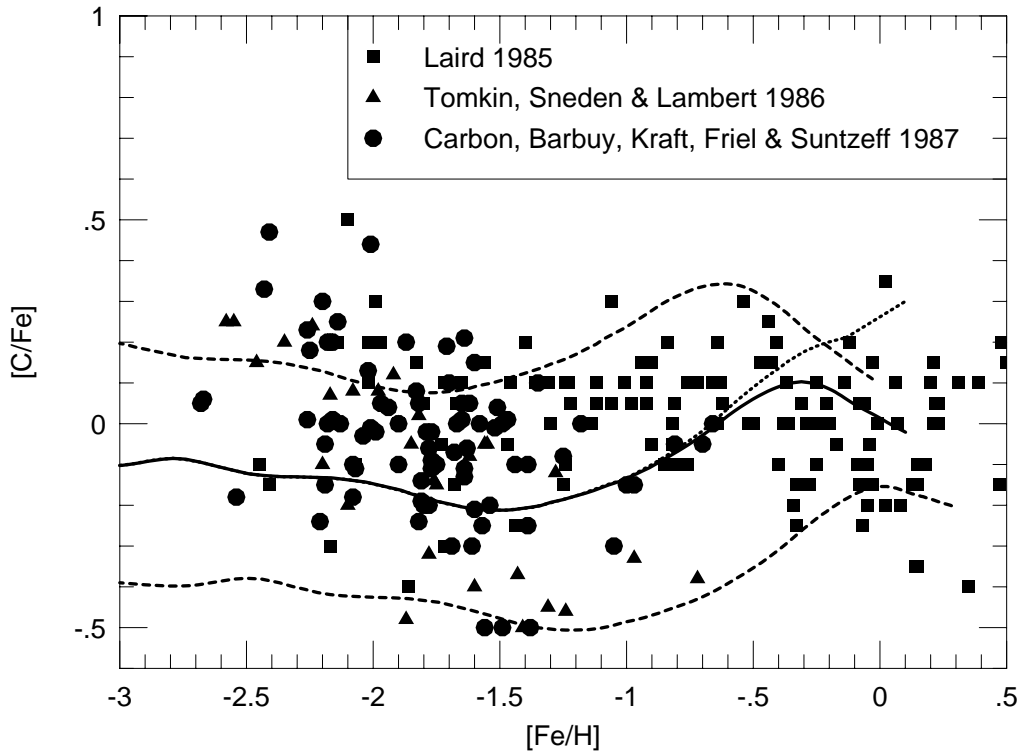

Fig. 13.— Evolution of the carbon to iron ratio [C/Fe] as a function of [Fe/H]. The solid line shows the calculation, the dashed lines indicate factors of two variation in the iron yields from massive stars and the dotted line shows the results when Type Ia supernovae are excluded. The observation of a roughly flat and solar [C/Fe] ratio is interesting because carbon and iron are synthesized by very different processes at different stages in the Galaxy's evolution. Massive star contributions to carbon and iron are sufficient to explain the metal-poor halo dwarf observations. Two competing sources come into play at [Fe/H] $\simeq$ -0.8 dex. Intermediate and low mass stars begin depositing large amounts of carbon but no iron, while Type Ia supernovae start injecting significant amounts of iron but no carbon. This effect is shown by comparing the solid line in the figure (which includes both Type II and Type Ia supernovae) with the dotted line (which excludes Type Ia supernovae). The interplay between these two sources, such that [C/Fe] $\simeq$ 0 dex, is a tight constraint for chemical evolution models. The calculation shown has intermediate mass stars raising the [C/Fe] ratio a little before Type Ia supernovae depress it, but the deflections are well within the star-to-star scatter of the observations. Explanation of the solar system abundance relies the nucleosynthesis of intermediate and low mass stars since carbon is underproduced in the Z $\geq$ 0.1 Z$_\odot$ massive star models by about a factor of 3.

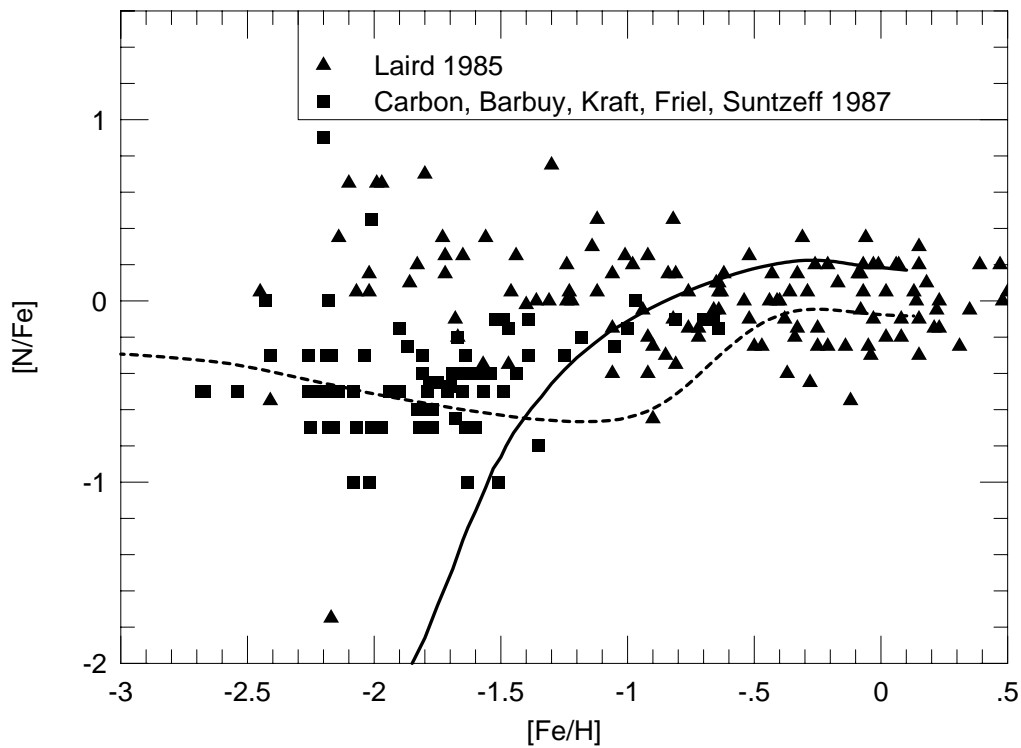

Fig. 14.— Evolution of the nitrogen to iron ratio [N/Fe] as a function of [Fe/H]. The solid line shows the calculation, and the dashed line shows a calculation with a large amount of convective overshoot. The apparent agreement between the two surveys shown surveys belie the great difficulty in determining [N/Fe] ratios in dwarfs (see text), but the main conclusion of these two surveys is that nitrogen has a strong primary component. The solid line shows that no primary nitrogen is produced in the standard set of massive stars. Explanation of the solar system abundance relies on the nucleosynthesis in intermediate and low mass stars since nitrogen is underproduced in the $Z \geq 0.1\ Z_\odot$ massive star models by about a factor of 3.

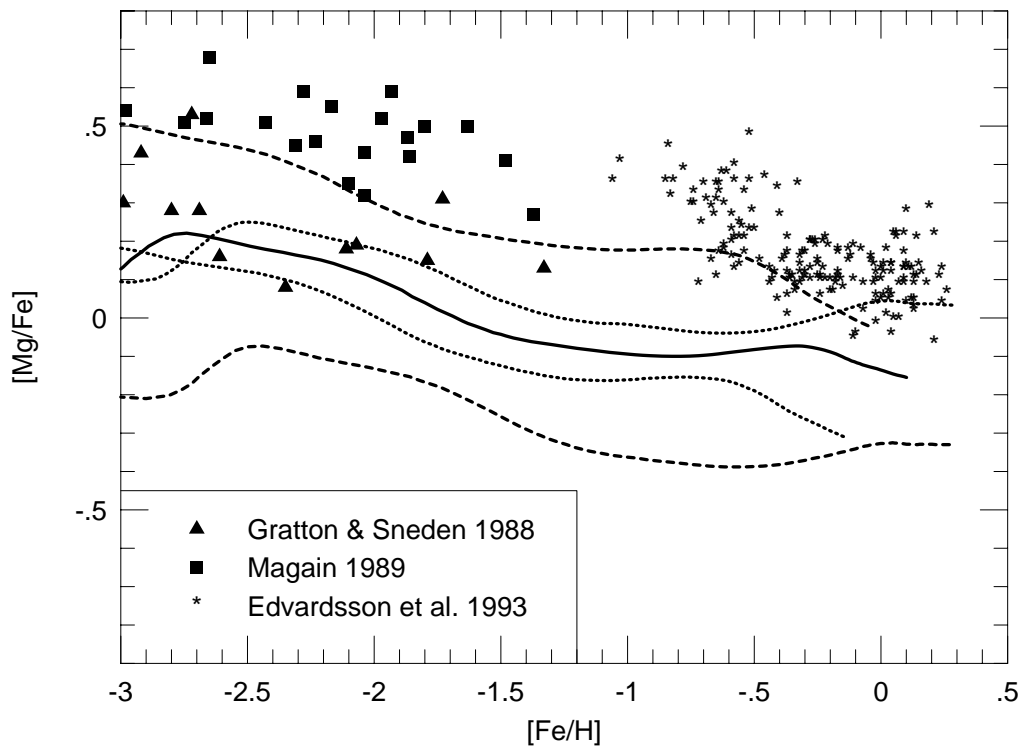

Fig. 15.— Evolution of the magnesium to iron ratio [Mg/Fe] as a function of [Fe/H]. The solid line shows the calculation, dashed lines shows variations of two in the iron yields from massive stars and the dotted lines show variations in the exponent of the initial mass function. The observations shown in the figure include dwarfs and field giants (see text). Magnesium production by massive stars can account for the halo dwarf observations. Intermediate or low mass star produce no magnesium or iron in the calculations shown, so the [Mg/Fe] ratio does not change when they dominant contributions back into the interstellar medium. The injection of iron from Type Ia supernovae becomes large enough to drive the [Mg/Fe] ratio down to the solar value. The best fit to the [Mg/Fe] observations may be a systematic reduction of the iron yields from massive stars by a factor of two. Factors of two increase in the iron yields are excluded. Even with a reduced iron yield the model tends to underestimate the [Mg/Fe] ratio in the disk, which may be indicative of a small magnesium contribution from intermediate and low mass stars, or a slightly enhanced magnesium yield from Type Ia supernovae. The figure shows the general property of the $\alpha$-chain elements that the sensitivity of the evolutions to variations in the exponent of the initial mass function (dotted lines; see text) much is less than the uncertainties associated with modeling the explosion mechanism.

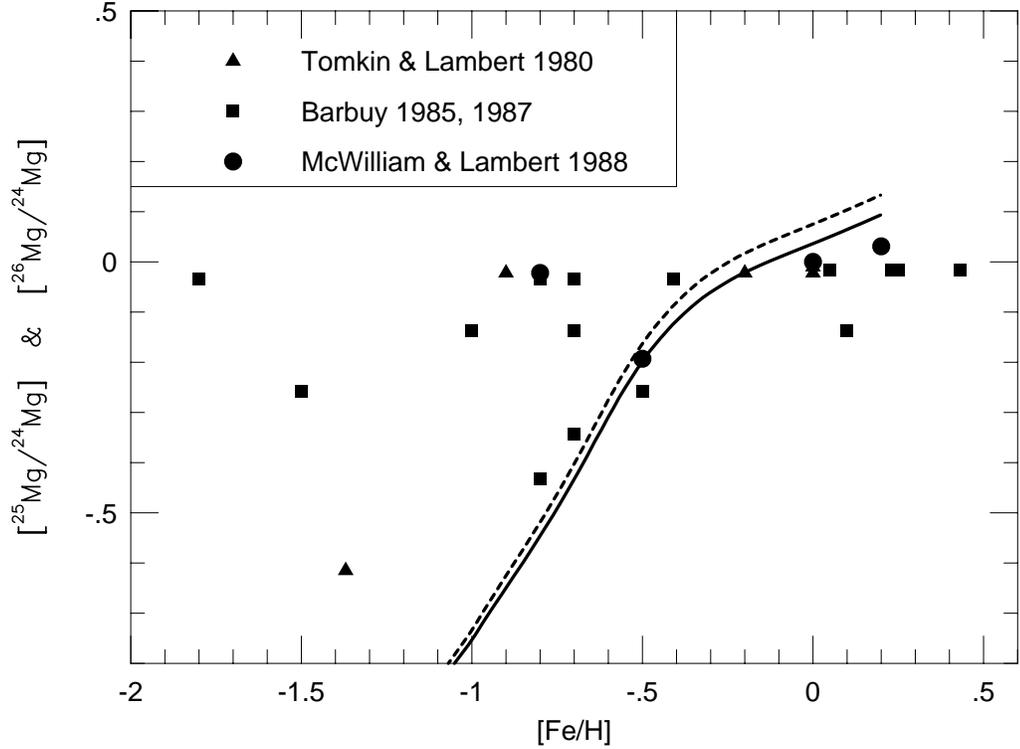

Fig. 16.— Evolution of the magnesium isotopes as a function of the metallicity [Fe/H]. The solid curve shows the [$^{25}$Mg/$^{24}$Mg] ratio, and the dashed curve shows the [$^{26}$Mg/$^{24}$Mg] ratio. Since almost all of the observations have some difficulty splitting the differences between the neutron rich species, only the observed [$^{25}$Mg/$^{24}$Mg] ratios are shown. Abundances of the magnesium isotopes are generally determined by the initial metallicity of the star, but weak interactions decrease this dependence. For [Fe/H] $\gtrsim$ -1.0 dex, the calculations agree fairly well with the magnesium isotope ratios found in disk dwarfs. The observed ratios in halo dwarfs are less cohesive, with some studies suggesting solar ratios while others indicate factors of 3 suppression. At metallicities of [Fe/H] $\simeq$ -1.0 dex, the model gives isotopic ratios that are about a factor 2 smaller than any of the observations would suggest. This may indicate that some contributions from another source is necessary.

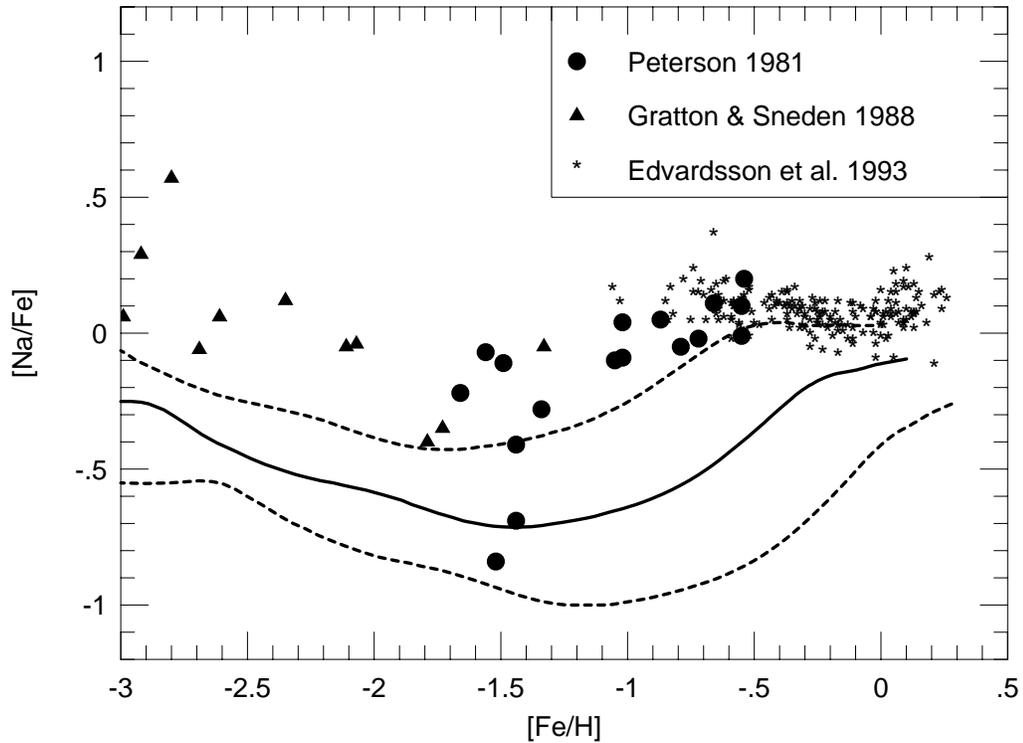

Fig. 17.— Evolution of the sodium to iron ratio [Na/Fe] as a function of [Fe/H]. The solid line shows the calculation, and the dashed lines show variations of two in the iron yields from massive stars. As discussed in the text, the Gratton & Sneden (1988) sodium abundances appear to systematically too large by about 0.3 dex. "Corrected" values have been used for the 3 most metal-rich stars of their survey. If this "corrected" trend continues, then the [Na/Fe] ratios shown in the figures for the most metal-poor stars should be lowered, and the observations would come into better agreement with the theoretical models. For metallicities larger than -1.0 dex the figure indicates that there might be a need for small additional sources of sodium.

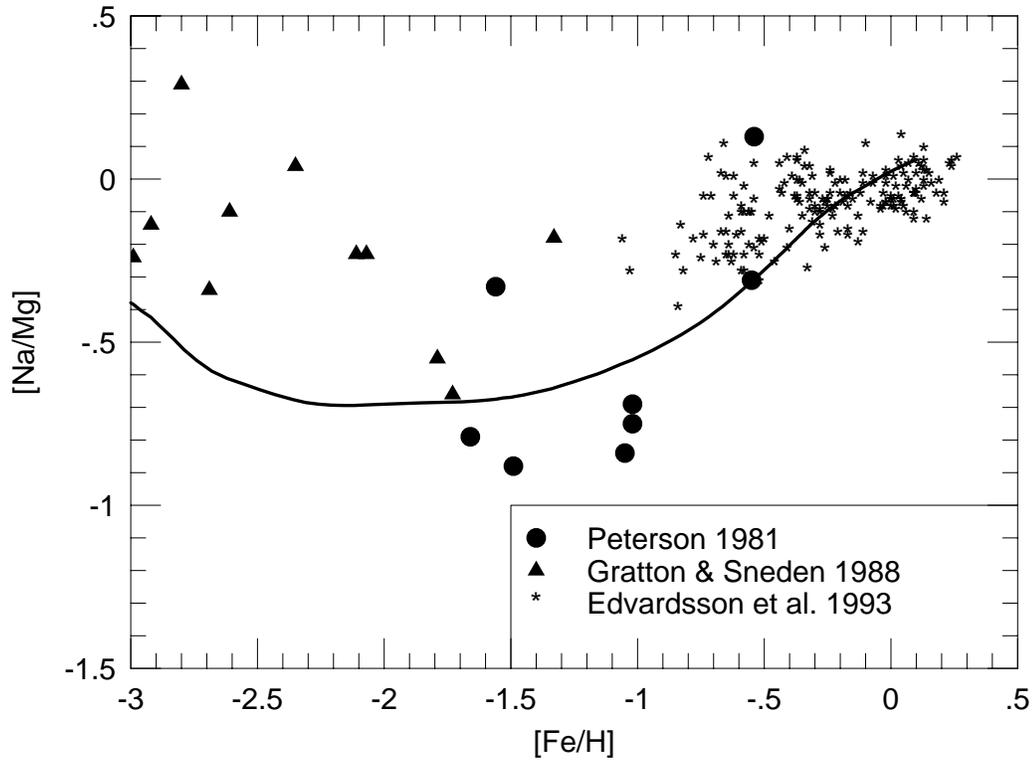

Fig. 18.— Evolution of the sodium to magnesium ratio [Na/Mg] as a function of [Fe/H]. The solid line shows the calculation, and the observations are discussed in the text. "Corrected" values have been used for the 3 most metal-rich stars of the Gratton & Sneden (1988) survey (see text). If this "corrected" trend continues, then the [Na/Mg] ratios shown in the figures for the most metal-poor stars should be lowered, bringing the observations into better agreement with the theoretical models. The calculated [Na/Mg] evolution is in excellent agreement with the disk dwarf observations, where a clear odd-even effect is present.

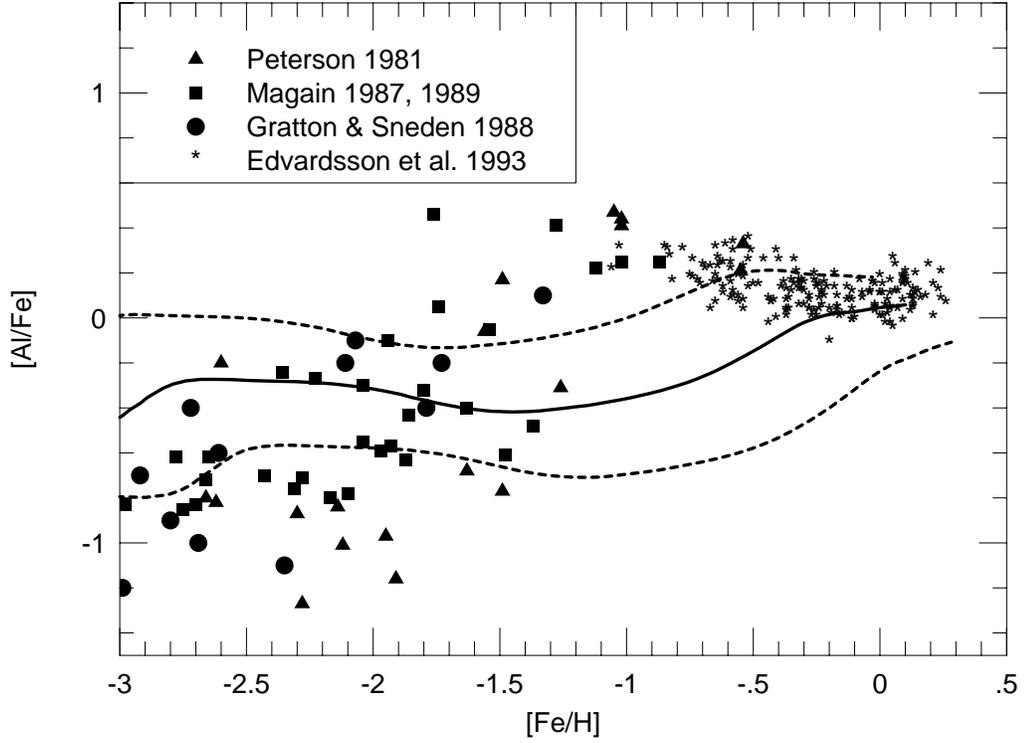

Fig. 19.— Evolution of the aluminum to iron ratio [Al/Fe] as a function of [Fe/H]. The solid line shows the calculation, and the dashed lines show variations of two in the iron yields from massive stars. The observations shown in the figure, and discussed in the text, suggests that the [Al/Fe] ratio is solar with a mild slope in Population I stars, followed by a precipitous drop in passing to the halo stars. The calculated evolution of [Al/Fe] is slightly larger than the halo star observations, does not rise as fast through the [Fe/H] $\simeq$ -1.0 dex halo-disk transition point, and is slightly lower than the disk star observations. For metallicities larger than -1.0 dex the figure indicates that there might be a need for small additional sources of aluminum.

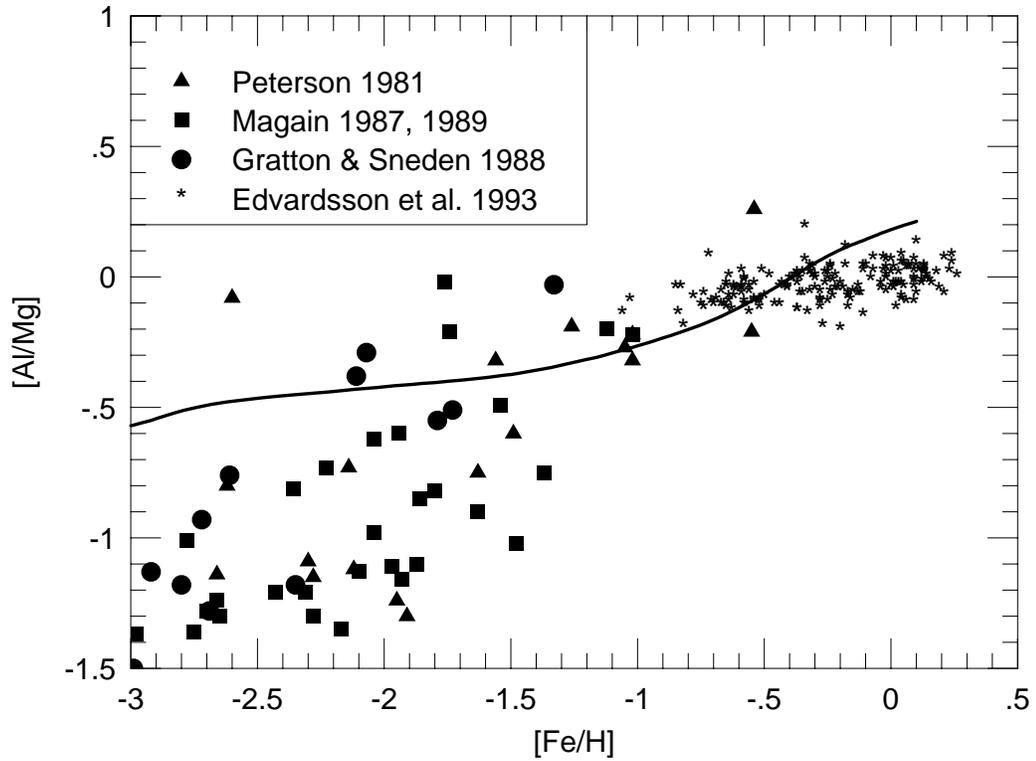

Fig. 20.— Evolution of the aluminum to magnesium ratio [Al/Mg] as a function of [Fe/H]. The solid line shows the calculation and the observations are discussed in detail the text. The observed [Al/Mg] ratio shows a distinct odd-even effect effect in the halo, and the ratio is more subsolar than that due to enhancement of [Mg/Fe] at these metallicities. A much milder (if at all) odd-even effect is present in disk dwarfs. The calculated evolution shows the same difficulties as the [Al/Fe] ratio of Figure 19.

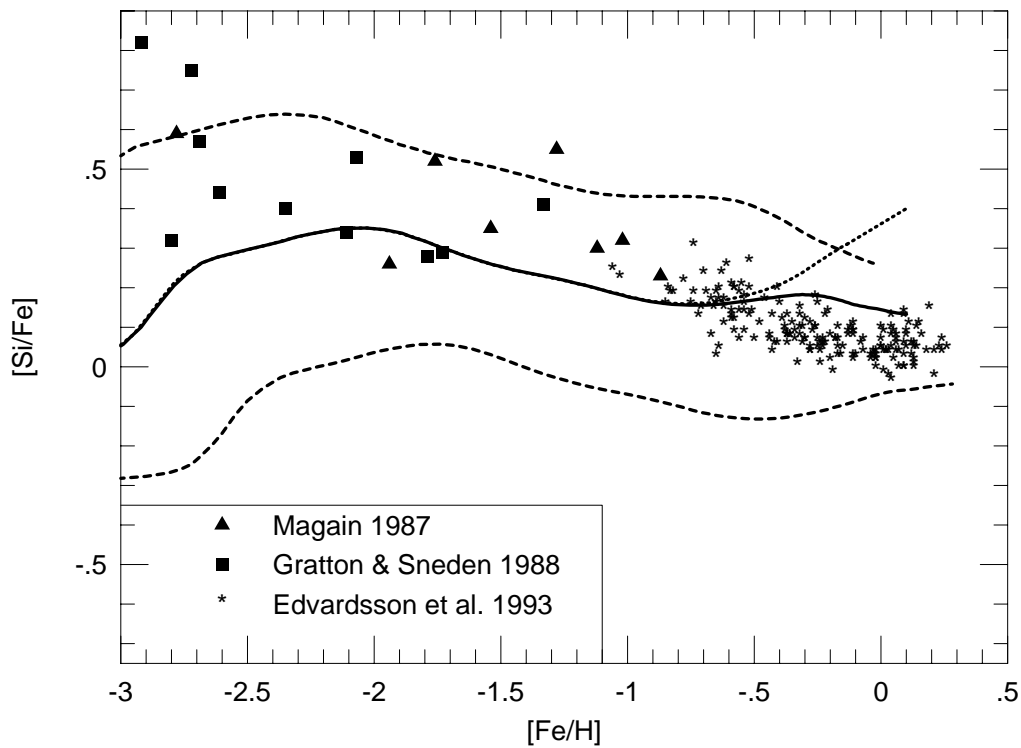

Fig. 21.— Evolution of the silicon to iron ratio [Si/Fe] as a function of [Fe/H]. The solid line shows the calculation; the dashed lines indicate factors of two variation in the iron yields from massive stars and the dotted line shows the results when Type Ia supernovae are excluded. The abundance determinations shown in the figure employ both dwarfs and field giants, and are primarily determined from neutral line transitions. Many of the observed trends of silicon are typical of the $\alpha$-chain nuclei – a factor of about three enhancement over the solar in the halo, the slow decline due to mass and metallicity effects, and the drop down to the solar ratio due to the injection of iron from Type Ia supernovae. This effect is shown by comparing the solid line in the figure (which includes both Type II and Type Ia supernovae) with the dotted line (which excludes Type Ia supernovae). The calculated evolution of the [Si/Fe] ratio is in overall agreement with the observations of this $\alpha$ element. The departure from classical $\alpha$ element abundances at [Fe/H] $\simeq$ -2.5 dex is primarily due to uncertainty in the M $\gtrsim$ 30 M$_\odot$ extremely metal-poor massive star models. Production of all the stable silicon isotopes by the massive star models is sufficient to explain the solar abundances (see Figure 4).

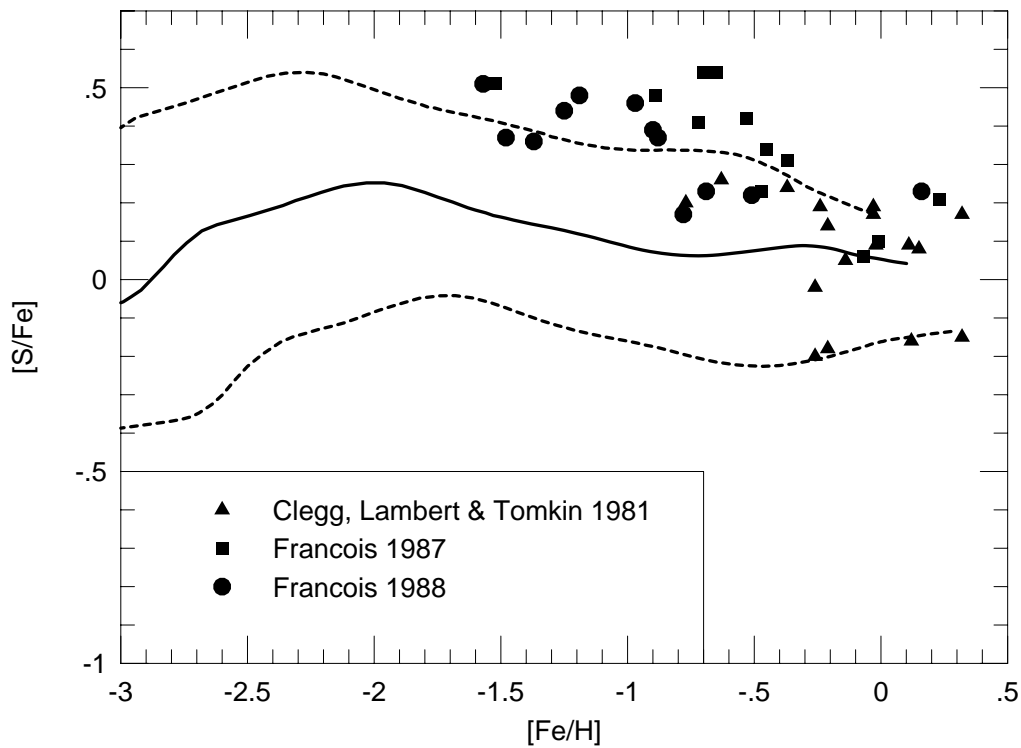

Fig. 22.— Evolution of the sulfur to iron ratio [S/Fe] as a function of [Fe/H]. The solid line shows the calculation; the dashed lines indicate factors of two variation in the iron yields from massive stars. Sulfur has seldom been studied primarily because potentially useful lines in the near-infrared portion of the spectrum are weak and often blended, but it appears to have many of the characteristics of an $\alpha$-chain element. Following a suggestion by Lambert (1989), the sulfur abundances of Francois (1987; 1988) have been "corrected" by -0.2 dex (see text). The calculated evolution of the [S/Fe] ratio is generally in agreement with the observations, but smaller at low values of [Fe/H] by about a factor of two . The depression to a solar value at [Fe/H] $\simeq$ -2.5 dex is primarily due to uncertainties in the M $\gtrsim$ 30 $M_\odot$ extremely metal-poor massive star models. Intermediate and low mass stars contribute no sulfur or iron in the calculations shown. The sulfur produced by the Z $\geq$ 0.1 $Z_\odot$ massive star models is balanced by the iron produced from Type Ia supernovae, which keeps the [S/Fe] ratio relatively flat in Population I stars. Production of all the stable sulfur isotopes by the massive star models is sufficient to explain the solar abundances (see Figure 4).

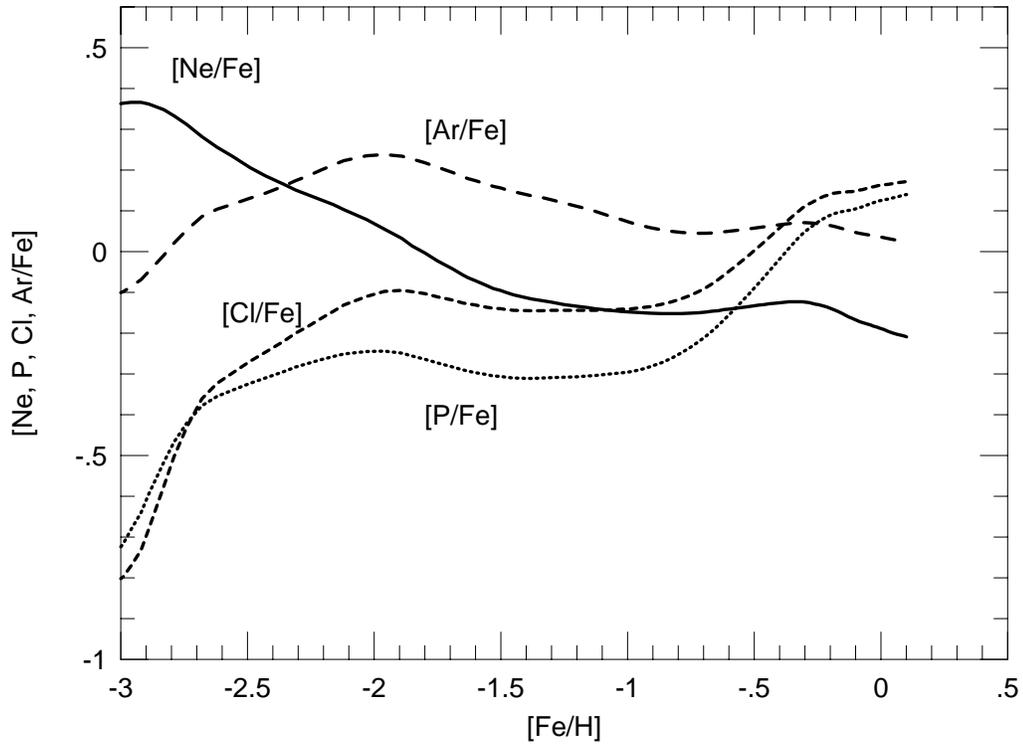

Fig. 23.— Predicted neon, phosphorus, chlorine, and argon abundances relative to iron as a function of [Fe/H]. This group of elements apparently have no determinations of the *ab initio* abundances in either dwarf or giant stars, because neon and argon are noble gases that only have optical transitions from very high excitation levels, and molecular forms of chlorine and phosphorus show only very weak lines in synthetic stellar spectra models. Both the odd-Z elements, phosphorus and chorine, show subsolar abundances at low metallicities that then slowly climb up to their solar values Neon and argon, being $\alpha$ elements, have almost the complete opposite behavior, showing supersolar values in the halo that slowly decrease to solar ratios in the disk.

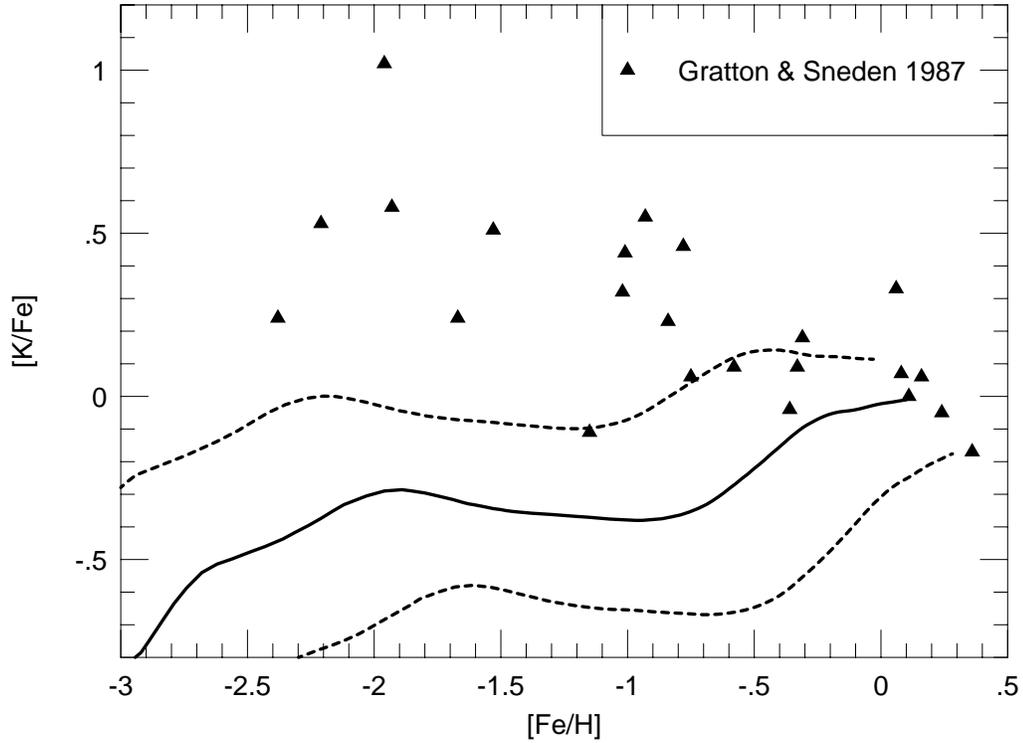

Fig. 24.— Evolution of the potassium to iron ratio [K/Fe] as a function of [Fe/H]. The solid line shows the calculation; the dashed lines indicate factors of two variation in the iron yields from massive stars. The calculated [K/Fe] evolution is in agreement with the observations of dwarfs and field giants for [Fe/H] $\gtrsim$ -0.6 dex. At smaller metallicities, the calculations show that one might expect [K/Fe] < 0.0 dex, in sharp contrast to the observations. Gratton & Sneden (1987) commented that the K I resonance lines are often heavily blended with extremely strong lines of atmospheric molecular oxygen. In addition, the low excitation energies of the K I lines may be strongly susceptible to overionization (non-LTE) or strong hyperfine structure effects. Production of all the stable potassium isotopes by the massive star models is sufficient to explain the solar abundances (see Figure 4).

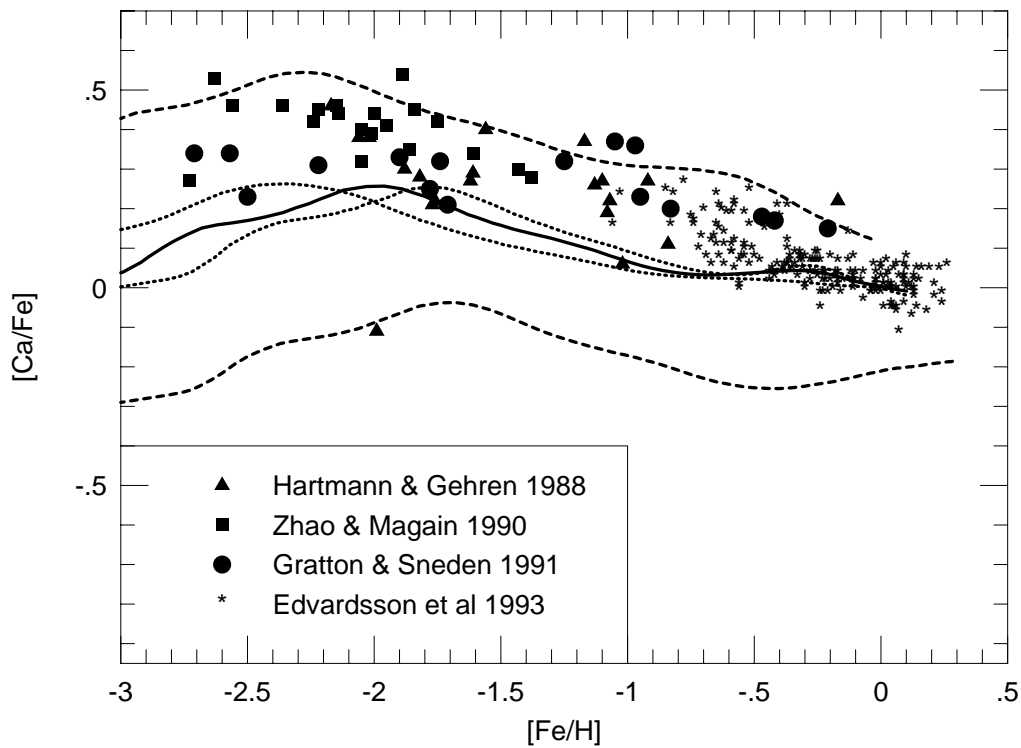

Fig. 25.— Evolution of the calcium to iron ratio [Ca/Fe] as a function of [Fe/H]. The solid line shows the calculation, the dashed lines indicate factors of two variation in the iron yields from massive stars and the dotted lines show variations in the efficiency of star formation. Many of the observed trends of calcium are typical of the $\alpha$-chain nuclei – a factor of about three enhancement over the solar in the halo, the slow decline due to mass and metallicity effects followed by a drop down to the solar ratio. The star HD 89499 from the Hartmann & Gehren (1988) survey, which has [Ca/Fe]=-0.11 dex and [Fe/H]=-2.0 dex, also posses peculiar abundances of magnesium, aluminum and scandium. The calculated evolution of the [Ca/Fe] ratio is in overall agreement with the observations of this $\alpha$ element. The departure from classical $\alpha$ element abundances at [Fe/H] $\simeq$ -2.5 dex is primarily due to uncertainties in the extremely low metallicity M $\gtrsim$ 30 M$_\odot$ massive star models. The calcium produced by the Z $\geq$ 0.1 Z$_\odot$ massive star models is nearly balanced with the iron produced from Type Ia supernovae, which keeps the [Ca/Fe] ratio relatively flat in Population I stars. The dotted lines show the evolutions for significant variations in the efficiency of star formation ($\nu$=0.8 and 5.8; see Table 1), and indicate the general property that the $\alpha$-chain element evolutions are very robust with respect to modifications in this free parameter.

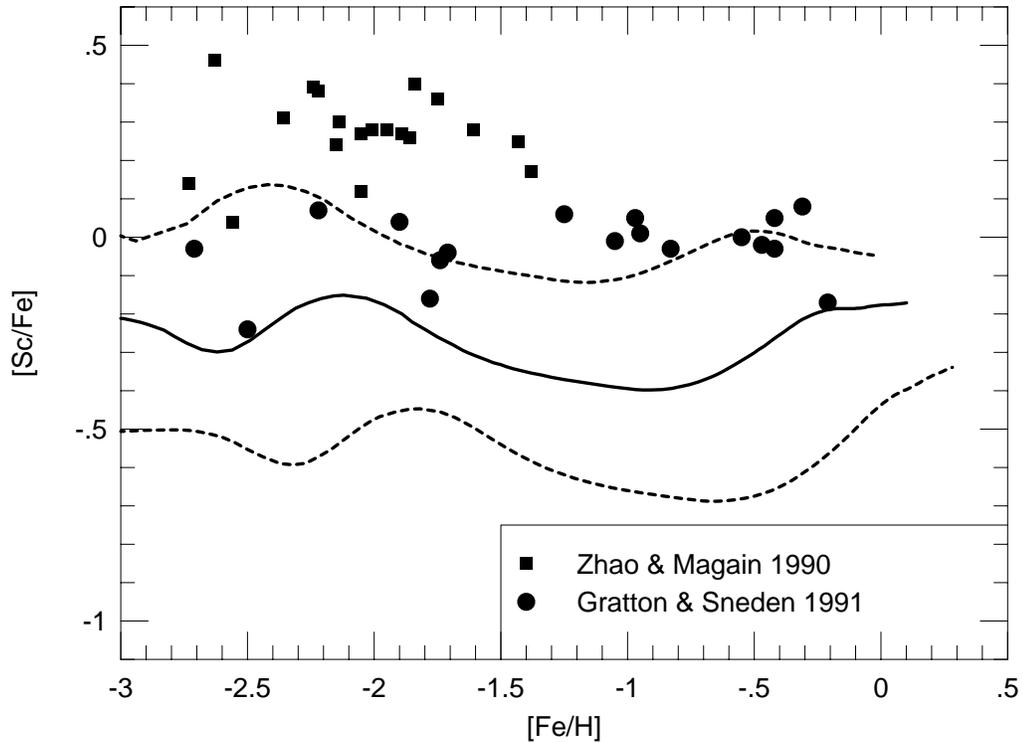

Fig. 26.— Evolution of the scandium to iron ratio [Sc/Fe] as a function of [Fe/H]. The solid line shows the calculation; the dashed lines indicate factors of two variation in the iron yields from massive stars. Scandium has only one stable isotope, and since it is an odd-Z element, it is important to include hyperfine structure effects during reduction of the spectroscopic data. Of the two surveys shown, the Gratton & Sneden (1991) values are probably the more accurate (see text). The calculated evolution is in good agreement the observations of a flat [Sc/Fe] ratio, although perhaps systematically smaller than the observations by about a factor of 1.5 at metallicities characteristic of the halo population. However, the solar abundance of $^{45}$Sc is also too small by the about same factor (see Figures 4 and 5). Intermediate and low mass stars contribute no scandium or iron in the calculations shown. The [Sc/Fe] ratio stays relatively flat at late times ([Fe/H] $\geq$ -1.0 dex) due to the balance between the scandium produced by the Z $\geq$ 0.1 $Z_\odot$ Type II supernovae models and the iron produced from Type Ia events.

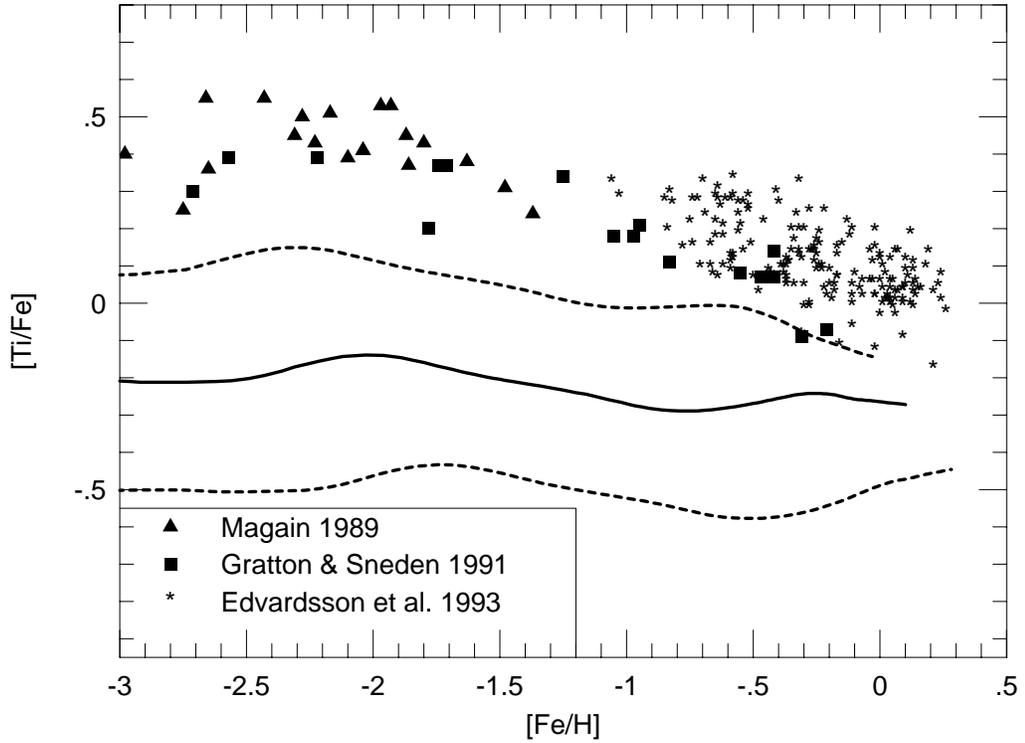

Fig. 27.— Evolution of the titanium to iron ratio [Ti/Fe] as a function of [Fe/H]. The solid line shows the calculation, and the dashed lines indicate factors of two variation in the iron yields from massive stars. The titanium abundance determinations shown in the figure employ both dwarfs and giants, and are primarily found from neutral line transitions. Observationally, titanium displays many of the $\alpha$ element hallmarks, while the calculations suggest it should scale with iron, especially since titanium is produced in the deepest layers of massive stars. In addition, the solar abundance of $^{48}$Ti, the most abundant titanium isotope, is deficient by a factor of 2 (see Figure 4). However, both the $^{48}$Ti yield and the ratio [Ti/Fe] are sensitive to the parameters of the explosion and the amount of material that falls back onto the neutron star.

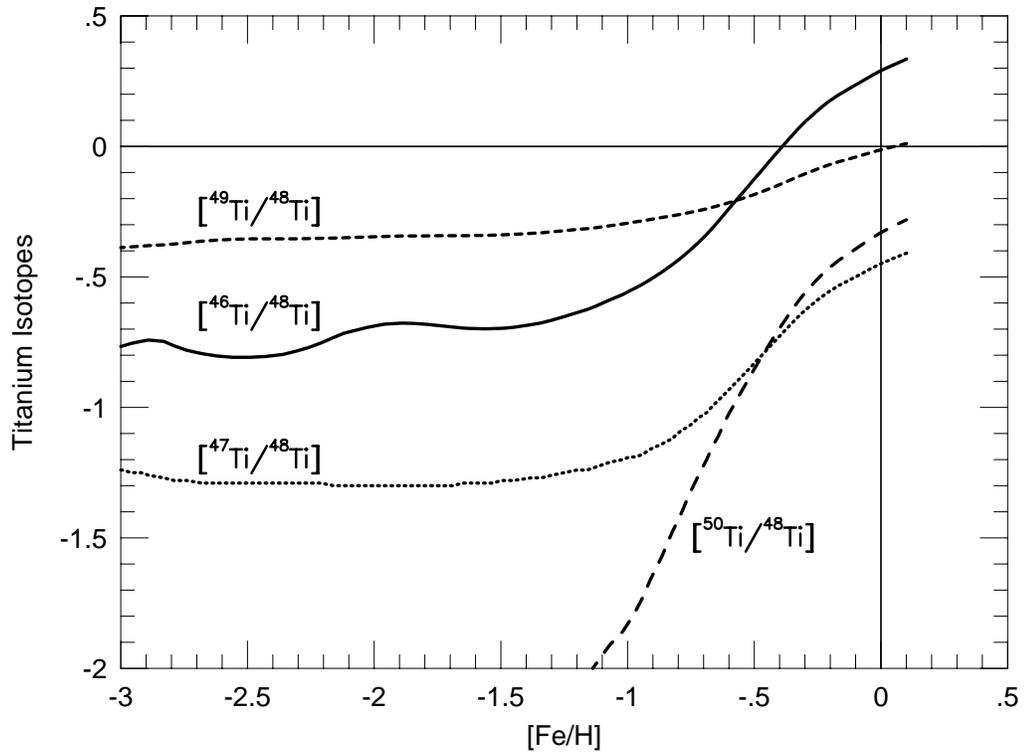

Fig. 28.— Evolution of the titanium isotopes as a function of [Fe/H]. The isotope $^{46}$Ti is produced in sufficient quantity by massive stars to explain the solar abundance, while $^{47}$Ti is underproduced by about a factor of 5.

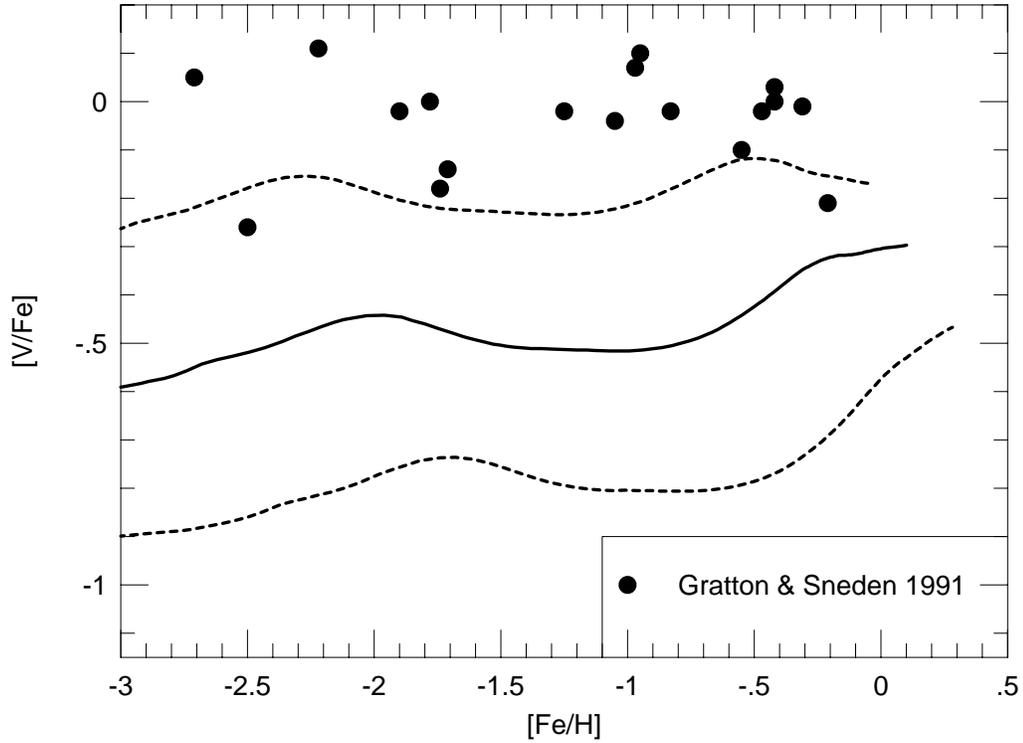

Fig. 29.— Evolution of the vanadium to iron ratio [V/Fe] as a function of [Fe/H]. Again the solid line shows the calculation; dashed lines indicate factors of two variation in the iron yields from massive stars. Vanadium is the least abundant of the iron group elements. The calculated evolution is in good agreement with the observations of a flat [V/Fe] ratio, but is systematically smaller than the observations by about a factor of three. However, elemental vanadium is dominated by the isotope $^{51}$V, whose solar abundance (Figures 4 and 5) is underproduced by slightly more than a factor of two. The probable origins of the isotope $^{51}$V are discussed in §3.1. Intermediate and low mass stars and Type Ia supernova contribute no vanadium or iron in the calculation shown. The [V/Fe] ratio is kept relatively flat in at late times due to the balance between the vanadium produced by the $Z \geq 0.1\ Z_\odot$ massive star models and the iron produced by Type Ia supernovae.

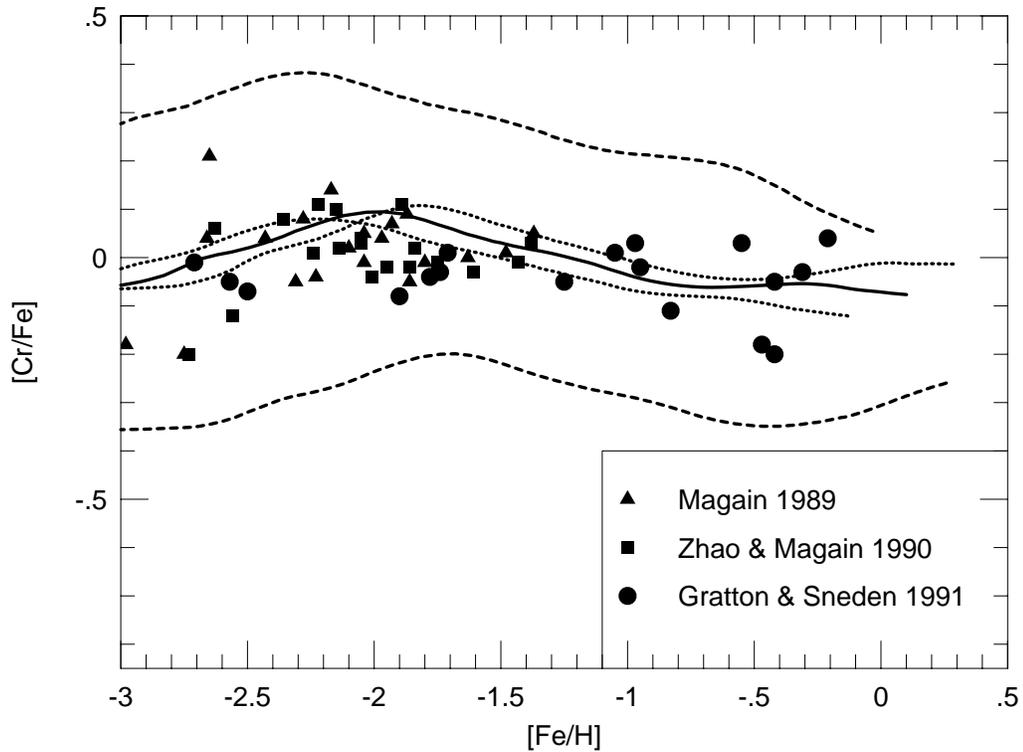

Fig. 30.— Evolution of the chromium to iron ratio [Cr/Fe] as a function of [Fe/H]. The solid line shows the calculation, the dashed lines indicate factors of two variation in the iron yields from massive stars and the dotted lines show variations in the exponent of the initial mass function. It is gratifying that the calculated evolution of this abundant iron group nucleide is in excellent agreement with observations. The chromium produced by the $Z \geq 0.1\ Z_\odot$ massive star models is balanced by the iron produced from Type Ia supernovae, which keeps the [Cr/Fe] ratio relatively flat in Population I stars. Inclusion of the nucleosynthesis from Type Ia supernovae also improves the fit to the solar abundances of the isotopes $^{50-53}$Cr (see Figures 4 and 5). The dotted line which lies below the standard calculation (see Table 1) for most of the [Fe/H] evolution corresponds to an initial mass function exponent of -1.61, while the dotted line that has an initial mass funtion exponent of -1.01 lies mostly above the standard calculation. The figure shows the general property that evolution of the iron peak elements results are weak functions of the initial mass function exponent.

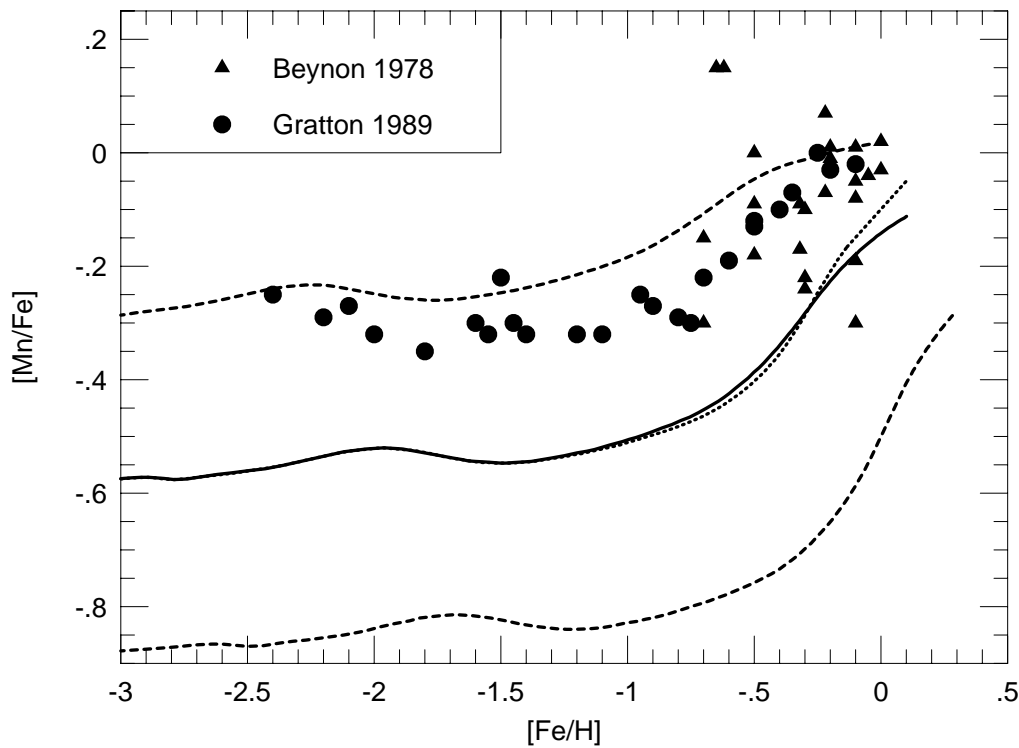

Fig. 31.— Evolution of the manganese to iron ratio [Mn/Fe] as a function of [Fe/H]. The solid line shows the calculation; dashed lines indicate factors of two variation in the iron yields from massive stars and the dotted line shows the results when Type Ia supernovae are excluded. The calculated [Mn/Fe] evolution is in good agreement with the shape of the observations, but is deficient at low metallicities by roughly a factor of two which may be due to the uncertainty in the extremely metal-poor massive star models. Intermediate and low mass stars contribute no manganese or iron in the calculations shown. The solar metallicity exploded massive star models have about a factor of 5 larger manganese yield than the models with smaller initial metallicities, which overwhelms the iron contributions from Type Ia supernovae and causes the increase in the computed [Mn/Fe] ratio for [Fe/H] > -1.0 dex. This effect is shown by comparing the solid line in the figure (which includes both Type II and Type Ia supernovae) with the dotted line (which excludes Type Ia supernovae). Inclusion of nucleosynthesis from Type Ia supernovae improves the fit to the solar abundances of magnesium (see Figures 4 and 5).

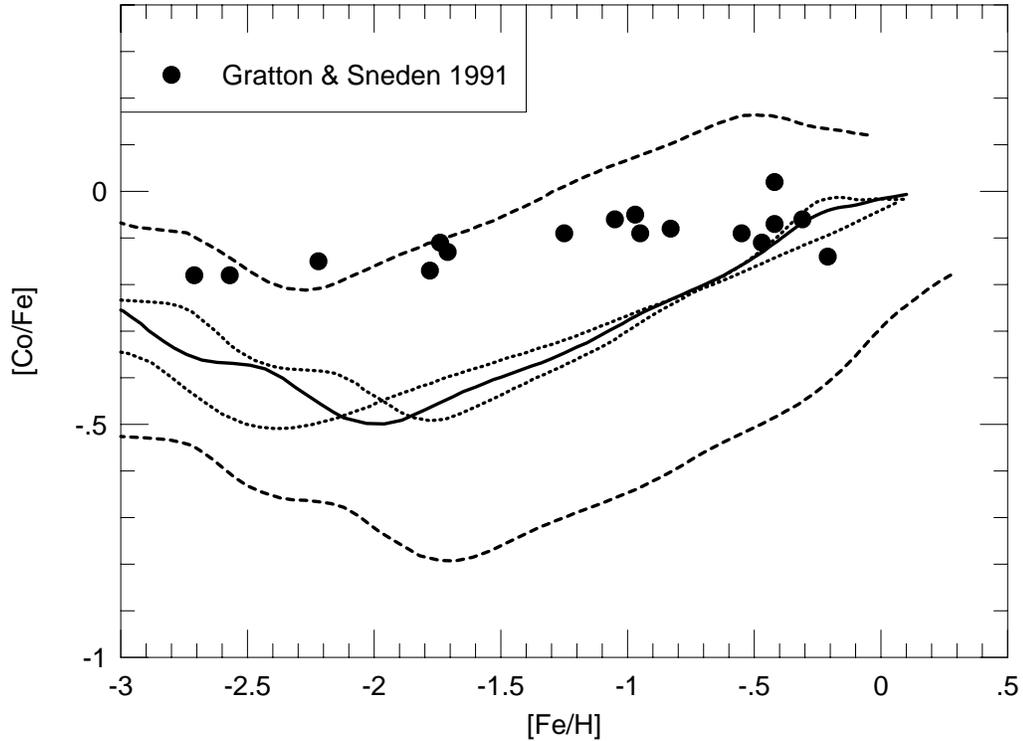

Fig. 32.— Evolution of the cobalt to iron ratio [Co/Fe] as a function of [Fe/H]. The solid line shows the standard calculation, the dashed lines indicate variations of two in the iron yields from massive stars and the dotted lines show variations in the efficiency of star formation. The calculated evolution is in reasonable agreement with the observations of a relatively flat [Co/Fe] ratio, although the computed history appears to be smaller than the observations by about a factor of two at particular low metallicities. The undulations of the curve shown in is due to mass and metallicity effects in the exploded massive star models. Intermediate and low mass stars contribute no cobalt or iron in the calculations, and the cobalt produced by the $Z \geq 0.1\,Z_\odot$ massive star models is balanced by the iron produced from Type Ia supernovae, which keeps the [Co/Fe] ratio relatively flat in Population I stars. Production of cobalt by massive stars is sufficient to explain the solar abundance (see Figure 4). The dotted lines show the evolutions for significant variations in the efficiency of star formation ($\nu$=0.8 and 5.8; see Table 1), and indicate the general property that the $\alpha$-chain element evolutions are very robust with respect to modifications in this free parameter.

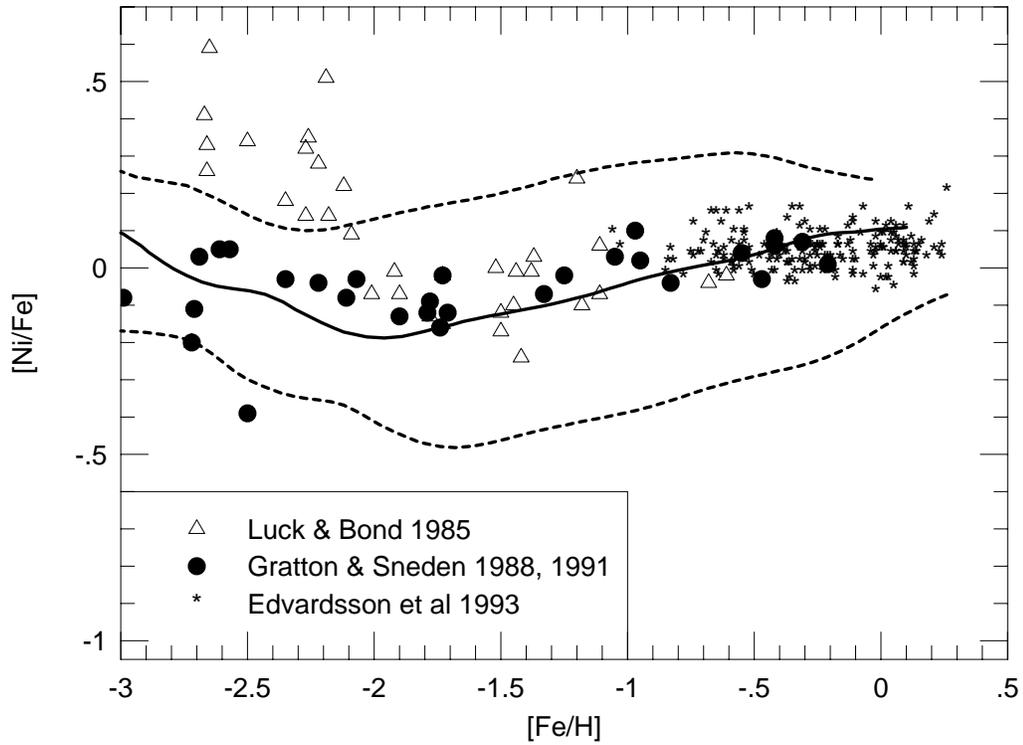

Fig. 33.— Evolution of the nickel to iron ratio [Ni/Fe] as a function of [Fe/H]. The solid line shows the calculation, and the dashed lines indicate factors of two variation in the iron yields from massive stars. All the lines of Ni I longward of 3831 Å are very weak in most metal-poor stars, and the distinct rise in [Ni/Fe] at low metallicities found in the Luck & Bond (1983, 1985) survey has been suggested to be caused by a systematic overestimate of the equivalent widths below the detection limit (Peterson & Carney 1989; Gratton & Sneden 1991). It is gratifying that the calculated evolution of this abundant iron group nucleide is in good agreement with observations.

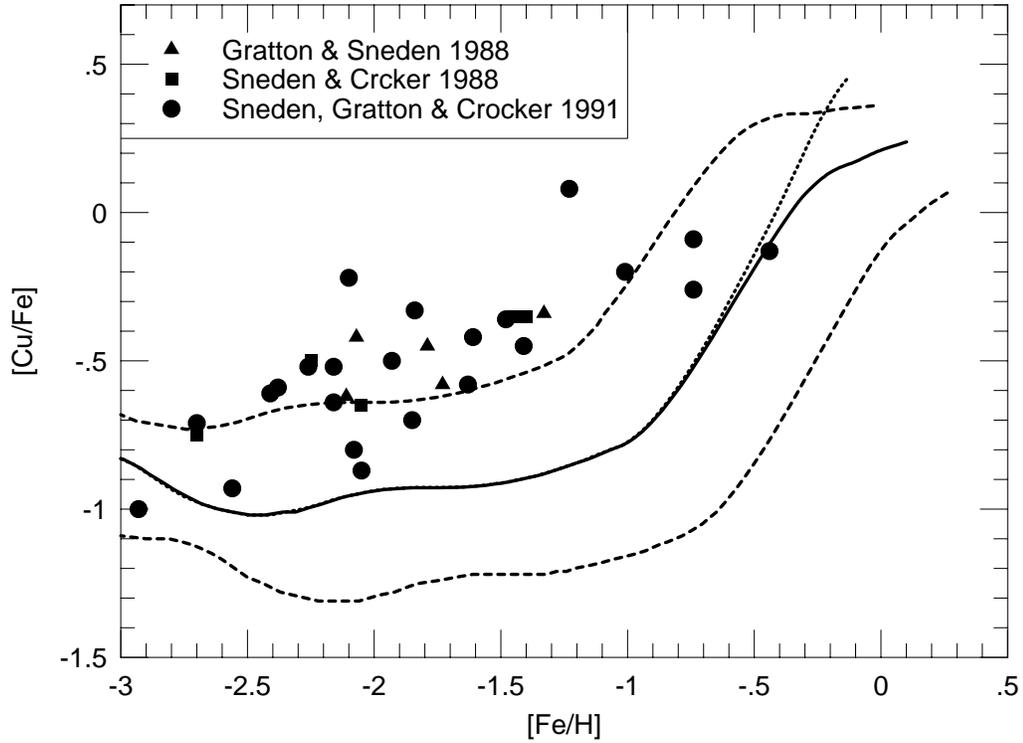

Fig. 34.— Evolution of the copper to iron ratio [Cu/Fe] as a function of [Fe/H]. The solid line shows the calculation; the dashed lines indicate factors of two variation in the iron yields from massive stars; and the dotted line shows the evolution when Type Ia supernovae are excluded. The calculated evolution is in reasonable agreement with the observations of a deficient [Cu/Fe] ratio in metal-poor stars, and then increasing past the disk-halo transition point at [Fe/H] $\simeq$ -1.0 dex. Intermediate - low mass stars and Type Ia supernova contribute little or no copper in the calculations shown. The solar metallicity exploded massive star models have about a factor of 5 larger copper yield than the models with smaller initial metallicities. This overwhelms the iron contributions from Type Ia supernovae, and causes the increase in the computed [Cu/Fe] ratio for [Fe/H] > -1.0 dex (compare the solid and dotted curves). The solar abundances of the two stable copper isotopes are well accounted for by the presupernova and exploded star models (see Figure 4 and 5).

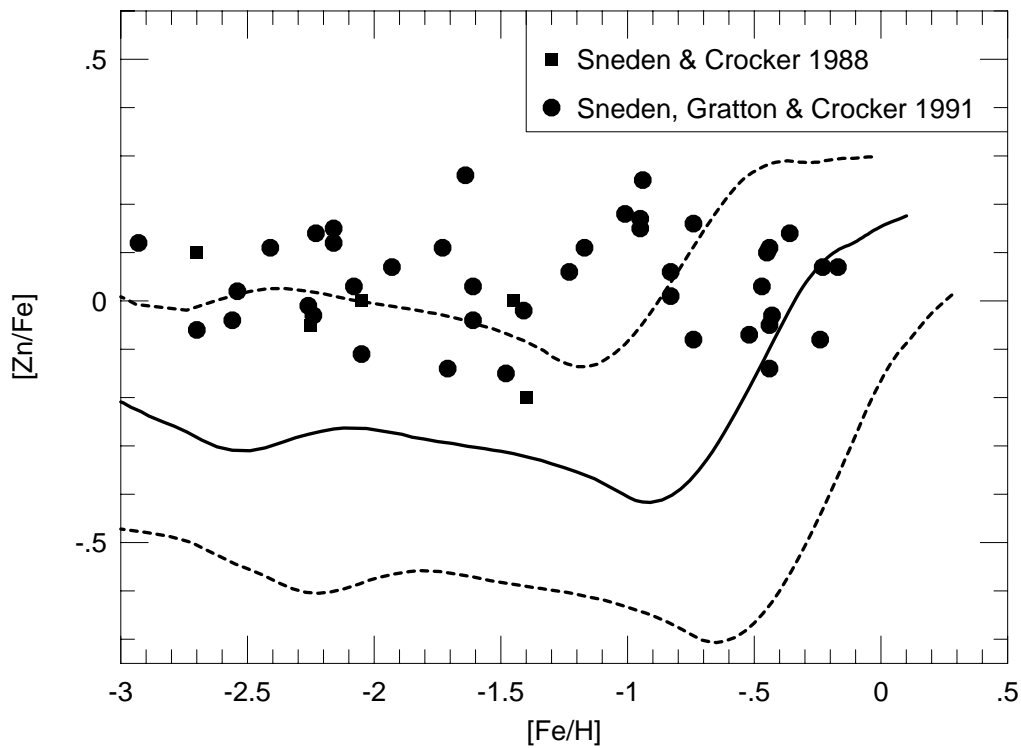

Fig. 35.— Evolution of the zinc to iron ratio [Zn/Fe] as a function of [Fe/H]. Again, the solid line shows the calculation, and the dashed lines indicate factors of two variation in the iron yields from massive stars. The calculated evolution is in reasonable agreement with the observations of a solar [Zn/Fe] ratio at low metallicities, although the computed history is smaller than the observations by about a factor of two. However, elemental zinc is dominated by the isotope $^{64}$Zn which is deficient (see Figures 4 and 5) in the exploded massive star models. This may be a consequence of the fact that the nuclear reaction network employed in the stellar evolution calculations was terminated at germanium, or an artifact of how the explosion was parameterized. Intermediate and low mass stars contribute no zinc or iron in the calculations shown, and standard carbon deflagration models for Type Ia supernovae produce little zinc. The solar metallicity massive star models have about a factor of 5 larger zinc yield than the models with smaller initial metallicities. This overwhelms the iron contributions from Type Ia supernovae, and causes the increase in the computed [Zn/Fe] ratio for [Fe/H] > -1.0 dex. The amplitude of the jump is decreased if more Type Ia supernovae are included.

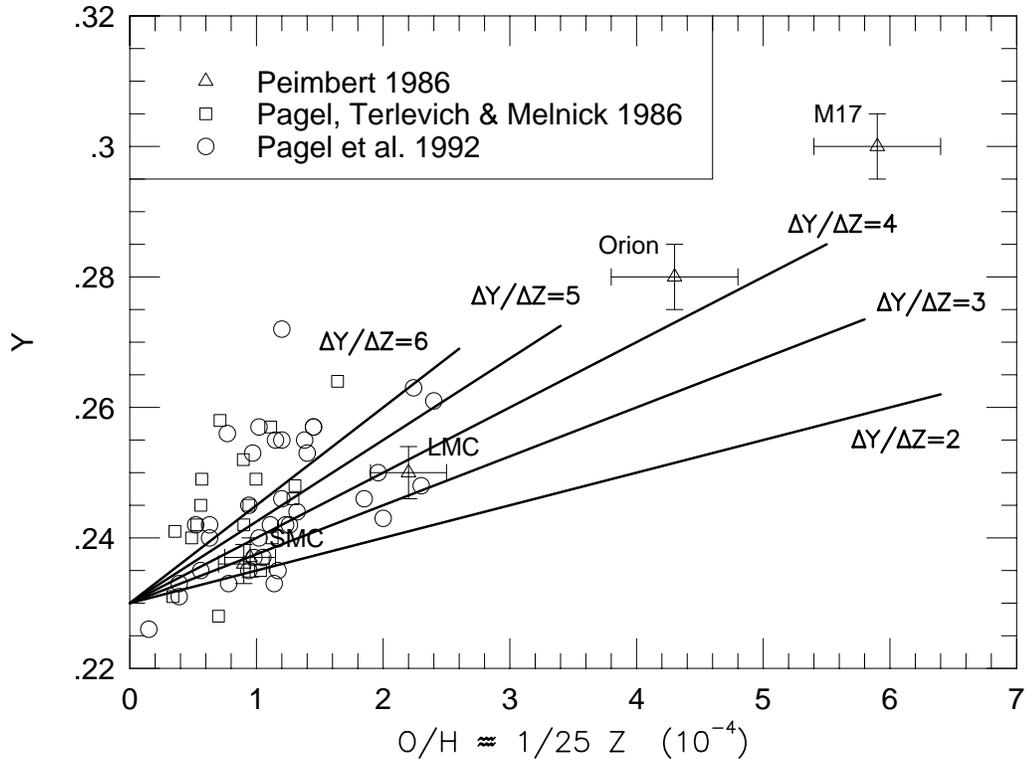

Fig. 36.— Observed helium abundances versus the O/H ratios found in Galactic and extragalactic H II regions. Only the error bars for the Peimbert (1986) survey are shown; the other surveys have similar error bars (see text). Reduction of the basic spectrographic data is especially demanding in these observations (e.g removal of Wolf-Rayet contamination and grain depletion). Transformation of the observed O/H ratios into total metallicity values is commonly done by assuming that oxygen constitutes some fraction the total metallicity. Also shown are the results of equation (20) for various slopes $\Delta Y/\Delta Z$ that that start from a primordial helium abundance of $Y_P$=0.23. Although the scatter is large, values of $\Delta Y/\Delta Z = 4.0 \pm 1.0$ are taken to be representative.

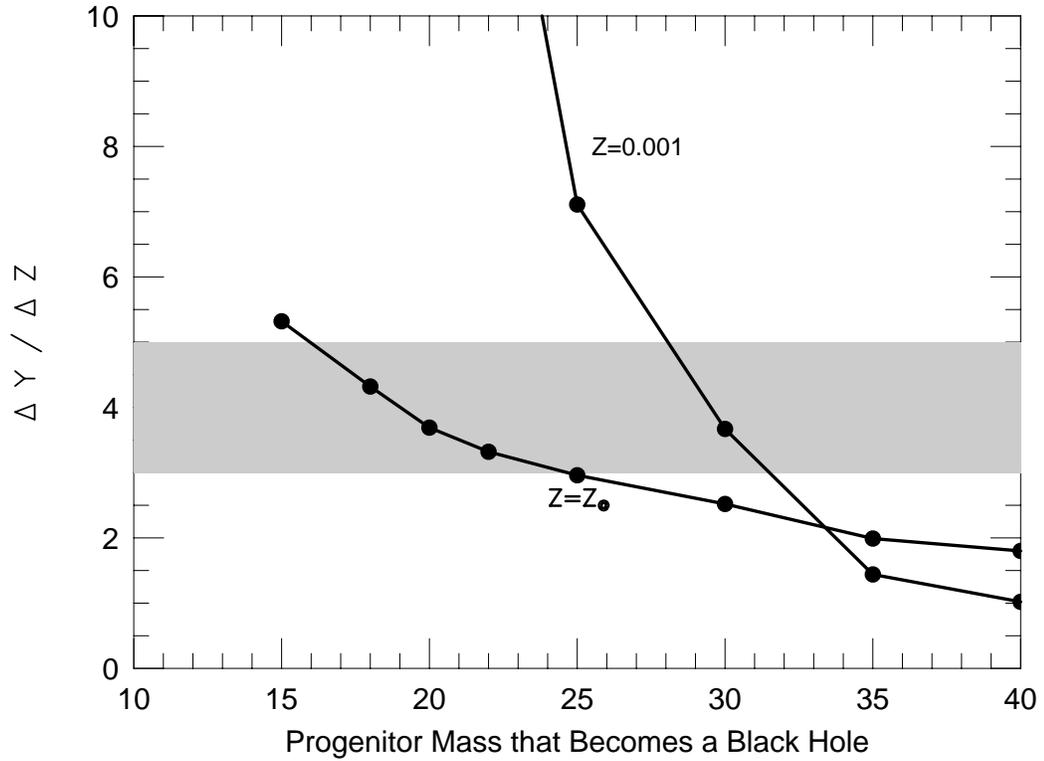

Fig. 37.— Helium to metal enrichment ratio as a function of the black hole mass limit. The observational limits inferred from Figure 36 are shown as the grey band. One curve is for a time very early on in the evolution when the total metallicity is $Z=10^{-3}$ $Z_\odot$ and another curve is at after 10.4 Gyr when $Z=Z_\odot$. The low metallicity curve, which is the cleanest constraints on the black hole mass cutoff (see text), suggests that the observed $\Delta Y/\Delta Z$ ratio is reproduced when stars above $\simeq 30$ $M_\odot$ become black holes. The solar metallicity curve indicates a much lower values of $\simeq 17$ $M_\odot$, but complications from the intervening 10.4 Gyr of chemical evolution place a large uncertainty on this value. The tentative conclusion, based on the low metallicity curve, is that the Galaxy may contain a large number of stellar mass black holes (see text for caveats).

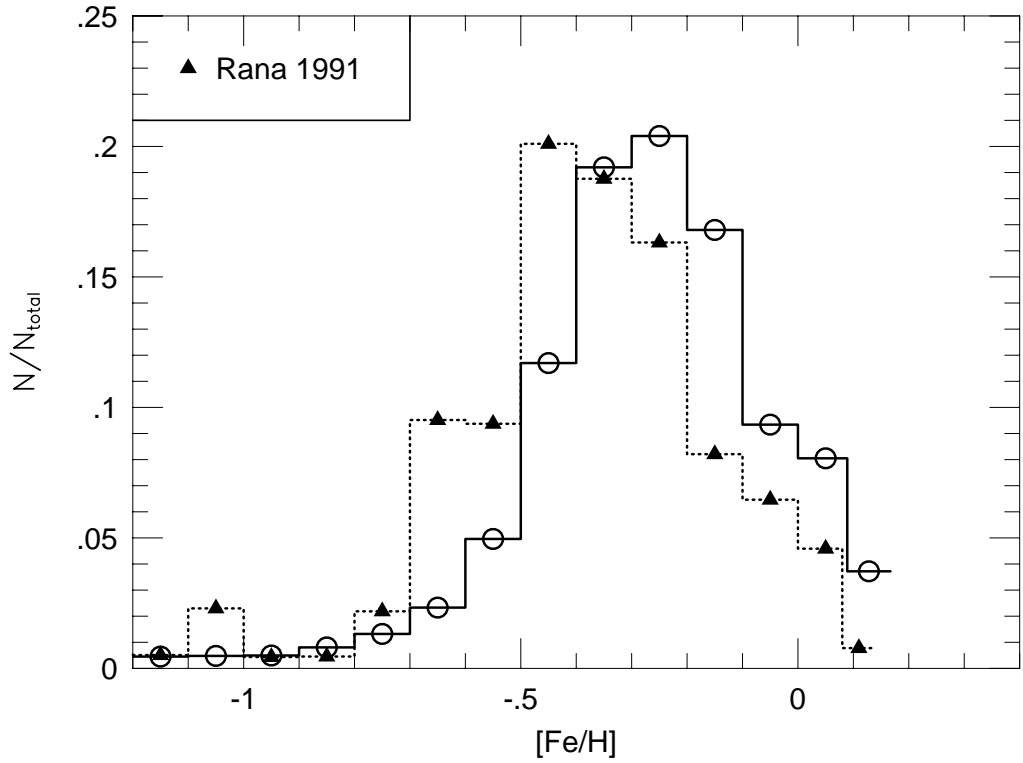

Fig. 38.— Cumulative solar neighborhood G-Dwarf distribution function. The dashed line histogram shows the cumulative G dwarf data reconstructed by Rana (1991), while the solid line histogram shows the calculation. Both histograms are normalized to the total number of G dwarf stars. The combination of infall of primordial or nearly unprocessed gas and dynamical heating of stellar populations satisfactorily reproduces the observed G dwarf distribution.

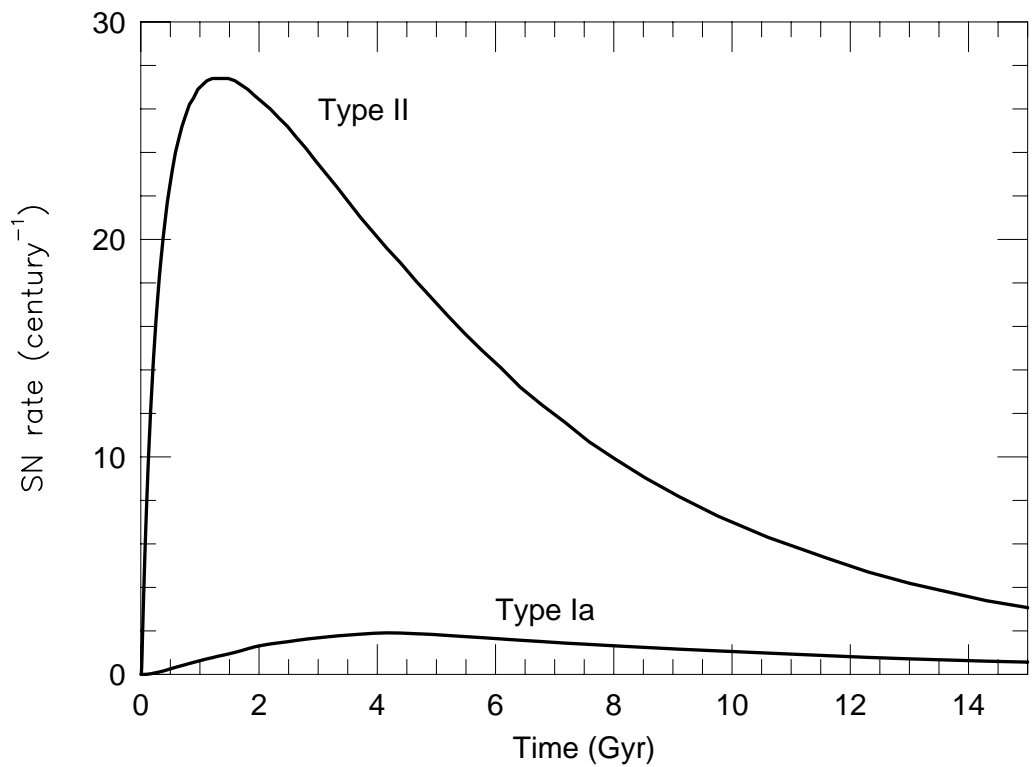

Fig. 39.— Number of supernova per century as a function of time. These rates are calculated by integrating in radius across the model Galaxy. As the evolution of the Galaxy begins, a large abundance of gas gets turned into a large number of stars. The death of these massive stars gives a correspondingly large Type II + Ib supernova rate. As the amount of gas decreases due to the formation of compact remnants and long-lived low mass stars the Type II +Ib supernova rate decreases with time. The onset of Type Ia supernovae is delayed due to the longer lifetime of the intermediate mass star which produces a white dwarf. After 15 Gyr, the calculated Type II + Ib supernova rate is 3.07 per century, and 0.53 for Type Ia supernova. For a total Galactic blue luminosity of $2.3 \times 10^{10}$ $L_\odot$, a Hubble constant of 75 km s$^{-1}$ Mpc$^{-1}$, and a Sbc Galactic morphology, these rates are in excellent agreement with the estimates of van den Bergh & Mcclure (1994).